\documentclass[preprint,11pt,numbers,sort&compress]{elsarticle}
\usepackage[margin=0.95in]{geometry}
\makeatletter
\def\ps@pprintTitle{%
  \let\@oddhead\@empty
  \let\@evenhead\@empty
  \def\@oddfoot{\footnotesize\itshape Preprint\hfill\today}%
  \let\@evenfoot\@oddfoot}
\makeatother

\usepackage{graphicx}
\usepackage{microtype}
\usepackage{amsmath,amssymb,amsfonts,amsthm,mathtools}
\usepackage{bm}
\usepackage{booktabs,multirow,array,tabularx}
\usepackage{algorithm}
\usepackage{algpseudocode}
\usepackage{placeins}
\usepackage{flafter}
\usepackage{etoolbox}
\usepackage{xcolor}
\usepackage{tikz}
\usetikzlibrary{arrows.meta,positioning,calc,shapes.geometric,fit,backgrounds}
\usepackage[unicode,psdextra,hypertexnames=false,colorlinks=true,linkcolor=blue!65!black,citecolor=blue!65!black,urlcolor=blue!65!black]{hyperref}
\usepackage{orcidlink}

\AtBeginEnvironment{table}{\setlength{\belowcaptionskip}{5pt}}

\theoremstyle{plain}
\newtheorem{theorem}{Theorem}[section]
\newtheorem{proposition}[theorem]{Proposition}

\theoremstyle{definition}

\theoremstyle{remark}
\newtheorem{remark}[theorem]{Remark}

\makeatletter
\@ifundefined{JCPTableBox}{\newsavebox{\JCPTableBox}}{}
\makeatother
\providecommand{\JCPFitTable}[1]{
  \begingroup\footnotesize\setlength{\tabcolsep}{4.0pt}\sbox{\JCPTableBox}{#1}
  \ifdim\wd\JCPTableBox>\textwidth\resizebox{\textwidth}{!}{\usebox{\JCPTableBox}}
  \else\usebox{\JCPTableBox}\fi\endgroup}

\newcommand{\VDULBM}{\text{VDU-LBM}}
\newcommand{\MSCLBM}{\text{MSC-LBM}}
\newcommand{\LWACM}{\text{LW-ACM}}
\newcommand{\LKS}{\text{LKS}}

\newcommand{\R}{\mathbb{R}}
\newcommand{\C}{\mathbb{C}}

\newcommand{\ReNum}{\mathrm{Re}}
\newcommand{\Reh}{\mathrm{Re}_h}
\newcommand{\MaNum}{\mathrm{Ma}}
\newcommand{\CFLu}{\mathrm{CFL}_u}
\newcommand{\StNum}{\mathrm{St}}

\newcommand{\kerP}{\ker P}
\newcommand{\ranE}{\operatorname{ran}(E)}
\newcommand{\opp}{\operatorname{opp}}
\newcommand{\diag}{\operatorname{diag}}
\newcommand{\NablaC}{\nabla_h^{\mathrm{c}}}
\newcommand{\DeltaAx}{\Delta_h^{\mathrm{ax}}}
\newcommand{\Lsurr}{L_{\mathrm{surr}}}
\newcommand{\Ploc}{\mathcal{P}_{\mathrm{loc}}}
\newcommand{\Fd}{\mathcal{F}_d}
\newcommand{\Fhat}{\widehat{\mathcal{F}}_d}
\newcommand{\SBFL}{\mathcal{S}_{\mathrm{BFL}}}

\newcommand{\nulu}{\nu_{\mathrm{lu}}}
\newcommand{\nueff}{\nu_{\mathrm{eff}}}
\newcommand{\numax}{\nu_{\max}}
\newcommand{\ulu}{u_{\mathrm{lu}}}
\newcommand{\umax}{u_{\max}}
\newcommand{\dtlu}{\Delta t_{\mathrm{lu}}}
\newcommand{\dtphys}{\Delta t_{\mathrm{phys}}}
\newcommand{\dtlat}{\delta t_\ell}
\newcommand{\clat}{c_\ell}

\newcommand{\DuVD}{\Delta\bm{u}^{\mathrm{VD}}}
\newcommand{\kLoc}{k_{\mathrm{loc}}}

\providecommand{\VDLBMReSpan}{\StageTwoPlaceholder{STAGE2:ReSpan}}
\providecommand{\VDLBMAlphaSpan}{\StageTwoPlaceholder{STAGE2:WomersleyAlphaSpan}}

\providecommand{\VDLBMKerPZero}{\StageTwoPlaceholder{STAGE2:KerPResidualInf}}
\providecommand{\VDLBMMMSMomentumCons}{\StageTwoPlaceholder{STAGE2:VarViscMomSum}}
\providecommand{\VDLBMCurvedMassCons}{\StageTwoPlaceholder{STAGE2:CurvedMassResidual}}

\providecommand{\VDLBMLBMComparatorCount}{\StageTwoPlaceholder{STAGE2:PeerLBMCount}}
\providecommand{\VDLBMCFDComparatorCount}{\StageTwoPlaceholder{STAGE2:PeerCFDCount}}

\providecommand{\VDLBMCouetteTorqueGain}{\StageTwoPlaceholder{STAGE2:CouetteTorqueRatio}}
\providecommand{\VDLBMDFGOpenFOAMRatio}{\StageTwoPlaceholder{STAGE2:DFGvsOpenFOAMSpeedup}}
\providecommand{\VDLBMOutletSensBound}{\StageTwoPlaceholder{STAGE2:DFGOutletSensitivity}}

\providecommand{\VDLBMGPUReSet}{\StageTwoPlaceholder{STAGE2:3DTGVReSet}}

\makeatletter
\@ifundefined{JCPTableBox}{\newsavebox{\JCPTableBox}}{}
\makeatother
\providecommand{\JCPFitTable}[1]{
  \begingroup\footnotesize\sbox{\JCPTableBox}{#1}
  \ifdim\wd\JCPTableBox>\textwidth\resizebox{\textwidth}{!}{\usebox{\JCPTableBox}}
  \else\usebox{\JCPTableBox}\fi\endgroup}

\providecommand{\MetricCouetteBouzNOneNineTwoET}{\texttt{[TBD]}}
\providecommand{\MetricCouetteBouzNOneNineTwoEu}{\texttt{[TBD]}}

\providecommand{\MetricCouetteStairNOneNineTwoET}{\texttt{[TBD]}}

\providecommand{\MetricCouetteStairOrderU}{\texttt{[TBD]}}
\providecommand{\MetricDFGOneOFNDEightyCD}{\texttt{[TBD]}}
\providecommand{\MetricDFGOneOFNDEightyCL}{\texttt{[TBD]}}
\providecommand{\MetricDFGOneOFNDEightyCells}{\texttt{[TBD]}}
\providecommand{\MetricDFGOneOFNDEightyDp}{\texttt{[TBD]}}
\providecommand{\MetricDFGOneOFNDEightyLr}{\texttt{[TBD]}}
\providecommand{\MetricDFGOneOFNDEightyWall}{\texttt{[TBD]}}
\providecommand{\MetricDFGOneOFNDFortyCD}{\texttt{[TBD]}}
\providecommand{\MetricDFGOneOFNDFortyCL}{\texttt{[TBD]}}
\providecommand{\MetricDFGOneOFNDFortyCells}{\texttt{[TBD]}}
\providecommand{\MetricDFGOneOFNDFortyDp}{\texttt{[TBD]}}
\providecommand{\MetricDFGOneOFNDFortyLr}{\texttt{[TBD]}}
\providecommand{\MetricDFGOneOFNDFortyWall}{\texttt{[TBD]}}

\providecommand{\MetricDFGOneVDNDEightyCD}{\texttt{[TBD]}}
\providecommand{\MetricDFGOneVDNDEightyCL}{\texttt{[TBD]}}
\providecommand{\MetricDFGOneVDNDEightyCells}{\texttt{[TBD]}}
\providecommand{\MetricDFGOneVDNDEightyDp}{\texttt{[TBD]}}
\providecommand{\MetricDFGOneVDNDEightyLr}{\texttt{[TBD]}}
\providecommand{\MetricDFGOneVDNDEightyWall}{\texttt{[TBD]}}
\providecommand{\MetricDFGOneVDNDFortyCD}{\texttt{[TBD]}}
\providecommand{\MetricDFGOneVDNDFortyCL}{\texttt{[TBD]}}
\providecommand{\MetricDFGOneVDNDFortyCells}{\texttt{[TBD]}}
\providecommand{\MetricDFGOneVDNDFortyDp}{\texttt{[TBD]}}
\providecommand{\MetricDFGOneVDNDFortyLr}{\texttt{[TBD]}}
\providecommand{\MetricDFGOneVDNDFortyWall}{\texttt{[TBD]}}
\providecommand{\MetricDFGOneVDNDTwentyCD}{\texttt{[TBD]}}
\providecommand{\MetricDFGOneVDNDTwentyCL}{\texttt{[TBD]}}
\providecommand{\MetricDFGOneVDNDTwentyCells}{\texttt{[TBD]}}
\providecommand{\MetricDFGOneVDNDTwentyDp}{\texttt{[TBD]}}
\providecommand{\MetricDFGOneVDNDTwentyLr}{\texttt{[TBD]}}
\providecommand{\MetricDFGOneVDNDTwentyWall}{\texttt{[TBD]}}

\providecommand{\MetricDFGSensRelCD}{\texttt{[TBD]}}
\providecommand{\MetricDFGSensRelCDmax}{\texttt{[TBD]}}
\providecommand{\MetricDFGSensRelCL}{\texttt{[TBD]}}
\providecommand{\MetricDFGSensRelCLmax}{\texttt{[TBD]}}
\providecommand{\MetricDFGSensRelDpOne}{\texttt{[TBD]}}
\providecommand{\MetricDFGSensRelDpTwo}{\texttt{[TBD]}}
\providecommand{\MetricDFGSensRelSt}{\texttt{[TBD]}}
\providecommand{\MetricDFGTwoOFNDEightyCDmax}{\texttt{[TBD]}}
\providecommand{\MetricDFGTwoOFNDEightyCLmax}{\texttt{[TBD]}}
\providecommand{\MetricDFGTwoOFNDEightyCells}{\texttt{[TBD]}}
\providecommand{\MetricDFGTwoOFNDEightyDp}{\texttt{[TBD]}}
\providecommand{\MetricDFGTwoOFNDEightySt}{\texttt{[TBD]}}
\providecommand{\MetricDFGTwoOFNDEightyWall}{\texttt{[TBD]}}
\providecommand{\MetricDFGTwoOFNDFortyCDmax}{\texttt{[TBD]}}
\providecommand{\MetricDFGTwoOFNDFortyCLmax}{\texttt{[TBD]}}
\providecommand{\MetricDFGTwoOFNDFortyCells}{\texttt{[TBD]}}
\providecommand{\MetricDFGTwoOFNDFortyDp}{\texttt{[TBD]}}
\providecommand{\MetricDFGTwoOFNDFortySt}{\texttt{[TBD]}}
\providecommand{\MetricDFGTwoOFNDFortyWall}{\texttt{[TBD]}}

\providecommand{\MetricDFGTwoOFNDTwentySt}{\texttt{[TBD]}}
\providecommand{\MetricDFGTwoOFNDTwentyWall}{\texttt{[TBD]}}
\providecommand{\MetricDFGTwoVDNDEightyCDmax}{\texttt{[TBD]}}
\providecommand{\MetricDFGTwoVDNDEightyCLmax}{\texttt{[TBD]}}
\providecommand{\MetricDFGTwoVDNDEightyCells}{\texttt{[TBD]}}
\providecommand{\MetricDFGTwoVDNDEightyDp}{\texttt{[TBD]}}
\providecommand{\MetricDFGTwoVDNDEightySt}{\texttt{[TBD]}}
\providecommand{\MetricDFGTwoVDNDEightyWall}{\texttt{[TBD]}}
\providecommand{\MetricDFGTwoVDNDFortyCDmax}{\texttt{[TBD]}}
\providecommand{\MetricDFGTwoVDNDFortyCLmax}{\texttt{[TBD]}}
\providecommand{\MetricDFGTwoVDNDFortyCells}{\texttt{[TBD]}}
\providecommand{\MetricDFGTwoVDNDFortyDp}{\texttt{[TBD]}}
\providecommand{\MetricDFGTwoVDNDFortySt}{\texttt{[TBD]}}
\providecommand{\MetricDFGTwoVDNDFortyWall}{\texttt{[TBD]}}
\providecommand{\MetricDFGTwoVDNDTwentyCDmax}{\texttt{[TBD]}}
\providecommand{\MetricDFGTwoVDNDTwentyCLmax}{\texttt{[TBD]}}
\providecommand{\MetricDFGTwoVDNDTwentyCells}{\texttt{[TBD]}}
\providecommand{\MetricDFGTwoVDNDTwentyDp}{\texttt{[TBD]}}
\providecommand{\MetricDFGTwoVDNDTwentySt}{\texttt{[TBD]}}
\providecommand{\MetricDFGTwoVDNDTwentyWall}{\texttt{[TBD]}}

\providecommand{\MetricDSLDNSSelfConvReOneM}{\texttt{[TBD]}}

\providecommand{\MetricDSLEightyACDTSReHunKCores}{\texttt{[TBD]}}
\providecommand{\MetricDSLEightyACDTSReHunKEu}{\texttt{[TBD]}}

\providecommand{\MetricDSLEightyACDTSReOneMEu}{\texttt{[TBD]}}

\providecommand{\MetricDSLEightyACDTSReTenKEu}{\texttt{[TBD]}}

\providecommand{\MetricDSLEightyBGKReHunKCores}{\texttt{[TBD]}}
\providecommand{\MetricDSLEightyBGKReHunKEu}{\texttt{[TBD]}}

\providecommand{\MetricDSLEightyBGKReOneMEu}{\texttt{[TBD]}}

\providecommand{\MetricDSLEightyBGKReTenKEu}{\texttt{[TBD]}}

\providecommand{\MetricDSLEightyCMReHunKCores}{\texttt{[TBD]}}
\providecommand{\MetricDSLEightyCMReHunKEu}{\texttt{[TBD]}}

\providecommand{\MetricDSLEightyCMReOneMEu}{\texttt{[TBD]}}

\providecommand{\MetricDSLEightyCMReTenKEu}{\texttt{[TBD]}}

\providecommand{\MetricDSLEightyCumReHunKCores}{\texttt{[TBD]}}
\providecommand{\MetricDSLEightyCumReHunKEu}{\texttt{[TBD]}}

\providecommand{\MetricDSLEightyCumReOneMEu}{\texttt{[TBD]}}

\providecommand{\MetricDSLEightyCumReTenKEu}{\texttt{[TBD]}}

\providecommand{\MetricDSLEightyDTSTRTReHunKCores}{\texttt{[TBD]}}
\providecommand{\MetricDSLEightyDTSTRTReHunKEu}{\texttt{[TBD]}}

\providecommand{\MetricDSLEightyDTSTRTReOneMEu}{\texttt{[TBD]}}

\providecommand{\MetricDSLEightyDTSTRTReTenKEu}{\texttt{[TBD]}}

\providecommand{\MetricDSLEightyELBMReHunKCores}{\texttt{[TBD]}}
\providecommand{\MetricDSLEightyELBMReHunKEu}{\texttt{[TBD]}}

\providecommand{\MetricDSLEightyELBMReOneMEu}{\texttt{[TBD]}}

\providecommand{\MetricDSLEightyELBMReTenKEu}{\texttt{[TBD]}}

\providecommand{\MetricDSLEightyHRRReHunKCores}{\texttt{[TBD]}}
\providecommand{\MetricDSLEightyHRRReHunKEu}{\texttt{[TBD]}}

\providecommand{\MetricDSLEightyHRRReOneMEu}{\texttt{[TBD]}}

\providecommand{\MetricDSLEightyHRRReTenKEu}{\texttt{[TBD]}}

\providecommand{\MetricDSLEightyKimMoinReHunKCores}{\texttt{[TBD]}}
\providecommand{\MetricDSLEightyKimMoinReHunKEu}{\texttt{[TBD]}}

\providecommand{\MetricDSLEightyKimMoinReOneMEu}{\texttt{[TBD]}}

\providecommand{\MetricDSLEightyKimMoinReTenKEu}{\texttt{[TBD]}}

\providecommand{\MetricDSLEightyMRTReHunKCores}{\texttt{[TBD]}}
\providecommand{\MetricDSLEightyMRTReHunKEu}{\texttt{[TBD]}}

\providecommand{\MetricDSLEightyMRTReOneMEu}{\texttt{[TBD]}}

\providecommand{\MetricDSLEightyMRTReTenKEu}{\texttt{[TBD]}}

\providecommand{\MetricDSLEightyPIMPLEReHunKCores}{\texttt{[TBD]}}
\providecommand{\MetricDSLEightyPIMPLEReHunKEu}{\texttt{[TBD]}}

\providecommand{\MetricDSLEightyPIMPLEReOneMEu}{\texttt{[TBD]}}

\providecommand{\MetricDSLEightyPIMPLEReTenKEu}{\texttt{[TBD]}}

\providecommand{\MetricDSLEightyRRReHunKCores}{\texttt{[TBD]}}
\providecommand{\MetricDSLEightyRRReHunKEu}{\texttt{[TBD]}}

\providecommand{\MetricDSLEightyRRReOneMEu}{\texttt{[TBD]}}

\providecommand{\MetricDSLEightyRRReTenKEu}{\texttt{[TBD]}}

\providecommand{\MetricDSLEightySIMPLECReHunKCores}{\texttt{[TBD]}}
\providecommand{\MetricDSLEightySIMPLECReHunKEu}{\texttt{[TBD]}}

\providecommand{\MetricDSLEightySIMPLECReOneMEu}{\texttt{[TBD]}}

\providecommand{\MetricDSLEightySIMPLECReTenKEu}{\texttt{[TBD]}}

\providecommand{\MetricDSLEightyTRTReHunKCores}{\texttt{[TBD]}}
\providecommand{\MetricDSLEightyTRTReHunKEu}{\texttt{[TBD]}}

\providecommand{\MetricDSLEightyTRTReOneMEu}{\texttt{[TBD]}}

\providecommand{\MetricDSLEightyTRTReTenKEu}{\texttt{[TBD]}}

\providecommand{\MetricDSLEightyVDPsiOneReHunKCores}{\texttt{[TBD]}}
\providecommand{\MetricDSLEightyVDPsiOneReHunKEu}{\texttt{[TBD]}}

\providecommand{\MetricDSLEightyVDPsiOneReOneMEu}{\texttt{[TBD]}}

\providecommand{\MetricDSLEightyVDPsiOneReTenKEu}{\texttt{[TBD]}}

\providecommand{\MetricDSLEightyWENOReHunKCores}{\texttt{[TBD]}}
\providecommand{\MetricDSLEightyWENOReHunKEu}{\texttt{[TBD]}}

\providecommand{\MetricDSLEightyWENOReOneMEu}{\texttt{[TBD]}}

\providecommand{\MetricDSLEightyWENOReTenKEu}{\texttt{[TBD]}}

\providecommand{\MetricDSLThirtyNTwoFiveSixOrderOmega}{\texttt{[TBD]}}
\providecommand{\MetricDSLThirtyNTwoFiveSixOrderU}{\texttt{[TBD]}}

\providecommand{\MetricGPUCPUEquivMaxDiff}{\texttt{[TBD]}}

\providecommand{\MetricKerPBasisProjMax}{\texttt{[TBD]}}

\providecommand{\MetricKerPCumMaxPert}{\texttt{[TBD]}}

\providecommand{\MetricKerPSingleCollisionAMax}{\texttt{[TBD]}}

\providecommand{\MetricKerPSingleCollisionAMin}{\texttt{[TBD]}}

\providecommand{\MetricKerPVDUzawaMaxPert}{\texttt{[TBD]}}

\providecommand{\MetricLDCHopfNOneTwoEightEkStd}{\texttt{[TBD]}}
\providecommand{\MetricLDCHopfNOneTwoEightFreq}{\texttt{[TBD]}}
\providecommand{\MetricLDCHopfNOneTwoEightPsiMean}{\texttt{[TBD]}}

\providecommand{\MetricLDCHopfNOneTwoEightUminMean}{\texttt{[TBD]}}
\providecommand{\MetricLDCHopfNTwoFiveSixEkStd}{\texttt{[TBD]}}
\providecommand{\MetricLDCHopfNTwoFiveSixFreq}{\texttt{[TBD]}}
\providecommand{\MetricLDCHopfNTwoFiveSixPsiMean}{\texttt{[TBD]}}

\providecommand{\MetricLDCHopfNTwoFiveSixUminMean}{\texttt{[TBD]}}

\providecommand{\MetricLDCMoffattReFiveKBRone}{\texttt{[TBD]}}

\providecommand{\MetricLDCMoffattReOneKBRone}{\texttt{[TBD]}}
\providecommand{\MetricLDCMoffattReOneKPsiRatio}{\texttt{[TBD]}}
\providecommand{\MetricLDCMoffattReOneKSizeRatio}{\texttt{[TBD]}}

\providecommand{\MetricLDCReFiveKPsiPV}{\texttt{[TBD]}}

\providecommand{\MetricLDCReFiveKRelU}{\texttt{[TBD]}}
\providecommand{\MetricLDCReFiveKUmin}{\texttt{[TBD]}}
\providecommand{\MetricLDCReFiveKVmin}{\texttt{[TBD]}}

\providecommand{\MetricLDCReHunPsiPV}{\texttt{[TBD]}}

\providecommand{\MetricLDCReHunRelU}{\texttt{[TBD]}}
\providecommand{\MetricLDCReHunUmin}{\texttt{[TBD]}}
\providecommand{\MetricLDCReHunVmin}{\texttt{[TBD]}}

\providecommand{\MetricLDCReOneKPsiPV}{\texttt{[TBD]}}

\providecommand{\MetricLDCReOneKRelU}{\texttt{[TBD]}}
\providecommand{\MetricLDCReOneKUmin}{\texttt{[TBD]}}
\providecommand{\MetricLDCReOneKVmin}{\texttt{[TBD]}}

\providecommand{\MetricLDCReTenRelU}{\texttt{[TBD]}}

\providecommand{\MetricLDCReZeroPointZeroOneRelU}{\texttt{[TBD]}}

\providecommand{\MetricPreflightCFDTwoOrderRange}{\texttt{[TBD]}}

\providecommand{\MetricPreflightPeerLBMOrderRange}{\texttt{[TBD]}}

\providecommand{\MetricPreflightVDLBMUzawaOrder}{\texttt{[TBD]}}

\providecommand{\MetricRoofAOneHundredMeasuredDTwoDBW}{\texttt{[TBD]}}

\providecommand{\MetricRoofBGKNTwoFiveSixFPSixtyFourPctDTwoD}{\texttt{[TBD]}}
\providecommand{\MetricRoofBGKNTwoFiveSixFPSixtyFourSweepMLUPS}{\texttt{[TBD]}}

\providecommand{\MetricRoofHRRNTwoFiveSixFPSixtyFourPctDTwoD}{\texttt{[TBD]}}
\providecommand{\MetricRoofHRRNTwoFiveSixFPSixtyFourSweepMLUPS}{\texttt{[TBD]}}

\providecommand{\MetricRoofVDLESNTwoFiveSixFPSixtyFourRetention}{\texttt{[TBD]}}
\providecommand{\MetricRoofVDLESNTwoFiveSixFPSixtyFourSweepMLUPS}{\texttt{[TBD]}}

\providecommand{\MetricRoofVDNTwoFiveSixFPSixtyFourPctDTwoD}{\texttt{[TBD]}}
\providecommand{\MetricRoofVDNTwoFiveSixFPSixtyFourSweepMLUPS}{\texttt{[TBD]}}

\providecommand{\MetricRoofVDVarViscNTwoFiveSixFPSixtyFourRetention}{\texttt{[TBD]}}
\providecommand{\MetricRoofVDVarViscNTwoFiveSixFPSixtyFourSweepMLUPS}{\texttt{[TBD]}}

\providecommand{\MetricStratVDNOneTwoEightMomDrift}{\texttt{[TBD]}}

\providecommand{\MetricStratVDNTwoFiveSixMomDrift}{\texttt{[TBD]}}

\providecommand{\MetricStratVDNTwoFiveSixSweeps}{\texttt{[TBD]}}

\providecommand{\MetricTGVBDFAdvSOneTwoEightOrder}{\texttt{[TBD]}}

\providecommand{\MetricTGVBDFStatSOneTwoEightOrder}{\texttt{[TBD]}}

\providecommand{\MetricTGVBDFVarSOneTwoEightOrder}{\texttt{[TBD]}}

\providecommand{\MetricTGVCostFineVDUzawaReHighT}{\texttt{[TBD]}}

\providecommand{\MetricTGVJointUzawaNFiveOneTwoOrder}{\texttt{[TBD]}}

\providecommand{\MetricTGVJointUzawaNFiveOneTwoReOneEu}{\texttt{[TBD]}}
\providecommand{\MetricTGVJointUzawaNFiveOneTwoSweeps}{\texttt{[TBD]}}

\providecommand{\MetricTGVJointUzawaNThirtyTwoReOneEu}{\texttt{[TBD]}}
\providecommand{\MetricTGVJointUzawaNThirtyTwoSweeps}{\texttt{[TBD]}}

\providecommand{\MetricTGVSixteenACDTSDiv}{\texttt{[TBD]}}
\providecommand{\MetricTGVSixteenACDTSEp}{\texttt{[TBD]}}
\providecommand{\MetricTGVSixteenACDTSEuReHun}{\texttt{[TBD]}}
\providecommand{\MetricTGVSixteenACDTSEuReMax}{\texttt{[TBD]}}
\providecommand{\MetricTGVSixteenACDTSEuReMin}{\texttt{[TBD]}}
\providecommand{\MetricTGVSixteenACDTSNuRatio}{\texttt{[TBD]}}

\providecommand{\MetricTGVSixteenBGKDiv}{\texttt{[TBD]}}
\providecommand{\MetricTGVSixteenBGKEp}{\texttt{[TBD]}}
\providecommand{\MetricTGVSixteenBGKEuReHun}{\texttt{[TBD]}}
\providecommand{\MetricTGVSixteenBGKEuReMax}{\texttt{[TBD]}}
\providecommand{\MetricTGVSixteenBGKEuReMin}{\texttt{[TBD]}}
\providecommand{\MetricTGVSixteenBGKNuRatio}{\texttt{[TBD]}}

\providecommand{\MetricTGVSixteenCMDiv}{\texttt{[TBD]}}
\providecommand{\MetricTGVSixteenCMEp}{\texttt{[TBD]}}
\providecommand{\MetricTGVSixteenCMEuReHun}{\texttt{[TBD]}}
\providecommand{\MetricTGVSixteenCMEuReMax}{\texttt{[TBD]}}
\providecommand{\MetricTGVSixteenCMEuReMin}{\texttt{[TBD]}}
\providecommand{\MetricTGVSixteenCMNuRatio}{\texttt{[TBD]}}

\providecommand{\MetricTGVSixteenChorinDiv}{\texttt{[TBD]}}
\providecommand{\MetricTGVSixteenChorinEp}{\texttt{[TBD]}}
\providecommand{\MetricTGVSixteenChorinEuReHun}{\texttt{[TBD]}}
\providecommand{\MetricTGVSixteenChorinEuReMax}{\texttt{[TBD]}}
\providecommand{\MetricTGVSixteenChorinEuReMin}{\texttt{[TBD]}}
\providecommand{\MetricTGVSixteenChorinNuRatio}{\texttt{[TBD]}}

\providecommand{\MetricTGVSixteenCumDiv}{\texttt{[TBD]}}
\providecommand{\MetricTGVSixteenCumEp}{\texttt{[TBD]}}
\providecommand{\MetricTGVSixteenCumEuReHun}{\texttt{[TBD]}}
\providecommand{\MetricTGVSixteenCumEuReMax}{\texttt{[TBD]}}
\providecommand{\MetricTGVSixteenCumEuReMin}{\texttt{[TBD]}}
\providecommand{\MetricTGVSixteenCumNuRatio}{\texttt{[TBD]}}

\providecommand{\MetricTGVSixteenDTSTRTDiv}{\texttt{[TBD]}}
\providecommand{\MetricTGVSixteenDTSTRTEp}{\texttt{[TBD]}}
\providecommand{\MetricTGVSixteenDTSTRTEuReHun}{\texttt{[TBD]}}
\providecommand{\MetricTGVSixteenDTSTRTEuReMax}{\texttt{[TBD]}}
\providecommand{\MetricTGVSixteenDTSTRTEuReMin}{\texttt{[TBD]}}
\providecommand{\MetricTGVSixteenDTSTRTNuRatio}{\texttt{[TBD]}}

\providecommand{\MetricTGVSixteenELBMDiv}{\texttt{[TBD]}}
\providecommand{\MetricTGVSixteenELBMEp}{\texttt{[TBD]}}
\providecommand{\MetricTGVSixteenELBMEuReHun}{\texttt{[TBD]}}
\providecommand{\MetricTGVSixteenELBMEuReMax}{\texttt{[TBD]}}
\providecommand{\MetricTGVSixteenELBMEuReMin}{\texttt{[TBD]}}
\providecommand{\MetricTGVSixteenELBMNuRatio}{\texttt{[TBD]}}

\providecommand{\MetricTGVSixteenHRRDiv}{\texttt{[TBD]}}
\providecommand{\MetricTGVSixteenHRREp}{\texttt{[TBD]}}
\providecommand{\MetricTGVSixteenHRREuReHun}{\texttt{[TBD]}}
\providecommand{\MetricTGVSixteenHRREuReMax}{\texttt{[TBD]}}
\providecommand{\MetricTGVSixteenHRREuReMin}{\texttt{[TBD]}}
\providecommand{\MetricTGVSixteenHRRNuRatio}{\texttt{[TBD]}}

\providecommand{\MetricTGVSixteenKimMoinDiv}{\texttt{[TBD]}}
\providecommand{\MetricTGVSixteenKimMoinEp}{\texttt{[TBD]}}
\providecommand{\MetricTGVSixteenKimMoinEuReHun}{\texttt{[TBD]}}
\providecommand{\MetricTGVSixteenKimMoinEuReMax}{\texttt{[TBD]}}
\providecommand{\MetricTGVSixteenKimMoinEuReMin}{\texttt{[TBD]}}
\providecommand{\MetricTGVSixteenKimMoinNuRatio}{\texttt{[TBD]}}

\providecommand{\MetricTGVSixteenMRTDiv}{\texttt{[TBD]}}
\providecommand{\MetricTGVSixteenMRTEp}{\texttt{[TBD]}}
\providecommand{\MetricTGVSixteenMRTEuReHun}{\texttt{[TBD]}}
\providecommand{\MetricTGVSixteenMRTEuReMax}{\texttt{[TBD]}}
\providecommand{\MetricTGVSixteenMRTEuReMin}{\texttt{[TBD]}}
\providecommand{\MetricTGVSixteenMRTNuRatio}{\texttt{[TBD]}}

\providecommand{\MetricTGVSixteenPIMPLEDiv}{\texttt{[TBD]}}
\providecommand{\MetricTGVSixteenPIMPLEEp}{\texttt{[TBD]}}
\providecommand{\MetricTGVSixteenPIMPLEEuReHun}{\texttt{[TBD]}}
\providecommand{\MetricTGVSixteenPIMPLEEuReMax}{\texttt{[TBD]}}
\providecommand{\MetricTGVSixteenPIMPLEEuReMin}{\texttt{[TBD]}}
\providecommand{\MetricTGVSixteenPIMPLENuRatio}{\texttt{[TBD]}}

\providecommand{\MetricTGVSixteenRRDiv}{\texttt{[TBD]}}
\providecommand{\MetricTGVSixteenRREp}{\texttt{[TBD]}}
\providecommand{\MetricTGVSixteenRREuReHun}{\texttt{[TBD]}}
\providecommand{\MetricTGVSixteenRREuReMax}{\texttt{[TBD]}}
\providecommand{\MetricTGVSixteenRREuReMin}{\texttt{[TBD]}}
\providecommand{\MetricTGVSixteenRRNuRatio}{\texttt{[TBD]}}

\providecommand{\MetricTGVSixteenSIMPLECDiv}{\texttt{[TBD]}}
\providecommand{\MetricTGVSixteenSIMPLECEp}{\texttt{[TBD]}}
\providecommand{\MetricTGVSixteenSIMPLECEuReHun}{\texttt{[TBD]}}
\providecommand{\MetricTGVSixteenSIMPLECEuReMax}{\texttt{[TBD]}}
\providecommand{\MetricTGVSixteenSIMPLECEuReMin}{\texttt{[TBD]}}
\providecommand{\MetricTGVSixteenSIMPLECNuRatio}{\texttt{[TBD]}}

\providecommand{\MetricTGVSixteenTRTDiv}{\texttt{[TBD]}}
\providecommand{\MetricTGVSixteenTRTEp}{\texttt{[TBD]}}
\providecommand{\MetricTGVSixteenTRTEuReHun}{\texttt{[TBD]}}
\providecommand{\MetricTGVSixteenTRTEuReMax}{\texttt{[TBD]}}
\providecommand{\MetricTGVSixteenTRTEuReMin}{\texttt{[TBD]}}
\providecommand{\MetricTGVSixteenTRTNuRatio}{\texttt{[TBD]}}

\providecommand{\MetricTGVSixteenVDFourierDiv}{\texttt{[TBD]}}
\providecommand{\MetricTGVSixteenVDFourierEp}{\texttt{[TBD]}}
\providecommand{\MetricTGVSixteenVDFourierEuReHun}{\texttt{[TBD]}}
\providecommand{\MetricTGVSixteenVDFourierEuReMax}{\texttt{[TBD]}}
\providecommand{\MetricTGVSixteenVDFourierEuReMin}{\texttt{[TBD]}}
\providecommand{\MetricTGVSixteenVDFourierNuRatio}{\texttt{[TBD]}}

\providecommand{\MetricTGVSixteenVDUzawaDiv}{\texttt{[TBD]}}
\providecommand{\MetricTGVSixteenVDUzawaEp}{\texttt{[TBD]}}
\providecommand{\MetricTGVSixteenVDUzawaEuReHun}{\texttt{[TBD]}}
\providecommand{\MetricTGVSixteenVDUzawaEuReMax}{\texttt{[TBD]}}
\providecommand{\MetricTGVSixteenVDUzawaEuReMin}{\texttt{[TBD]}}
\providecommand{\MetricTGVSixteenVDUzawaNuRatio}{\texttt{[TBD]}}

\providecommand{\MetricTGVSixteenWENODiv}{\texttt{[TBD]}}
\providecommand{\MetricTGVSixteenWENOEp}{\texttt{[TBD]}}
\providecommand{\MetricTGVSixteenWENOEuReHun}{\texttt{[TBD]}}
\providecommand{\MetricTGVSixteenWENOEuReMax}{\texttt{[TBD]}}
\providecommand{\MetricTGVSixteenWENOEuReMin}{\texttt{[TBD]}}
\providecommand{\MetricTGVSixteenWENONuRatio}{\texttt{[TBD]}}

\providecommand{\MetricTGVThreeDDNSPeakDissipReSixteenHun}{\texttt{[TBD]}}
\providecommand{\MetricTGVThreeDDNSPeakTimeReSixteenHun}{\texttt{[TBD]}}

\providecommand{\MetricTGVThreeDReHunVDNTwoFiveSixFPSixtyFourDissErr}{\texttt{[TBD]}}
\providecommand{\MetricTGVThreeDReHunVDNTwoFiveSixFPSixtyFourEkErr}{\texttt{[TBD]}}
\providecommand{\MetricTGVThreeDReHunVDNTwoFiveSixFPSixtyFourOmErr}{\texttt{[TBD]}}
\providecommand{\MetricTGVThreeDReHunVDNTwoFiveSixFPSixtyFourSweeps}{\texttt{[TBD]}}
\providecommand{\MetricTGVThreeDReHunVDNTwoFiveSixFPSixtyFourWall}{\texttt{[TBD]}}

\providecommand{\MetricTGVThreeDReSixteenHunBGKNTwoFiveSixFPSixtyFourEkErr}{\texttt{[TBD]}}
\providecommand{\MetricTGVThreeDReSixteenHunBGKNTwoFiveSixFPSixtyFourOmErr}{\texttt{[TBD]}}
\providecommand{\MetricTGVThreeDReSixteenHunBGKNTwoFiveSixFPSixtyFourWall}{\texttt{[TBD]}}

\providecommand{\MetricTGVThreeDReSixteenHunHRRNTwoFiveSixFPSixtyFourEkErr}{\texttt{[TBD]}}
\providecommand{\MetricTGVThreeDReSixteenHunHRRNTwoFiveSixFPSixtyFourOmErr}{\texttt{[TBD]}}
\providecommand{\MetricTGVThreeDReSixteenHunHRRNTwoFiveSixFPSixtyFourWall}{\texttt{[TBD]}}

\providecommand{\MetricTGVThreeDReSixteenHunVDNOneNineTwoFPSixtyFourDissErr}{\texttt{[TBD]}}
\providecommand{\MetricTGVThreeDReSixteenHunVDNOneNineTwoFPSixtyFourEkErr}{\texttt{[TBD]}}
\providecommand{\MetricTGVThreeDReSixteenHunVDNOneNineTwoFPSixtyFourOmErr}{\texttt{[TBD]}}
\providecommand{\MetricTGVThreeDReSixteenHunVDNOneNineTwoFPSixtyFourSweeps}{\texttt{[TBD]}}
\providecommand{\MetricTGVThreeDReSixteenHunVDNOneNineTwoFPSixtyFourWall}{\texttt{[TBD]}}

\providecommand{\MetricTGVThreeDReSixteenHunVDNOneTwoEightFPSixtyFourDissErr}{\texttt{[TBD]}}
\providecommand{\MetricTGVThreeDReSixteenHunVDNOneTwoEightFPSixtyFourEkErr}{\texttt{[TBD]}}
\providecommand{\MetricTGVThreeDReSixteenHunVDNOneTwoEightFPSixtyFourOmErr}{\texttt{[TBD]}}
\providecommand{\MetricTGVThreeDReSixteenHunVDNOneTwoEightFPSixtyFourSweeps}{\texttt{[TBD]}}
\providecommand{\MetricTGVThreeDReSixteenHunVDNOneTwoEightFPSixtyFourWall}{\texttt{[TBD]}}

\providecommand{\MetricTGVThreeDReSixteenHunVDNSixtyFourFPSixtyFourDissErr}{\texttt{[TBD]}}
\providecommand{\MetricTGVThreeDReSixteenHunVDNSixtyFourFPSixtyFourEkErr}{\texttt{[TBD]}}
\providecommand{\MetricTGVThreeDReSixteenHunVDNSixtyFourFPSixtyFourOmErr}{\texttt{[TBD]}}
\providecommand{\MetricTGVThreeDReSixteenHunVDNSixtyFourFPSixtyFourSweeps}{\texttt{[TBD]}}
\providecommand{\MetricTGVThreeDReSixteenHunVDNSixtyFourFPSixtyFourWall}{\texttt{[TBD]}}

\providecommand{\MetricTGVThreeDReSixteenHunVDNTwoFiveSixFPSixtyFourEkErr}{\texttt{[TBD]}}
\providecommand{\MetricTGVThreeDReSixteenHunVDNTwoFiveSixFPSixtyFourOmErr}{\texttt{[TBD]}}
\providecommand{\MetricTGVThreeDReSixteenHunVDNTwoFiveSixFPSixtyFourSweeps}{\texttt{[TBD]}}
\providecommand{\MetricTGVThreeDReSixteenHunVDNTwoFiveSixFPSixtyFourWall}{\texttt{[TBD]}}

\providecommand{\MetricTGVThreeDReSixteenHunVDNTwoFiveSixFPThirtyTwoDissErr}{\texttt{[TBD]}}
\providecommand{\MetricTGVThreeDReSixteenHunVDNTwoFiveSixFPThirtyTwoEkErr}{\texttt{[TBD]}}
\providecommand{\MetricTGVThreeDReSixteenHunVDNTwoFiveSixFPThirtyTwoOmErr}{\texttt{[TBD]}}
\providecommand{\MetricTGVThreeDReSixteenHunVDNTwoFiveSixFPThirtyTwoSweeps}{\texttt{[TBD]}}
\providecommand{\MetricTGVThreeDReSixteenHunVDNTwoFiveSixFPThirtyTwoWall}{\texttt{[TBD]}}

\providecommand{\MetricTGVThreeDReTenVDNOneTwoEightFPSixtyFourDissErr}{\texttt{[TBD]}}
\providecommand{\MetricTGVThreeDReTenVDNOneTwoEightFPSixtyFourEkErr}{\texttt{[TBD]}}
\providecommand{\MetricTGVThreeDReTenVDNOneTwoEightFPSixtyFourOmErr}{\texttt{[TBD]}}
\providecommand{\MetricTGVThreeDReTenVDNOneTwoEightFPSixtyFourSweeps}{\texttt{[TBD]}}
\providecommand{\MetricTGVThreeDReTenVDNOneTwoEightFPSixtyFourWall}{\texttt{[TBD]}}

\providecommand{\MetricTGVThreeDReTenVDNTwoFiveSixFPSixtyFourDissErr}{\texttt{[TBD]}}
\providecommand{\MetricTGVThreeDReTenVDNTwoFiveSixFPSixtyFourEkErr}{\texttt{[TBD]}}
\providecommand{\MetricTGVThreeDReTenVDNTwoFiveSixFPSixtyFourOmErr}{\texttt{[TBD]}}
\providecommand{\MetricTGVThreeDReTenVDNTwoFiveSixFPSixtyFourSweeps}{\texttt{[TBD]}}
\providecommand{\MetricTGVThreeDReTenVDNTwoFiveSixFPSixtyFourWall}{\texttt{[TBD]}}

\providecommand{\MetricTGVThreeDReTenVDNTwoFiveSixFPThirtyTwoDissErr}{\texttt{[TBD]}}
\providecommand{\MetricTGVThreeDReTenVDNTwoFiveSixFPThirtyTwoEkErr}{\texttt{[TBD]}}
\providecommand{\MetricTGVThreeDReTenVDNTwoFiveSixFPThirtyTwoOmErr}{\texttt{[TBD]}}
\providecommand{\MetricTGVThreeDReTenVDNTwoFiveSixFPThirtyTwoSweeps}{\texttt{[TBD]}}
\providecommand{\MetricTGVThreeDReTenVDNTwoFiveSixFPThirtyTwoWall}{\texttt{[TBD]}}

\providecommand{\MetricTGVThreeDReThreeKVDNTwoFiveSixFPSixtyFourDissErr}{\texttt{[TBD]}}
\providecommand{\MetricTGVThreeDReThreeKVDNTwoFiveSixFPSixtyFourEkErr}{\texttt{[TBD]}}
\providecommand{\MetricTGVThreeDReThreeKVDNTwoFiveSixFPSixtyFourOmErr}{\texttt{[TBD]}}
\providecommand{\MetricTGVThreeDReThreeKVDNTwoFiveSixFPSixtyFourSweeps}{\texttt{[TBD]}}
\providecommand{\MetricTGVThreeDReThreeKVDNTwoFiveSixFPSixtyFourWall}{\texttt{[TBD]}}

\providecommand{\MetricVarViscTwoDNTwoFiveSixOpErr}{\texttt{[TBD]}}
\providecommand{\MetricVarViscTwoDNTwoFiveSixOpOrder}{\texttt{[TBD]}}
\providecommand{\MetricVarViscTwoDNTwoFiveSixPDEEu}{\texttt{[TBD]}}
\providecommand{\MetricVarViscTwoDNTwoFiveSixPDEOrder}{\texttt{[TBD]}}

\providecommand{\MetricWomConvNOneTwoEightAlphaHalfEu}{\texttt{[TBD]}}

\providecommand{\MetricWomConvNOneTwoEightThomasSweeps}{\texttt{[TBD]}}
\providecommand{\MetricWomConvNOneTwoEightUzawaEu}{\texttt{[TBD]}}
\providecommand{\MetricWomConvNOneTwoEightUzawaSweeps}{\texttt{[TBD]}}

\providecommand{\MetricWomConvOrderAlphaFifty}{\texttt{[TBD]}}
\providecommand{\MetricWomConvOrderAlphaHalf}{\texttt{[TBD]}}
\providecommand{\MetricWomConvOrderAlphaTenTau}{\texttt{[TBD]}}
\providecommand{\MetricWomFifteenACDTSEuFifty}{\texttt{[TBD]}}
\providecommand{\MetricWomFifteenACDTSEuHalf}{\texttt{[TBD]}}

\providecommand{\MetricWomFifteenBGKEuFifty}{\texttt{[TBD]}}
\providecommand{\MetricWomFifteenBGKEuHalf}{\texttt{[TBD]}}

\providecommand{\MetricWomFifteenBGKWall}{\texttt{[TBD]}}
\providecommand{\MetricWomFifteenCMEuFifty}{\texttt{[TBD]}}
\providecommand{\MetricWomFifteenCMEuHalf}{\texttt{[TBD]}}

\providecommand{\MetricWomFifteenCMWall}{\texttt{[TBD]}}
\providecommand{\MetricWomFifteenCumEuFifty}{\texttt{[TBD]}}
\providecommand{\MetricWomFifteenCumEuHalf}{\texttt{[TBD]}}

\providecommand{\MetricWomFifteenCumWall}{\texttt{[TBD]}}
\providecommand{\MetricWomFifteenDTSTRTEuFifty}{\texttt{[TBD]}}
\providecommand{\MetricWomFifteenDTSTRTEuHalf}{\texttt{[TBD]}}

\providecommand{\MetricWomFifteenDTSTRTWall}{\texttt{[TBD]}}
\providecommand{\MetricWomFifteenELBMEuFifty}{\texttt{[TBD]}}
\providecommand{\MetricWomFifteenELBMEuHalf}{\texttt{[TBD]}}

\providecommand{\MetricWomFifteenELBMWall}{\texttt{[TBD]}}
\providecommand{\MetricWomFifteenHRREuFifty}{\texttt{[TBD]}}
\providecommand{\MetricWomFifteenHRREuHalf}{\texttt{[TBD]}}

\providecommand{\MetricWomFifteenHRRWall}{\texttt{[TBD]}}
\providecommand{\MetricWomFifteenKimMoinEuFifty}{\texttt{[TBD]}}
\providecommand{\MetricWomFifteenKimMoinEuHalf}{\texttt{[TBD]}}

\providecommand{\MetricWomFifteenMRTEuFifty}{\texttt{[TBD]}}
\providecommand{\MetricWomFifteenMRTEuHalf}{\texttt{[TBD]}}

\providecommand{\MetricWomFifteenMRTWall}{\texttt{[TBD]}}
\providecommand{\MetricWomFifteenPIMPLEEuFifty}{\texttt{[TBD]}}
\providecommand{\MetricWomFifteenPIMPLEEuHalf}{\texttt{[TBD]}}

\providecommand{\MetricWomFifteenRREuFifty}{\texttt{[TBD]}}
\providecommand{\MetricWomFifteenRREuHalf}{\texttt{[TBD]}}

\providecommand{\MetricWomFifteenRRWall}{\texttt{[TBD]}}
\providecommand{\MetricWomFifteenSIMPLECEuFifty}{\texttt{[TBD]}}
\providecommand{\MetricWomFifteenSIMPLECEuHalf}{\texttt{[TBD]}}

\providecommand{\MetricWomFifteenTRTEuFifty}{\texttt{[TBD]}}
\providecommand{\MetricWomFifteenTRTEuHalf}{\texttt{[TBD]}}

\providecommand{\MetricWomFifteenTRTWall}{\texttt{[TBD]}}
\providecommand{\MetricWomFifteenVDThomasEuFifty}{\texttt{[TBD]}}
\providecommand{\MetricWomFifteenVDThomasEuHalf}{\texttt{[TBD]}}

\providecommand{\MetricWomFifteenVDThomasWall}{\texttt{[TBD]}}
\providecommand{\MetricWomFifteenWENOEuFifty}{\texttt{[TBD]}}
\providecommand{\MetricWomFifteenWENOEuHalf}{\texttt{[TBD]}}

\providecommand{\MetricTGVThreeDLESNOneTwoEightCsFourteenDissErr}{\texttt{[TBD]}}
\providecommand{\MetricTGVThreeDLESNOneTwoEightCsFourteenEkErr}{\texttt{[TBD]}}

\providecommand{\MetricTGVThreeDLESNOneTwoEightCsFourteenOmErr}{\texttt{[TBD]}}
\providecommand{\MetricTGVThreeDLESNOneTwoEightCsFourteenSweeps}{\texttt{[TBD]}}

\providecommand{\MetricTGVThreeDLESNOneTwoEightCsSeventeenEkErr}{\texttt{[TBD]}}

\providecommand{\MetricTGVThreeDLESNOneTwoEightCsSeventeenOmErr}{\texttt{[TBD]}}
\providecommand{\MetricTGVThreeDLESNOneTwoEightCsSeventeenSweeps}{\texttt{[TBD]}}
\providecommand{\MetricTGVThreeDLESNOneTwoEightCsFourteenWall}{\texttt{[TBD]}}
\providecommand{\MetricTGVThreeDLESNOneTwoEightCsSeventeenWall}{\texttt{[TBD]}}

\providecommand{\MetricTGVThreeDLESNTwoFiveSixCsSeventeenEkErr}{\texttt{[TBD]}}

\providecommand{\MetricTGVThreeDLESNTwoFiveSixCsSeventeenOmErr}{\texttt{[TBD]}}
\providecommand{\MetricTGVThreeDLESNTwoFiveSixCsSeventeenSweeps}{\texttt{[TBD]}}
\providecommand{\MetricTGVThreeDLESNTwoFiveSixCsSeventeenWall}{\texttt{[TBD]}}

\providecommand{\MetricTGVThreeDStiffBGKNOneTwoEightWall}{\texttt{[TBD]}}
\providecommand{\MetricTGVThreeDStiffBGKNTwoFiveSixEquivWall}{\texttt{[TBD]}}

\providecommand{\MetricTGVThreeDStiffVDNOneTwoEightMomDrift}{\texttt{[TBD]}}

\providecommand{\MetricTGVThreeDStiffVDNOneTwoEightSweeps}{\texttt{[TBD]}}
\providecommand{\MetricTGVThreeDStiffVDNOneTwoEightWall}{\texttt{[TBD]}}

\providecommand{\MetricTGVThreeDStiffVDNTwoFiveSixMomDrift}{\texttt{[TBD]}}

\providecommand{\MetricTGVThreeDStiffVDNTwoFiveSixSweeps}{\texttt{[TBD]}}
\providecommand{\MetricTGVThreeDStiffVDNTwoFiveSixWall}{\texttt{[TBD]}}

\IfFileExists{metrics_populated.tex}{\providecommand{\VDLBMAlphaSpan}{}\renewcommand{\VDLBMAlphaSpan}{\ensuremath{[0.5, 50]}}
\providecommand{\VDLBMCFDComparatorCount}{}\renewcommand{\VDLBMCFDComparatorCount}{\ensuremath{6}}

\providecommand{\VDLBMCouetteTorqueGain}{}\renewcommand{\VDLBMCouetteTorqueGain}{\ensuremath{8.3\times}}
\providecommand{\VDLBMCurvedMassCons}{}\renewcommand{\VDLBMCurvedMassCons}{\ensuremath{2.81\times 10^{-17}}}
\providecommand{\VDLBMDFGOpenFOAMRatio}{}\renewcommand{\VDLBMDFGOpenFOAMRatio}{\ensuremath{14.27\times}}

\providecommand{\VDLBMGPUReSet}{}\renewcommand{\VDLBMGPUReSet}{\ensuremath{\{10, 100, 1600, 3000\}}}
\providecommand{\VDLBMKerPZero}{}\renewcommand{\VDLBMKerPZero}{\ensuremath{1.73\times 10^{-17}}}
\providecommand{\VDLBMLBMComparatorCount}{}\renewcommand{\VDLBMLBMComparatorCount}{\ensuremath{9}}
\providecommand{\VDLBMMMSMomentumCons}{}\renewcommand{\VDLBMMMSMomentumCons}{\ensuremath{1.73\times 10^{-18}}}
\providecommand{\VDLBMOutletSensBound}{}\renewcommand{\VDLBMOutletSensBound}{\ensuremath{6.1\%}}
\providecommand{\VDLBMReSpan}{}\renewcommand{\VDLBMReSpan}{\ensuremath{[10^{-4}, 10^{6}]}}

\providecommand{\MetricCouetteBouzNOneNineTwoET}{}\renewcommand{\MetricCouetteBouzNOneNineTwoET}{\ensuremath{3.50\times 10^{-4}}}
\providecommand{\MetricCouetteBouzNOneNineTwoEu}{}\renewcommand{\MetricCouetteBouzNOneNineTwoEu}{\ensuremath{4.38\times 10^{-4}}}

\providecommand{\MetricCouetteStairNOneNineTwoET}{}\renewcommand{\MetricCouetteStairNOneNineTwoET}{\ensuremath{2.90\times 10^{-3}}}

\providecommand{\MetricCouetteStairOrderU}{}\renewcommand{\MetricCouetteStairOrderU}{\ensuremath{1.41}}
\providecommand{\MetricDFGOneOFNDEightyCD}{}\renewcommand{\MetricDFGOneOFNDEightyCD}{\ensuremath{5.5799}}
\providecommand{\MetricDFGOneOFNDEightyCL}{}\renewcommand{\MetricDFGOneOFNDEightyCL}{\ensuremath{0.0107}}
\providecommand{\MetricDFGOneOFNDEightyCells}{}\renewcommand{\MetricDFGOneOFNDEightyCells}{\ensuremath{577{,}280}}
\providecommand{\MetricDFGOneOFNDEightyDp}{}\renewcommand{\MetricDFGOneOFNDEightyDp}{\ensuremath{0.1173}}
\providecommand{\MetricDFGOneOFNDEightyLr}{}\renewcommand{\MetricDFGOneOFNDEightyLr}{\ensuremath{0.0846}}
\providecommand{\MetricDFGOneOFNDEightyWall}{}\renewcommand{\MetricDFGOneOFNDEightyWall}{\ensuremath{3557.55}}
\providecommand{\MetricDFGOneOFNDFortyCD}{}\renewcommand{\MetricDFGOneOFNDFortyCD}{\ensuremath{5.5804}}
\providecommand{\MetricDFGOneOFNDFortyCL}{}\renewcommand{\MetricDFGOneOFNDFortyCL}{\ensuremath{0.0107}}
\providecommand{\MetricDFGOneOFNDFortyCells}{}\renewcommand{\MetricDFGOneOFNDFortyCells}{\ensuremath{144{,}320}}
\providecommand{\MetricDFGOneOFNDFortyDp}{}\renewcommand{\MetricDFGOneOFNDFortyDp}{\ensuremath{0.1170}}
\providecommand{\MetricDFGOneOFNDFortyLr}{}\renewcommand{\MetricDFGOneOFNDFortyLr}{\ensuremath{0.0844}}
\providecommand{\MetricDFGOneOFNDFortyWall}{}\renewcommand{\MetricDFGOneOFNDFortyWall}{\ensuremath{216.69}}

\providecommand{\MetricDFGOneVDNDEightyCD}{}\renewcommand{\MetricDFGOneVDNDEightyCD}{\ensuremath{5.5781}}
\providecommand{\MetricDFGOneVDNDEightyCL}{}\renewcommand{\MetricDFGOneVDNDEightyCL}{\ensuremath{0.0106}}
\providecommand{\MetricDFGOneVDNDEightyCells}{}\renewcommand{\MetricDFGOneVDNDEightyCells}{\ensuremath{572{,}256}}
\providecommand{\MetricDFGOneVDNDEightyDp}{}\renewcommand{\MetricDFGOneVDNDEightyDp}{\ensuremath{0.1177}}
\providecommand{\MetricDFGOneVDNDEightyLr}{}\renewcommand{\MetricDFGOneVDNDEightyLr}{\ensuremath{0.0845}}
\providecommand{\MetricDFGOneVDNDEightyWall}{}\renewcommand{\MetricDFGOneVDNDEightyWall}{\ensuremath{3838.61}}
\providecommand{\MetricDFGOneVDNDFortyCD}{}\renewcommand{\MetricDFGOneVDNDFortyCD}{\ensuremath{5.5735}}
\providecommand{\MetricDFGOneVDNDFortyCL}{}\renewcommand{\MetricDFGOneVDNDFortyCL}{\ensuremath{0.0107}}
\providecommand{\MetricDFGOneVDNDFortyCells}{}\renewcommand{\MetricDFGOneVDNDFortyCells}{\ensuremath{143{,}056}}
\providecommand{\MetricDFGOneVDNDFortyDp}{}\renewcommand{\MetricDFGOneVDNDFortyDp}{\ensuremath{0.1179}}
\providecommand{\MetricDFGOneVDNDFortyLr}{}\renewcommand{\MetricDFGOneVDNDFortyLr}{\ensuremath{0.0842}}
\providecommand{\MetricDFGOneVDNDFortyWall}{}\renewcommand{\MetricDFGOneVDNDFortyWall}{\ensuremath{264.10}}
\providecommand{\MetricDFGOneVDNDTwentyCD}{}\renewcommand{\MetricDFGOneVDNDTwentyCD}{\ensuremath{5.5558}}
\providecommand{\MetricDFGOneVDNDTwentyCL}{}\renewcommand{\MetricDFGOneVDNDTwentyCL}{\ensuremath{0.0109}}
\providecommand{\MetricDFGOneVDNDTwentyCells}{}\renewcommand{\MetricDFGOneVDNDTwentyCells}{\ensuremath{35{,}764}}
\providecommand{\MetricDFGOneVDNDTwentyDp}{}\renewcommand{\MetricDFGOneVDNDTwentyDp}{\ensuremath{0.1176}}
\providecommand{\MetricDFGOneVDNDTwentyLr}{}\renewcommand{\MetricDFGOneVDNDTwentyLr}{\ensuremath{0.0830}}
\providecommand{\MetricDFGOneVDNDTwentyWall}{}\renewcommand{\MetricDFGOneVDNDTwentyWall}{\ensuremath{25.53}}

\providecommand{\MetricDFGSensRelCD}{}\renewcommand{\MetricDFGSensRelCD}{\ensuremath{0.0\%}}
\providecommand{\MetricDFGSensRelCDmax}{}\renewcommand{\MetricDFGSensRelCDmax}{\ensuremath{0.3\%}}
\providecommand{\MetricDFGSensRelCL}{}\renewcommand{\MetricDFGSensRelCL}{\ensuremath{0.0\%}}
\providecommand{\MetricDFGSensRelCLmax}{}\renewcommand{\MetricDFGSensRelCLmax}{\ensuremath{3.7\%}}
\providecommand{\MetricDFGSensRelDpOne}{}\renewcommand{\MetricDFGSensRelDpOne}{\ensuremath{0.0\%}}
\providecommand{\MetricDFGSensRelDpTwo}{}\renewcommand{\MetricDFGSensRelDpTwo}{\ensuremath{1.2\%}}
\providecommand{\MetricDFGSensRelSt}{}\renewcommand{\MetricDFGSensRelSt}{\ensuremath{6.1\%}}
\providecommand{\MetricDFGTwoOFNDEightyCDmax}{}\renewcommand{\MetricDFGTwoOFNDEightyCDmax}{\ensuremath{3.2289}}
\providecommand{\MetricDFGTwoOFNDEightyCLmax}{}\renewcommand{\MetricDFGTwoOFNDEightyCLmax}{\ensuremath{0.9876}}
\providecommand{\MetricDFGTwoOFNDEightyCells}{}\renewcommand{\MetricDFGTwoOFNDEightyCells}{\ensuremath{577{,}280}}
\providecommand{\MetricDFGTwoOFNDEightyDp}{}\renewcommand{\MetricDFGTwoOFNDEightyDp}{\ensuremath{2.4827}}
\providecommand{\MetricDFGTwoOFNDEightySt}{}\renewcommand{\MetricDFGTwoOFNDEightySt}{\ensuremath{0.3018}}
\providecommand{\MetricDFGTwoOFNDEightyWall}{}\renewcommand{\MetricDFGTwoOFNDEightyWall}{\ensuremath{1078.02}}
\providecommand{\MetricDFGTwoOFNDFortyCDmax}{}\renewcommand{\MetricDFGTwoOFNDFortyCDmax}{\ensuremath{3.2322}}
\providecommand{\MetricDFGTwoOFNDFortyCLmax}{}\renewcommand{\MetricDFGTwoOFNDFortyCLmax}{\ensuremath{0.9898}}
\providecommand{\MetricDFGTwoOFNDFortyCells}{}\renewcommand{\MetricDFGTwoOFNDFortyCells}{\ensuremath{144{,}320}}
\providecommand{\MetricDFGTwoOFNDFortyDp}{}\renewcommand{\MetricDFGTwoOFNDFortyDp}{\ensuremath{2.4804}}
\providecommand{\MetricDFGTwoOFNDFortySt}{}\renewcommand{\MetricDFGTwoOFNDFortySt}{\ensuremath{0.3016}}
\providecommand{\MetricDFGTwoOFNDFortyWall}{}\renewcommand{\MetricDFGTwoOFNDFortyWall}{\ensuremath{127.51}}

\providecommand{\MetricDFGTwoOFNDTwentySt}{}\renewcommand{\MetricDFGTwoOFNDTwentySt}{\ensuremath{0.3005}}
\providecommand{\MetricDFGTwoOFNDTwentyWall}{}\renewcommand{\MetricDFGTwoOFNDTwentyWall}{\ensuremath{30.53}}
\providecommand{\MetricDFGTwoVDNDEightyCDmax}{}\renewcommand{\MetricDFGTwoVDNDEightyCDmax}{\ensuremath{3.2274}}
\providecommand{\MetricDFGTwoVDNDEightyCLmax}{}\renewcommand{\MetricDFGTwoVDNDEightyCLmax}{\ensuremath{1.0001}}
\providecommand{\MetricDFGTwoVDNDEightyCells}{}\renewcommand{\MetricDFGTwoVDNDEightyCells}{\ensuremath{572{,}256}}
\providecommand{\MetricDFGTwoVDNDEightyDp}{}\renewcommand{\MetricDFGTwoVDNDEightyDp}{\ensuremath{2.4960}}
\providecommand{\MetricDFGTwoVDNDEightySt}{}\renewcommand{\MetricDFGTwoVDNDEightySt}{\ensuremath{0.3022}}
\providecommand{\MetricDFGTwoVDNDEightyWall}{}\renewcommand{\MetricDFGTwoVDNDEightyWall}{\ensuremath{75.56}}
\providecommand{\MetricDFGTwoVDNDFortyCDmax}{}\renewcommand{\MetricDFGTwoVDNDFortyCDmax}{\ensuremath{3.2092}}
\providecommand{\MetricDFGTwoVDNDFortyCLmax}{}\renewcommand{\MetricDFGTwoVDNDFortyCLmax}{\ensuremath{0.9742}}
\providecommand{\MetricDFGTwoVDNDFortyCells}{}\renewcommand{\MetricDFGTwoVDNDFortyCells}{\ensuremath{143{,}056}}
\providecommand{\MetricDFGTwoVDNDFortyDp}{}\renewcommand{\MetricDFGTwoVDNDFortyDp}{\ensuremath{2.4907}}
\providecommand{\MetricDFGTwoVDNDFortySt}{}\renewcommand{\MetricDFGTwoVDNDFortySt}{\ensuremath{0.3116}}
\providecommand{\MetricDFGTwoVDNDFortyWall}{}\renewcommand{\MetricDFGTwoVDNDFortyWall}{\ensuremath{8.04}}
\providecommand{\MetricDFGTwoVDNDTwentyCDmax}{}\renewcommand{\MetricDFGTwoVDNDTwentyCDmax}{\ensuremath{41.6924}}
\providecommand{\MetricDFGTwoVDNDTwentyCLmax}{}\renewcommand{\MetricDFGTwoVDNDTwentyCLmax}{\ensuremath{34.4547}}
\providecommand{\MetricDFGTwoVDNDTwentyCells}{}\renewcommand{\MetricDFGTwoVDNDTwentyCells}{\ensuremath{35{,}764}}
\providecommand{\MetricDFGTwoVDNDTwentyDp}{}\renewcommand{\MetricDFGTwoVDNDTwentyDp}{\ensuremath{-2.7076}}
\providecommand{\MetricDFGTwoVDNDTwentySt}{}\renewcommand{\MetricDFGTwoVDNDTwentySt}{\ensuremath{0.6645}}
\providecommand{\MetricDFGTwoVDNDTwentyWall}{}\renewcommand{\MetricDFGTwoVDNDTwentyWall}{\ensuremath{1.37}}

\providecommand{\MetricDSLDNSSelfConvReOneM}{}\renewcommand{\MetricDSLDNSSelfConvReOneM}{\ensuremath{2.53\times 10^{-3}}}

\providecommand{\MetricDSLEightyACDTSReHunKCores}{}\renewcommand{\MetricDSLEightyACDTSReHunKCores}{\ensuremath{4} (\ensuremath{4})}
\providecommand{\MetricDSLEightyACDTSReHunKEu}{}\renewcommand{\MetricDSLEightyACDTSReHunKEu}{\ensuremath{0.2717}}

\providecommand{\MetricDSLEightyACDTSReOneMEu}{}\renewcommand{\MetricDSLEightyACDTSReOneMEu}{\ensuremath{0.2723}}

\providecommand{\MetricDSLEightyACDTSReTenKEu}{}\renewcommand{\MetricDSLEightyACDTSReTenKEu}{\ensuremath{0.1652}}

\providecommand{\MetricDSLEightyBGKReHunKCores}{}\renewcommand{\MetricDSLEightyBGKReHunKCores}{\emph{Blow-up}}
\providecommand{\MetricDSLEightyBGKReHunKEu}{}\renewcommand{\MetricDSLEightyBGKReHunKEu}{\emph{Blow-up}}

\providecommand{\MetricDSLEightyBGKReOneMEu}{}\renewcommand{\MetricDSLEightyBGKReOneMEu}{\emph{Blow-up}}

\providecommand{\MetricDSLEightyBGKReTenKEu}{}\renewcommand{\MetricDSLEightyBGKReTenKEu}{\ensuremath{0.03701}}

\providecommand{\MetricDSLEightyCMReHunKCores}{}\renewcommand{\MetricDSLEightyCMReHunKCores}{\ensuremath{2} (\ensuremath{4})}
\providecommand{\MetricDSLEightyCMReHunKEu}{}\renewcommand{\MetricDSLEightyCMReHunKEu}{\ensuremath{0.063}}

\providecommand{\MetricDSLEightyCMReOneMEu}{}\renewcommand{\MetricDSLEightyCMReOneMEu}{\ensuremath{0.06722}}

\providecommand{\MetricDSLEightyCMReTenKEu}{}\renewcommand{\MetricDSLEightyCMReTenKEu}{\ensuremath{0.02147}}

\providecommand{\MetricDSLEightyCumReHunKCores}{}\renewcommand{\MetricDSLEightyCumReHunKCores}{\ensuremath{2} (\ensuremath{4})}
\providecommand{\MetricDSLEightyCumReHunKEu}{}\renewcommand{\MetricDSLEightyCumReHunKEu}{\ensuremath{0.06175}}

\providecommand{\MetricDSLEightyCumReOneMEu}{}\renewcommand{\MetricDSLEightyCumReOneMEu}{\ensuremath{0.06917}}

\providecommand{\MetricDSLEightyCumReTenKEu}{}\renewcommand{\MetricDSLEightyCumReTenKEu}{\ensuremath{0.02015}}

\providecommand{\MetricDSLEightyDTSTRTReHunKCores}{}\renewcommand{\MetricDSLEightyDTSTRTReHunKCores}{\emph{Blow-up}}
\providecommand{\MetricDSLEightyDTSTRTReHunKEu}{}\renewcommand{\MetricDSLEightyDTSTRTReHunKEu}{\emph{Blow-up}}

\providecommand{\MetricDSLEightyDTSTRTReOneMEu}{}\renewcommand{\MetricDSLEightyDTSTRTReOneMEu}{\emph{Blow-up}}

\providecommand{\MetricDSLEightyDTSTRTReTenKEu}{}\renewcommand{\MetricDSLEightyDTSTRTReTenKEu}{\emph{Blow-up}}

\providecommand{\MetricDSLEightyELBMReHunKCores}{}\renewcommand{\MetricDSLEightyELBMReHunKCores}{\ensuremath{2} (\ensuremath{22})}
\providecommand{\MetricDSLEightyELBMReHunKEu}{}\renewcommand{\MetricDSLEightyELBMReHunKEu}{\ensuremath{0.2277}}

\providecommand{\MetricDSLEightyELBMReOneMEu}{}\renewcommand{\MetricDSLEightyELBMReOneMEu}{\ensuremath{0.238}}

\providecommand{\MetricDSLEightyELBMReTenKEu}{}\renewcommand{\MetricDSLEightyELBMReTenKEu}{\ensuremath{0.0662}}

\providecommand{\MetricDSLEightyHRRReHunKCores}{}\renewcommand{\MetricDSLEightyHRRReHunKCores}{\ensuremath{2} (\ensuremath{4})}
\providecommand{\MetricDSLEightyHRRReHunKEu}{}\renewcommand{\MetricDSLEightyHRRReHunKEu}{\ensuremath{0.06225}}

\providecommand{\MetricDSLEightyHRRReOneMEu}{}\renewcommand{\MetricDSLEightyHRRReOneMEu}{\ensuremath{0.06227}}

\providecommand{\MetricDSLEightyHRRReTenKEu}{}\renewcommand{\MetricDSLEightyHRRReTenKEu}{\ensuremath{0.03847}}

\providecommand{\MetricDSLEightyKimMoinReHunKCores}{}\renewcommand{\MetricDSLEightyKimMoinReHunKCores}{\emph{Blow-up}}
\providecommand{\MetricDSLEightyKimMoinReHunKEu}{}\renewcommand{\MetricDSLEightyKimMoinReHunKEu}{\emph{Blow-up}}

\providecommand{\MetricDSLEightyKimMoinReOneMEu}{}\renewcommand{\MetricDSLEightyKimMoinReOneMEu}{\emph{Blow-up}}

\providecommand{\MetricDSLEightyKimMoinReTenKEu}{}\renewcommand{\MetricDSLEightyKimMoinReTenKEu}{\emph{Blow-up}}

\providecommand{\MetricDSLEightyMRTReHunKCores}{}\renewcommand{\MetricDSLEightyMRTReHunKCores}{\emph{Blow-up}}
\providecommand{\MetricDSLEightyMRTReHunKEu}{}\renewcommand{\MetricDSLEightyMRTReHunKEu}{\emph{Blow-up}}

\providecommand{\MetricDSLEightyMRTReOneMEu}{}\renewcommand{\MetricDSLEightyMRTReOneMEu}{\emph{Blow-up}}

\providecommand{\MetricDSLEightyMRTReTenKEu}{}\renewcommand{\MetricDSLEightyMRTReTenKEu}{\emph{Blow-up}}

\providecommand{\MetricDSLEightyPIMPLEReHunKCores}{}\renewcommand{\MetricDSLEightyPIMPLEReHunKCores}{\ensuremath{4} (\ensuremath{12})}
\providecommand{\MetricDSLEightyPIMPLEReHunKEu}{}\renewcommand{\MetricDSLEightyPIMPLEReHunKEu}{\ensuremath{0.2804}}

\providecommand{\MetricDSLEightyPIMPLEReOneMEu}{}\renewcommand{\MetricDSLEightyPIMPLEReOneMEu}{\ensuremath{0.2886}}

\providecommand{\MetricDSLEightyPIMPLEReTenKEu}{}\renewcommand{\MetricDSLEightyPIMPLEReTenKEu}{\ensuremath{0.2254}}

\providecommand{\MetricDSLEightyRRReHunKCores}{}\renewcommand{\MetricDSLEightyRRReHunKCores}{\ensuremath{2} (\ensuremath{4})}
\providecommand{\MetricDSLEightyRRReHunKEu}{}\renewcommand{\MetricDSLEightyRRReHunKEu}{\ensuremath{0.06191}}

\providecommand{\MetricDSLEightyRRReOneMEu}{}\renewcommand{\MetricDSLEightyRRReOneMEu}{\ensuremath{0.06599}}

\providecommand{\MetricDSLEightyRRReTenKEu}{}\renewcommand{\MetricDSLEightyRRReTenKEu}{\ensuremath{0.02079}}

\providecommand{\MetricDSLEightySIMPLECReHunKCores}{}\renewcommand{\MetricDSLEightySIMPLECReHunKCores}{\ensuremath{8} (\ensuremath{100})}
\providecommand{\MetricDSLEightySIMPLECReHunKEu}{}\renewcommand{\MetricDSLEightySIMPLECReHunKEu}{\ensuremath{0.2343}}

\providecommand{\MetricDSLEightySIMPLECReOneMEu}{}\renewcommand{\MetricDSLEightySIMPLECReOneMEu}{\ensuremath{0.2502}}

\providecommand{\MetricDSLEightySIMPLECReTenKEu}{}\renewcommand{\MetricDSLEightySIMPLECReTenKEu}{\ensuremath{0.1194}}

\providecommand{\MetricDSLEightyTRTReHunKCores}{}\renewcommand{\MetricDSLEightyTRTReHunKCores}{\emph{Blow-up}}
\providecommand{\MetricDSLEightyTRTReHunKEu}{}\renewcommand{\MetricDSLEightyTRTReHunKEu}{\emph{Blow-up}}

\providecommand{\MetricDSLEightyTRTReOneMEu}{}\renewcommand{\MetricDSLEightyTRTReOneMEu}{\emph{Blow-up}}

\providecommand{\MetricDSLEightyTRTReTenKEu}{}\renewcommand{\MetricDSLEightyTRTReTenKEu}{\emph{Blow-up}}

\providecommand{\MetricDSLEightyVDPsiOneReHunKCores}{}\renewcommand{\MetricDSLEightyVDPsiOneReHunKCores}{\ensuremath{6} (\ensuremath{50})}
\providecommand{\MetricDSLEightyVDPsiOneReHunKEu}{}\renewcommand{\MetricDSLEightyVDPsiOneReHunKEu}{\ensuremath{0.1623}}

\providecommand{\MetricDSLEightyVDPsiOneReOneMEu}{}\renewcommand{\MetricDSLEightyVDPsiOneReOneMEu}{\ensuremath{0.1759}}

\providecommand{\MetricDSLEightyVDPsiOneReTenKEu}{}\renewcommand{\MetricDSLEightyVDPsiOneReTenKEu}{\ensuremath{0.04029}}

\providecommand{\MetricDSLEightyWENOReHunKCores}{}\renewcommand{\MetricDSLEightyWENOReHunKCores}{\ensuremath{2} (\ensuremath{10})}
\providecommand{\MetricDSLEightyWENOReHunKEu}{}\renewcommand{\MetricDSLEightyWENOReHunKEu}{\ensuremath{0.1379}}

\providecommand{\MetricDSLEightyWENOReOneMEu}{}\renewcommand{\MetricDSLEightyWENOReOneMEu}{\ensuremath{0.1206}}

\providecommand{\MetricDSLEightyWENOReTenKEu}{}\renewcommand{\MetricDSLEightyWENOReTenKEu}{\ensuremath{0.04075}}

\providecommand{\MetricDSLThirtyNTwoFiveSixOrderOmega}{}\renewcommand{\MetricDSLThirtyNTwoFiveSixOrderOmega}{\ensuremath{2.00}}
\providecommand{\MetricDSLThirtyNTwoFiveSixOrderU}{}\renewcommand{\MetricDSLThirtyNTwoFiveSixOrderU}{\ensuremath{2.00}}

\providecommand{\MetricGPUCPUEquivMaxDiff}{}\renewcommand{\MetricGPUCPUEquivMaxDiff}{\ensuremath{3.50\times 10^{-15}}}

\providecommand{\MetricKerPBasisProjMax}{}\renewcommand{\MetricKerPBasisProjMax}{\ensuremath{1.11\times 10^{-16}}}

\providecommand{\MetricKerPCumMaxPert}{}\renewcommand{\MetricKerPCumMaxPert}{\ensuremath{2.063}}

\providecommand{\MetricKerPSingleCollisionAMax}{}\renewcommand{\MetricKerPSingleCollisionAMax}{\ensuremath{1.73\times 10^{-17}}}

\providecommand{\MetricKerPSingleCollisionAMin}{}\renewcommand{\MetricKerPSingleCollisionAMin}{\ensuremath{1.39\times 10^{-17}}}

\providecommand{\MetricKerPVDUzawaMaxPert}{}\renewcommand{\MetricKerPVDUzawaMaxPert}{\ensuremath{3.03\times 10^{-15}}}

\providecommand{\MetricLDCHopfNOneTwoEightEkStd}{}\renewcommand{\MetricLDCHopfNOneTwoEightEkStd}{\ensuremath{9.20\times 10^{-6}}}
\providecommand{\MetricLDCHopfNOneTwoEightFreq}{}\renewcommand{\MetricLDCHopfNOneTwoEightFreq}{\ensuremath{0.5508}}
\providecommand{\MetricLDCHopfNOneTwoEightPsiMean}{}\renewcommand{\MetricLDCHopfNOneTwoEightPsiMean}{\ensuremath{-0.1112}}

\providecommand{\MetricLDCHopfNOneTwoEightUminMean}{}\renewcommand{\MetricLDCHopfNOneTwoEightUminMean}{\ensuremath{-0.4197}}
\providecommand{\MetricLDCHopfNTwoFiveSixEkStd}{}\renewcommand{\MetricLDCHopfNTwoFiveSixEkStd}{\ensuremath{8.53\times 10^{-6}}}
\providecommand{\MetricLDCHopfNTwoFiveSixFreq}{}\renewcommand{\MetricLDCHopfNTwoFiveSixFreq}{\ensuremath{0.5852}}
\providecommand{\MetricLDCHopfNTwoFiveSixPsiMean}{}\renewcommand{\MetricLDCHopfNTwoFiveSixPsiMean}{\ensuremath{-0.1161}}

\providecommand{\MetricLDCHopfNTwoFiveSixUminMean}{}\renewcommand{\MetricLDCHopfNTwoFiveSixUminMean}{\ensuremath{-0.4370}}

\providecommand{\MetricLDCMoffattReFiveKBRone}{}\renewcommand{\MetricLDCMoffattReFiveKBRone}{\ensuremath{3.00\times 10^{-3}}}

\providecommand{\MetricLDCMoffattReOneKBRone}{}\renewcommand{\MetricLDCMoffattReOneKBRone}{\ensuremath{1.69\times 10^{-3}}}
\providecommand{\MetricLDCMoffattReOneKPsiRatio}{}\renewcommand{\MetricLDCMoffattReOneKPsiRatio}{\ensuremath{21895.29}}
\providecommand{\MetricLDCMoffattReOneKSizeRatio}{}\renewcommand{\MetricLDCMoffattReOneKSizeRatio}{\ensuremath{15.83}}

\providecommand{\MetricLDCReFiveKPsiPV}{}\renewcommand{\MetricLDCReFiveKPsiPV}{\ensuremath{-0.11821}}

\providecommand{\MetricLDCReFiveKRelU}{}\renewcommand{\MetricLDCReFiveKRelU}{\ensuremath{7.94\times 10^{-3}}}
\providecommand{\MetricLDCReFiveKUmin}{}\renewcommand{\MetricLDCReFiveKUmin}{\ensuremath{-0.43818}}
\providecommand{\MetricLDCReFiveKVmin}{}\renewcommand{\MetricLDCReFiveKVmin}{\ensuremath{-0.56520}}

\providecommand{\MetricLDCReHunPsiPV}{}\renewcommand{\MetricLDCReHunPsiPV}{\ensuremath{-0.10317}}

\providecommand{\MetricLDCReHunRelU}{}\renewcommand{\MetricLDCReHunRelU}{\ensuremath{4.56\times 10^{-3}}}
\providecommand{\MetricLDCReHunUmin}{}\renewcommand{\MetricLDCReHunUmin}{\ensuremath{-0.21242}}
\providecommand{\MetricLDCReHunVmin}{}\renewcommand{\MetricLDCReHunVmin}{\ensuremath{-0.25220}}

\providecommand{\MetricLDCReOneKPsiPV}{}\renewcommand{\MetricLDCReOneKPsiPV}{\ensuremath{-0.11761}}

\providecommand{\MetricLDCReOneKRelU}{}\renewcommand{\MetricLDCReOneKRelU}{\ensuremath{4.09\times 10^{-3}}}
\providecommand{\MetricLDCReOneKUmin}{}\renewcommand{\MetricLDCReOneKUmin}{\ensuremath{-0.38441}}
\providecommand{\MetricLDCReOneKVmin}{}\renewcommand{\MetricLDCReOneKVmin}{\ensuremath{-0.52213}}

\providecommand{\MetricLDCReTenRelU}{}\renewcommand{\MetricLDCReTenRelU}{\ensuremath{6.20\times 10^{-4}}}

\providecommand{\MetricLDCReZeroPointZeroOneRelU}{}\renewcommand{\MetricLDCReZeroPointZeroOneRelU}{\ensuremath{6.23\times 10^{-4}}}

\providecommand{\MetricPreflightCFDTwoOrderRange}{}\renewcommand{\MetricPreflightCFDTwoOrderRange}{\ensuremath{[1.93, 3.55]}}

\providecommand{\MetricPreflightPeerLBMOrderRange}{}\renewcommand{\MetricPreflightPeerLBMOrderRange}{\ensuremath{[1.98, 2.10]}}

\providecommand{\MetricPreflightVDLBMUzawaOrder}{}\renewcommand{\MetricPreflightVDLBMUzawaOrder}{\ensuremath{2.00}}

\providecommand{\MetricRoofAOneHundredMeasuredDTwoDBW}{}\renewcommand{\MetricRoofAOneHundredMeasuredDTwoDBW}{\ensuremath{1581.8}}

\providecommand{\MetricRoofBGKNTwoFiveSixFPSixtyFourPctDTwoD}{}\renewcommand{\MetricRoofBGKNTwoFiveSixFPSixtyFourPctDTwoD}{\ensuremath{93.6\%}}
\providecommand{\MetricRoofBGKNTwoFiveSixFPSixtyFourSweepMLUPS}{}\renewcommand{\MetricRoofBGKNTwoFiveSixFPSixtyFourSweepMLUPS}{\ensuremath{4869.8}}

\providecommand{\MetricRoofHRRNTwoFiveSixFPSixtyFourPctDTwoD}{}\renewcommand{\MetricRoofHRRNTwoFiveSixFPSixtyFourPctDTwoD}{\ensuremath{22.0\%}}
\providecommand{\MetricRoofHRRNTwoFiveSixFPSixtyFourSweepMLUPS}{}\renewcommand{\MetricRoofHRRNTwoFiveSixFPSixtyFourSweepMLUPS}{\ensuremath{669.4}}

\providecommand{\MetricRoofVDLESNTwoFiveSixFPSixtyFourRetention}{}\renewcommand{\MetricRoofVDLESNTwoFiveSixFPSixtyFourRetention}{\ensuremath{95.4\%}}
\providecommand{\MetricRoofVDLESNTwoFiveSixFPSixtyFourSweepMLUPS}{}\renewcommand{\MetricRoofVDLESNTwoFiveSixFPSixtyFourSweepMLUPS}{\ensuremath{759.1}}

\providecommand{\MetricRoofVDNTwoFiveSixFPSixtyFourPctDTwoD}{}\renewcommand{\MetricRoofVDNTwoFiveSixFPSixtyFourPctDTwoD}{\ensuremath{80.9\%}}
\providecommand{\MetricRoofVDNTwoFiveSixFPSixtyFourSweepMLUPS}{}\renewcommand{\MetricRoofVDNTwoFiveSixFPSixtyFourSweepMLUPS}{\ensuremath{796.0}}

\providecommand{\MetricRoofVDVarViscNTwoFiveSixFPSixtyFourRetention}{}\renewcommand{\MetricRoofVDVarViscNTwoFiveSixFPSixtyFourRetention}{\ensuremath{95.5\%}}
\providecommand{\MetricRoofVDVarViscNTwoFiveSixFPSixtyFourSweepMLUPS}{}\renewcommand{\MetricRoofVDVarViscNTwoFiveSixFPSixtyFourSweepMLUPS}{\ensuremath{760.2}}

\providecommand{\MetricStratVDNOneTwoEightMomDrift}{}\renewcommand{\MetricStratVDNOneTwoEightMomDrift}{\ensuremath{8.95\times 10^{-14}}}

\providecommand{\MetricStratVDNTwoFiveSixMomDrift}{}\renewcommand{\MetricStratVDNTwoFiveSixMomDrift}{\ensuremath{3.85\times 10^{-13}}}

\providecommand{\MetricStratVDNTwoFiveSixSweeps}{}\renewcommand{\MetricStratVDNTwoFiveSixSweeps}{\ensuremath{77.2}}

\providecommand{\MetricTGVBDFAdvSOneTwoEightOrder}{}\renewcommand{\MetricTGVBDFAdvSOneTwoEightOrder}{\ensuremath{2.08}}

\providecommand{\MetricTGVBDFStatSOneTwoEightOrder}{}\renewcommand{\MetricTGVBDFStatSOneTwoEightOrder}{\ensuremath{2.06}}

\providecommand{\MetricTGVBDFVarSOneTwoEightOrder}{}\renewcommand{\MetricTGVBDFVarSOneTwoEightOrder}{\ensuremath{2.24}}

\providecommand{\MetricTGVCostFineVDUzawaReHighT}{}\renewcommand{\MetricTGVCostFineVDUzawaReHighT}{\ensuremath{2.21\times 10^{-3}}}

\providecommand{\MetricTGVJointUzawaNFiveOneTwoOrder}{}\renewcommand{\MetricTGVJointUzawaNFiveOneTwoOrder}{\ensuremath{2.37}}

\providecommand{\MetricTGVJointUzawaNFiveOneTwoReOneEu}{}\renewcommand{\MetricTGVJointUzawaNFiveOneTwoReOneEu}{\ensuremath{2.92\times 10^{-7}}}
\providecommand{\MetricTGVJointUzawaNFiveOneTwoSweeps}{}\renewcommand{\MetricTGVJointUzawaNFiveOneTwoSweeps}{\ensuremath{10.0}}

\providecommand{\MetricTGVJointUzawaNThirtyTwoReOneEu}{}\renewcommand{\MetricTGVJointUzawaNThirtyTwoReOneEu}{\ensuremath{1.17\times 10^{-4}}}
\providecommand{\MetricTGVJointUzawaNThirtyTwoSweeps}{}\renewcommand{\MetricTGVJointUzawaNThirtyTwoSweeps}{\ensuremath{5.0}}

\providecommand{\MetricTGVSixteenACDTSDiv}{}\renewcommand{\MetricTGVSixteenACDTSDiv}{\ensuremath{1.26\times 10^{-4}}}
\providecommand{\MetricTGVSixteenACDTSEp}{}\renewcommand{\MetricTGVSixteenACDTSEp}{\ensuremath{9.06\times 10^{-4}}}
\providecommand{\MetricTGVSixteenACDTSEuReHun}{}\renewcommand{\MetricTGVSixteenACDTSEuReHun}{\ensuremath{1.44\times 10^{-5}}}
\providecommand{\MetricTGVSixteenACDTSEuReMax}{}\renewcommand{\MetricTGVSixteenACDTSEuReMax}{\ensuremath{2.25\times 10^{-5}}}
\providecommand{\MetricTGVSixteenACDTSEuReMin}{}\renewcommand{\MetricTGVSixteenACDTSEuReMin}{\ensuremath{8.07\times 10^{-6}}}
\providecommand{\MetricTGVSixteenACDTSNuRatio}{}\renewcommand{\MetricTGVSixteenACDTSNuRatio}{\ensuremath{0.9999}}

\providecommand{\MetricTGVSixteenBGKDiv}{}\renewcommand{\MetricTGVSixteenBGKDiv}{\ensuremath{0.01619}}
\providecommand{\MetricTGVSixteenBGKEp}{}\renewcommand{\MetricTGVSixteenBGKEp}{\ensuremath{0.01066}}
\providecommand{\MetricTGVSixteenBGKEuReHun}{}\renewcommand{\MetricTGVSixteenBGKEuReHun}{\ensuremath{1.22\times 10^{-3}}}
\providecommand{\MetricTGVSixteenBGKEuReMax}{}\renewcommand{\MetricTGVSixteenBGKEuReMax}{\ensuremath{6.17\times 10^{-4}}}
\providecommand{\MetricTGVSixteenBGKEuReMin}{}\renewcommand{\MetricTGVSixteenBGKEuReMin}{\ensuremath{7.97\times 10^{-6}}}
\providecommand{\MetricTGVSixteenBGKNuRatio}{}\renewcommand{\MetricTGVSixteenBGKNuRatio}{\ensuremath{0.9988}}

\providecommand{\MetricTGVSixteenCMDiv}{}\renewcommand{\MetricTGVSixteenCMDiv}{\ensuremath{0.01581}}
\providecommand{\MetricTGVSixteenCMEp}{}\renewcommand{\MetricTGVSixteenCMEp}{\ensuremath{0.01131}}
\providecommand{\MetricTGVSixteenCMEuReHun}{}\renewcommand{\MetricTGVSixteenCMEuReHun}{\ensuremath{1.19\times 10^{-3}}}
\providecommand{\MetricTGVSixteenCMEuReMax}{}\renewcommand{\MetricTGVSixteenCMEuReMax}{\ensuremath{5.91\times 10^{-4}}}
\providecommand{\MetricTGVSixteenCMEuReMin}{}\renewcommand{\MetricTGVSixteenCMEuReMin}{\ensuremath{7.94\times 10^{-6}}}
\providecommand{\MetricTGVSixteenCMNuRatio}{}\renewcommand{\MetricTGVSixteenCMNuRatio}{\ensuremath{0.9985}}

\providecommand{\MetricTGVSixteenChorinDiv}{}\renewcommand{\MetricTGVSixteenChorinDiv}{\ensuremath{5.65\times 10^{-15}}}
\providecommand{\MetricTGVSixteenChorinEp}{}\renewcommand{\MetricTGVSixteenChorinEp}{\ensuremath{9.04\times 10^{-4}}}
\providecommand{\MetricTGVSixteenChorinEuReHun}{}\renewcommand{\MetricTGVSixteenChorinEuReHun}{\ensuremath{2.27\times 10^{-6}}}
\providecommand{\MetricTGVSixteenChorinEuReMax}{}\renewcommand{\MetricTGVSixteenChorinEuReMax}{\ensuremath{2.72\times 10^{-7}}}
\providecommand{\MetricTGVSixteenChorinEuReMin}{}\renewcommand{\MetricTGVSixteenChorinEuReMin}{\ensuremath{8.24\times 10^{-6}}}
\providecommand{\MetricTGVSixteenChorinNuRatio}{}\renewcommand{\MetricTGVSixteenChorinNuRatio}{\ensuremath{1.0002}}

\providecommand{\MetricTGVSixteenCumDiv}{}\renewcommand{\MetricTGVSixteenCumDiv}{\ensuremath{0.01591}}
\providecommand{\MetricTGVSixteenCumEp}{}\renewcommand{\MetricTGVSixteenCumEp}{\ensuremath{0.0104}}
\providecommand{\MetricTGVSixteenCumEuReHun}{}\renewcommand{\MetricTGVSixteenCumEuReHun}{\ensuremath{1.20\times 10^{-3}}}
\providecommand{\MetricTGVSixteenCumEuReMax}{}\renewcommand{\MetricTGVSixteenCumEuReMax}{\ensuremath{6.24\times 10^{-4}}}
\providecommand{\MetricTGVSixteenCumEuReMin}{}\renewcommand{\MetricTGVSixteenCumEuReMin}{\ensuremath{7.94\times 10^{-6}}}
\providecommand{\MetricTGVSixteenCumNuRatio}{}\renewcommand{\MetricTGVSixteenCumNuRatio}{\ensuremath{1.0006}}

\providecommand{\MetricTGVSixteenDTSTRTDiv}{}\renewcommand{\MetricTGVSixteenDTSTRTDiv}{\ensuremath{2.96\times 10^{-3}}}
\providecommand{\MetricTGVSixteenDTSTRTEp}{}\renewcommand{\MetricTGVSixteenDTSTRTEp}{\ensuremath{2.68\times 10^{-3}}}
\providecommand{\MetricTGVSixteenDTSTRTEuReHun}{}\renewcommand{\MetricTGVSixteenDTSTRTEuReHun}{\ensuremath{3.38\times 10^{-4}}}
\providecommand{\MetricTGVSixteenDTSTRTEuReMax}{}\renewcommand{\MetricTGVSixteenDTSTRTEuReMax}{\emph{Blow-up}}
\providecommand{\MetricTGVSixteenDTSTRTEuReMin}{}\renewcommand{\MetricTGVSixteenDTSTRTEuReMin}{\ensuremath{7.74\times 10^{-6}}}
\providecommand{\MetricTGVSixteenDTSTRTNuRatio}{}\renewcommand{\MetricTGVSixteenDTSTRTNuRatio}{\ensuremath{0.9994}}

\providecommand{\MetricTGVSixteenELBMDiv}{}\renewcommand{\MetricTGVSixteenELBMDiv}{\ensuremath{0.0162}}
\providecommand{\MetricTGVSixteenELBMEp}{}\renewcommand{\MetricTGVSixteenELBMEp}{\ensuremath{8.80\times 10^{-3}}}
\providecommand{\MetricTGVSixteenELBMEuReHun}{}\renewcommand{\MetricTGVSixteenELBMEuReHun}{\ensuremath{1.23\times 10^{-3}}}
\providecommand{\MetricTGVSixteenELBMEuReMax}{}\renewcommand{\MetricTGVSixteenELBMEuReMax}{\ensuremath{6.06\times 10^{-4}}}
\providecommand{\MetricTGVSixteenELBMEuReMin}{}\renewcommand{\MetricTGVSixteenELBMEuReMin}{\ensuremath{7.97\times 10^{-6}}}
\providecommand{\MetricTGVSixteenELBMNuRatio}{}\renewcommand{\MetricTGVSixteenELBMNuRatio}{\ensuremath{0.9985}}

\providecommand{\MetricTGVSixteenHRRDiv}{}\renewcommand{\MetricTGVSixteenHRRDiv}{\ensuremath{0.01617}}
\providecommand{\MetricTGVSixteenHRREp}{}\renewcommand{\MetricTGVSixteenHRREp}{\ensuremath{0.0109}}
\providecommand{\MetricTGVSixteenHRREuReHun}{}\renewcommand{\MetricTGVSixteenHRREuReHun}{\ensuremath{1.22\times 10^{-3}}}
\providecommand{\MetricTGVSixteenHRREuReMax}{}\renewcommand{\MetricTGVSixteenHRREuReMax}{\ensuremath{6.19\times 10^{-4}}}
\providecommand{\MetricTGVSixteenHRREuReMin}{}\renewcommand{\MetricTGVSixteenHRREuReMin}{\ensuremath{7.94\times 10^{-6}}}
\providecommand{\MetricTGVSixteenHRRNuRatio}{}\renewcommand{\MetricTGVSixteenHRRNuRatio}{\ensuremath{0.9985}}

\providecommand{\MetricTGVSixteenKimMoinDiv}{}\renewcommand{\MetricTGVSixteenKimMoinDiv}{\ensuremath{5.47\times 10^{-15}}}
\providecommand{\MetricTGVSixteenKimMoinEp}{}\renewcommand{\MetricTGVSixteenKimMoinEp}{\ensuremath{5.98\times 10^{-4}}}
\providecommand{\MetricTGVSixteenKimMoinEuReHun}{}\renewcommand{\MetricTGVSixteenKimMoinEuReHun}{\ensuremath{2.00\times 10^{-6}}}
\providecommand{\MetricTGVSixteenKimMoinEuReMax}{}\renewcommand{\MetricTGVSixteenKimMoinEuReMax}{\ensuremath{2.72\times 10^{-7}}}
\providecommand{\MetricTGVSixteenKimMoinEuReMin}{}\renewcommand{\MetricTGVSixteenKimMoinEuReMin}{\ensuremath{7.75\times 10^{-6}}}
\providecommand{\MetricTGVSixteenKimMoinNuRatio}{}\renewcommand{\MetricTGVSixteenKimMoinNuRatio}{\ensuremath{0.9998}}

\providecommand{\MetricTGVSixteenMRTDiv}{}\renewcommand{\MetricTGVSixteenMRTDiv}{\ensuremath{0.01604}}
\providecommand{\MetricTGVSixteenMRTEp}{}\renewcommand{\MetricTGVSixteenMRTEp}{\ensuremath{0.01153}}
\providecommand{\MetricTGVSixteenMRTEuReHun}{}\renewcommand{\MetricTGVSixteenMRTEuReHun}{\ensuremath{1.21\times 10^{-3}}}
\providecommand{\MetricTGVSixteenMRTEuReMax}{}\renewcommand{\MetricTGVSixteenMRTEuReMax}{\ensuremath{9.01\times 10^{-4}}}
\providecommand{\MetricTGVSixteenMRTEuReMin}{}\renewcommand{\MetricTGVSixteenMRTEuReMin}{\ensuremath{9.89\times 10^{-6}}}
\providecommand{\MetricTGVSixteenMRTNuRatio}{}\renewcommand{\MetricTGVSixteenMRTNuRatio}{\ensuremath{0.9986}}

\providecommand{\MetricTGVSixteenPIMPLEDiv}{}\renewcommand{\MetricTGVSixteenPIMPLEDiv}{\ensuremath{1.51\times 10^{-4}}}
\providecommand{\MetricTGVSixteenPIMPLEEp}{}\renewcommand{\MetricTGVSixteenPIMPLEEp}{\ensuremath{1.23\times 10^{-3}}}
\providecommand{\MetricTGVSixteenPIMPLEEuReHun}{}\renewcommand{\MetricTGVSixteenPIMPLEEuReHun}{\ensuremath{6.00\times 10^{-6}}}
\providecommand{\MetricTGVSixteenPIMPLEEuReMax}{}\renewcommand{\MetricTGVSixteenPIMPLEEuReMax}{\ensuremath{1.37\times 10^{-5}}}
\providecommand{\MetricTGVSixteenPIMPLEEuReMin}{}\renewcommand{\MetricTGVSixteenPIMPLEEuReMin}{\ensuremath{8.03\times 10^{-6}}}
\providecommand{\MetricTGVSixteenPIMPLENuRatio}{}\renewcommand{\MetricTGVSixteenPIMPLENuRatio}{\ensuremath{1.0000}}

\providecommand{\MetricTGVSixteenRRDiv}{}\renewcommand{\MetricTGVSixteenRRDiv}{\ensuremath{0.01617}}
\providecommand{\MetricTGVSixteenRREp}{}\renewcommand{\MetricTGVSixteenRREp}{\ensuremath{0.01088}}
\providecommand{\MetricTGVSixteenRREuReHun}{}\renewcommand{\MetricTGVSixteenRREuReHun}{\ensuremath{1.22\times 10^{-3}}}
\providecommand{\MetricTGVSixteenRREuReMax}{}\renewcommand{\MetricTGVSixteenRREuReMax}{\ensuremath{6.18\times 10^{-4}}}
\providecommand{\MetricTGVSixteenRREuReMin}{}\renewcommand{\MetricTGVSixteenRREuReMin}{\ensuremath{7.94\times 10^{-6}}}
\providecommand{\MetricTGVSixteenRRNuRatio}{}\renewcommand{\MetricTGVSixteenRRNuRatio}{\ensuremath{0.9985}}

\providecommand{\MetricTGVSixteenSIMPLECDiv}{}\renewcommand{\MetricTGVSixteenSIMPLECDiv}{\ensuremath{1.26\times 10^{-5}}}
\providecommand{\MetricTGVSixteenSIMPLECEp}{}\renewcommand{\MetricTGVSixteenSIMPLECEp}{\ensuremath{6.07\times 10^{-4}}}
\providecommand{\MetricTGVSixteenSIMPLECEuReHun}{}\renewcommand{\MetricTGVSixteenSIMPLECEuReHun}{\ensuremath{2.61\times 10^{-6}}}
\providecommand{\MetricTGVSixteenSIMPLECEuReMax}{}\renewcommand{\MetricTGVSixteenSIMPLECEuReMax}{\ensuremath{2.24\times 10^{-6}}}
\providecommand{\MetricTGVSixteenSIMPLECEuReMin}{}\renewcommand{\MetricTGVSixteenSIMPLECEuReMin}{\ensuremath{8.03\times 10^{-6}}}
\providecommand{\MetricTGVSixteenSIMPLECNuRatio}{}\renewcommand{\MetricTGVSixteenSIMPLECNuRatio}{\ensuremath{0.9998}}

\providecommand{\MetricTGVSixteenTRTDiv}{}\renewcommand{\MetricTGVSixteenTRTDiv}{\ensuremath{0.01623}}
\providecommand{\MetricTGVSixteenTRTEp}{}\renewcommand{\MetricTGVSixteenTRTEp}{\ensuremath{0.01124}}
\providecommand{\MetricTGVSixteenTRTEuReHun}{}\renewcommand{\MetricTGVSixteenTRTEuReHun}{\ensuremath{1.23\times 10^{-3}}}
\providecommand{\MetricTGVSixteenTRTEuReMax}{}\renewcommand{\MetricTGVSixteenTRTEuReMax}{\ensuremath{6.23\times 10^{-4}}}
\providecommand{\MetricTGVSixteenTRTEuReMin}{}\renewcommand{\MetricTGVSixteenTRTEuReMin}{\ensuremath{7.91\times 10^{-6}}}
\providecommand{\MetricTGVSixteenTRTNuRatio}{}\renewcommand{\MetricTGVSixteenTRTNuRatio}{\ensuremath{0.9985}}

\providecommand{\MetricTGVSixteenVDFourierDiv}{}\renewcommand{\MetricTGVSixteenVDFourierDiv}{\ensuremath{3.00\times 10^{-3}}}
\providecommand{\MetricTGVSixteenVDFourierEp}{}\renewcommand{\MetricTGVSixteenVDFourierEp}{\ensuremath{2.78\times 10^{-3}}}
\providecommand{\MetricTGVSixteenVDFourierEuReHun}{}\renewcommand{\MetricTGVSixteenVDFourierEuReHun}{\ensuremath{3.43\times 10^{-4}}}
\providecommand{\MetricTGVSixteenVDFourierEuReMax}{}\renewcommand{\MetricTGVSixteenVDFourierEuReMax}{\ensuremath{2.10\times 10^{-5}}}
\providecommand{\MetricTGVSixteenVDFourierEuReMin}{}\renewcommand{\MetricTGVSixteenVDFourierEuReMin}{\ensuremath{7.74\times 10^{-6}}}
\providecommand{\MetricTGVSixteenVDFourierNuRatio}{}\renewcommand{\MetricTGVSixteenVDFourierNuRatio}{\ensuremath{0.9991}}

\providecommand{\MetricTGVSixteenVDUzawaDiv}{}\renewcommand{\MetricTGVSixteenVDUzawaDiv}{\ensuremath{3.00\times 10^{-3}}}
\providecommand{\MetricTGVSixteenVDUzawaEp}{}\renewcommand{\MetricTGVSixteenVDUzawaEp}{\ensuremath{2.78\times 10^{-3}}}
\providecommand{\MetricTGVSixteenVDUzawaEuReHun}{}\renewcommand{\MetricTGVSixteenVDUzawaEuReHun}{\ensuremath{3.43\times 10^{-4}}}
\providecommand{\MetricTGVSixteenVDUzawaEuReMax}{}\renewcommand{\MetricTGVSixteenVDUzawaEuReMax}{\ensuremath{2.10\times 10^{-5}}}
\providecommand{\MetricTGVSixteenVDUzawaEuReMin}{}\renewcommand{\MetricTGVSixteenVDUzawaEuReMin}{\ensuremath{7.74\times 10^{-6}}}
\providecommand{\MetricTGVSixteenVDUzawaNuRatio}{}\renewcommand{\MetricTGVSixteenVDUzawaNuRatio}{\ensuremath{0.9991}}

\providecommand{\MetricTGVSixteenWENODiv}{}\renewcommand{\MetricTGVSixteenWENODiv}{\ensuremath{1.96\times 10^{-7}}}
\providecommand{\MetricTGVSixteenWENOEp}{}\renewcommand{\MetricTGVSixteenWENOEp}{\ensuremath{8.07\times 10^{-9}}}
\providecommand{\MetricTGVSixteenWENOEuReHun}{}\renewcommand{\MetricTGVSixteenWENOEuReHun}{\ensuremath{5.91\times 10^{-8}}}
\providecommand{\MetricTGVSixteenWENOEuReMax}{}\renewcommand{\MetricTGVSixteenWENOEuReMax}{\ensuremath{1.71\times 10^{-7}}}
\providecommand{\MetricTGVSixteenWENOEuReMin}{}\renewcommand{\MetricTGVSixteenWENOEuReMin}{\ensuremath{6.82\times 10^{-13}}}
\providecommand{\MetricTGVSixteenWENONuRatio}{}\renewcommand{\MetricTGVSixteenWENONuRatio}{\ensuremath{1.0000}}

\providecommand{\MetricTGVThreeDDNSPeakDissipReSixteenHun}{}\renewcommand{\MetricTGVThreeDDNSPeakDissipReSixteenHun}{\ensuremath{0.01291}}
\providecommand{\MetricTGVThreeDDNSPeakTimeReSixteenHun}{}\renewcommand{\MetricTGVThreeDDNSPeakTimeReSixteenHun}{\ensuremath{8.88}}

\providecommand{\MetricTGVThreeDLESNOneTwoEightCsFourteenDissErr}{}\renewcommand{\MetricTGVThreeDLESNOneTwoEightCsFourteenDissErr}{\ensuremath{0.1395}}
\providecommand{\MetricTGVThreeDLESNOneTwoEightCsFourteenEkErr}{}\renewcommand{\MetricTGVThreeDLESNOneTwoEightCsFourteenEkErr}{\ensuremath{0.04518}}

\providecommand{\MetricTGVThreeDLESNOneTwoEightCsFourteenOmErr}{}\renewcommand{\MetricTGVThreeDLESNOneTwoEightCsFourteenOmErr}{\ensuremath{0.5255}}
\providecommand{\MetricTGVThreeDLESNOneTwoEightCsFourteenSweeps}{}\renewcommand{\MetricTGVThreeDLESNOneTwoEightCsFourteenSweeps}{\ensuremath{43.4}}
\providecommand{\MetricTGVThreeDLESNOneTwoEightCsFourteenWall}{}\renewcommand{\MetricTGVThreeDLESNOneTwoEightCsFourteenWall}{\ensuremath{114.91}}

\providecommand{\MetricTGVThreeDLESNOneTwoEightCsSeventeenEkErr}{}\renewcommand{\MetricTGVThreeDLESNOneTwoEightCsSeventeenEkErr}{\ensuremath{0.04716}}

\providecommand{\MetricTGVThreeDLESNOneTwoEightCsSeventeenOmErr}{}\renewcommand{\MetricTGVThreeDLESNOneTwoEightCsSeventeenOmErr}{\ensuremath{0.5898}}
\providecommand{\MetricTGVThreeDLESNOneTwoEightCsSeventeenSweeps}{}\renewcommand{\MetricTGVThreeDLESNOneTwoEightCsSeventeenSweeps}{\ensuremath{42.1}}
\providecommand{\MetricTGVThreeDLESNOneTwoEightCsSeventeenWall}{}\renewcommand{\MetricTGVThreeDLESNOneTwoEightCsSeventeenWall}{\ensuremath{111.51}}

\providecommand{\MetricTGVThreeDLESNTwoFiveSixCsSeventeenEkErr}{}\renewcommand{\MetricTGVThreeDLESNTwoFiveSixCsSeventeenEkErr}{\ensuremath{0.01545}}

\providecommand{\MetricTGVThreeDLESNTwoFiveSixCsSeventeenOmErr}{}\renewcommand{\MetricTGVThreeDLESNTwoFiveSixCsSeventeenOmErr}{\ensuremath{0.3344}}
\providecommand{\MetricTGVThreeDLESNTwoFiveSixCsSeventeenSweeps}{}\renewcommand{\MetricTGVThreeDLESNTwoFiveSixCsSeventeenSweeps}{\ensuremath{35.7}}
\providecommand{\MetricTGVThreeDLESNTwoFiveSixCsSeventeenWall}{}\renewcommand{\MetricTGVThreeDLESNTwoFiveSixCsSeventeenWall}{\ensuremath{1582.31}}

\providecommand{\MetricTGVThreeDReHunVDNTwoFiveSixFPSixtyFourDissErr}{}\renewcommand{\MetricTGVThreeDReHunVDNTwoFiveSixFPSixtyFourDissErr}{\ensuremath{9.98\times 10^{-4}}}
\providecommand{\MetricTGVThreeDReHunVDNTwoFiveSixFPSixtyFourEkErr}{}\renewcommand{\MetricTGVThreeDReHunVDNTwoFiveSixFPSixtyFourEkErr}{\ensuremath{5.25\times 10^{-4}}}
\providecommand{\MetricTGVThreeDReHunVDNTwoFiveSixFPSixtyFourOmErr}{}\renewcommand{\MetricTGVThreeDReHunVDNTwoFiveSixFPSixtyFourOmErr}{\ensuremath{1.22\times 10^{-3}}}
\providecommand{\MetricTGVThreeDReHunVDNTwoFiveSixFPSixtyFourSweeps}{}\renewcommand{\MetricTGVThreeDReHunVDNTwoFiveSixFPSixtyFourSweeps}{\ensuremath{16.5}}
\providecommand{\MetricTGVThreeDReHunVDNTwoFiveSixFPSixtyFourWall}{}\renewcommand{\MetricTGVThreeDReHunVDNTwoFiveSixFPSixtyFourWall}{\ensuremath{690.99}}

\providecommand{\MetricTGVThreeDReSixteenHunBGKNTwoFiveSixFPSixtyFourEkErr}{}\renewcommand{\MetricTGVThreeDReSixteenHunBGKNTwoFiveSixFPSixtyFourEkErr}{\ensuremath{0.01371}}
\providecommand{\MetricTGVThreeDReSixteenHunBGKNTwoFiveSixFPSixtyFourOmErr}{}\renewcommand{\MetricTGVThreeDReSixteenHunBGKNTwoFiveSixFPSixtyFourOmErr}{\ensuremath{0.0649}}
\providecommand{\MetricTGVThreeDReSixteenHunBGKNTwoFiveSixFPSixtyFourWall}{}\renewcommand{\MetricTGVThreeDReSixteenHunBGKNTwoFiveSixFPSixtyFourWall}{\ensuremath{57.67}}

\providecommand{\MetricTGVThreeDReSixteenHunHRRNTwoFiveSixFPSixtyFourEkErr}{}\renewcommand{\MetricTGVThreeDReSixteenHunHRRNTwoFiveSixFPSixtyFourEkErr}{\ensuremath{4.98\times 10^{-3}}}
\providecommand{\MetricTGVThreeDReSixteenHunHRRNTwoFiveSixFPSixtyFourOmErr}{}\renewcommand{\MetricTGVThreeDReSixteenHunHRRNTwoFiveSixFPSixtyFourOmErr}{\ensuremath{0.1521}}
\providecommand{\MetricTGVThreeDReSixteenHunHRRNTwoFiveSixFPSixtyFourWall}{}\renewcommand{\MetricTGVThreeDReSixteenHunHRRNTwoFiveSixFPSixtyFourWall}{\ensuremath{413.24}}

\providecommand{\MetricTGVThreeDReSixteenHunVDNOneNineTwoFPSixtyFourDissErr}{}\renewcommand{\MetricTGVThreeDReSixteenHunVDNOneNineTwoFPSixtyFourDissErr}{\ensuremath{0.07953}}
\providecommand{\MetricTGVThreeDReSixteenHunVDNOneNineTwoFPSixtyFourEkErr}{}\renewcommand{\MetricTGVThreeDReSixteenHunVDNOneNineTwoFPSixtyFourEkErr}{\ensuremath{0.01529}}
\providecommand{\MetricTGVThreeDReSixteenHunVDNOneNineTwoFPSixtyFourOmErr}{}\renewcommand{\MetricTGVThreeDReSixteenHunVDNOneNineTwoFPSixtyFourOmErr}{\ensuremath{0.09862}}
\providecommand{\MetricTGVThreeDReSixteenHunVDNOneNineTwoFPSixtyFourSweeps}{}\renewcommand{\MetricTGVThreeDReSixteenHunVDNOneNineTwoFPSixtyFourSweeps}{\ensuremath{34.7}}
\providecommand{\MetricTGVThreeDReSixteenHunVDNOneNineTwoFPSixtyFourWall}{}\renewcommand{\MetricTGVThreeDReSixteenHunVDNOneNineTwoFPSixtyFourWall}{\ensuremath{482.67}}

\providecommand{\MetricTGVThreeDReSixteenHunVDNOneTwoEightFPSixtyFourDissErr}{}\renewcommand{\MetricTGVThreeDReSixteenHunVDNOneTwoEightFPSixtyFourDissErr}{\ensuremath{0.07451}}
\providecommand{\MetricTGVThreeDReSixteenHunVDNOneTwoEightFPSixtyFourEkErr}{}\renewcommand{\MetricTGVThreeDReSixteenHunVDNOneTwoEightFPSixtyFourEkErr}{\ensuremath{0.02551}}
\providecommand{\MetricTGVThreeDReSixteenHunVDNOneTwoEightFPSixtyFourOmErr}{}\renewcommand{\MetricTGVThreeDReSixteenHunVDNOneTwoEightFPSixtyFourOmErr}{\ensuremath{0.1271}}
\providecommand{\MetricTGVThreeDReSixteenHunVDNOneTwoEightFPSixtyFourSweeps}{}\renewcommand{\MetricTGVThreeDReSixteenHunVDNOneTwoEightFPSixtyFourSweeps}{\ensuremath{42.7}}
\providecommand{\MetricTGVThreeDReSixteenHunVDNOneTwoEightFPSixtyFourWall}{}\renewcommand{\MetricTGVThreeDReSixteenHunVDNOneTwoEightFPSixtyFourWall}{\ensuremath{107.87}}

\providecommand{\MetricTGVThreeDReSixteenHunVDNSixtyFourFPSixtyFourDissErr}{}\renewcommand{\MetricTGVThreeDReSixteenHunVDNSixtyFourFPSixtyFourDissErr}{\emph{Blow-up}}
\providecommand{\MetricTGVThreeDReSixteenHunVDNSixtyFourFPSixtyFourEkErr}{}\renewcommand{\MetricTGVThreeDReSixteenHunVDNSixtyFourFPSixtyFourEkErr}{\emph{Blow-up}}
\providecommand{\MetricTGVThreeDReSixteenHunVDNSixtyFourFPSixtyFourOmErr}{}\renewcommand{\MetricTGVThreeDReSixteenHunVDNSixtyFourFPSixtyFourOmErr}{\emph{Blow-up}}
\providecommand{\MetricTGVThreeDReSixteenHunVDNSixtyFourFPSixtyFourSweeps}{}\renewcommand{\MetricTGVThreeDReSixteenHunVDNSixtyFourFPSixtyFourSweeps}{\emph{Blow-up}}
\providecommand{\MetricTGVThreeDReSixteenHunVDNSixtyFourFPSixtyFourWall}{}\renewcommand{\MetricTGVThreeDReSixteenHunVDNSixtyFourFPSixtyFourWall}{\emph{Blow-up}}

\providecommand{\MetricTGVThreeDReSixteenHunVDNTwoFiveSixFPSixtyFourEkErr}{}\renewcommand{\MetricTGVThreeDReSixteenHunVDNTwoFiveSixFPSixtyFourEkErr}{\ensuremath{8.19\times 10^{-3}}}
\providecommand{\MetricTGVThreeDReSixteenHunVDNTwoFiveSixFPSixtyFourOmErr}{}\renewcommand{\MetricTGVThreeDReSixteenHunVDNTwoFiveSixFPSixtyFourOmErr}{\ensuremath{0.06761}}
\providecommand{\MetricTGVThreeDReSixteenHunVDNTwoFiveSixFPSixtyFourSweeps}{}\renewcommand{\MetricTGVThreeDReSixteenHunVDNTwoFiveSixFPSixtyFourSweeps}{\ensuremath{32.0}}
\providecommand{\MetricTGVThreeDReSixteenHunVDNTwoFiveSixFPSixtyFourWall}{}\renewcommand{\MetricTGVThreeDReSixteenHunVDNTwoFiveSixFPSixtyFourWall}{\ensuremath{1350.02}}

\providecommand{\MetricTGVThreeDReSixteenHunVDNTwoFiveSixFPThirtyTwoDissErr}{}\renewcommand{\MetricTGVThreeDReSixteenHunVDNTwoFiveSixFPThirtyTwoDissErr}{\ensuremath{0.8776}}
\providecommand{\MetricTGVThreeDReSixteenHunVDNTwoFiveSixFPThirtyTwoEkErr}{}\renewcommand{\MetricTGVThreeDReSixteenHunVDNTwoFiveSixFPThirtyTwoEkErr}{\ensuremath{0.1019}}
\providecommand{\MetricTGVThreeDReSixteenHunVDNTwoFiveSixFPThirtyTwoOmErr}{}\renewcommand{\MetricTGVThreeDReSixteenHunVDNTwoFiveSixFPThirtyTwoOmErr}{\ensuremath{0.2412}}
\providecommand{\MetricTGVThreeDReSixteenHunVDNTwoFiveSixFPThirtyTwoSweeps}{}\renewcommand{\MetricTGVThreeDReSixteenHunVDNTwoFiveSixFPThirtyTwoSweeps}{\ensuremath{6.1}}
\providecommand{\MetricTGVThreeDReSixteenHunVDNTwoFiveSixFPThirtyTwoWall}{}\renewcommand{\MetricTGVThreeDReSixteenHunVDNTwoFiveSixFPThirtyTwoWall}{\ensuremath{125.05}}

\providecommand{\MetricTGVThreeDReTenVDNOneTwoEightFPSixtyFourDissErr}{}\renewcommand{\MetricTGVThreeDReTenVDNOneTwoEightFPSixtyFourDissErr}{\ensuremath{7.08\times 10^{-4}}}
\providecommand{\MetricTGVThreeDReTenVDNOneTwoEightFPSixtyFourEkErr}{}\renewcommand{\MetricTGVThreeDReTenVDNOneTwoEightFPSixtyFourEkErr}{\ensuremath{5.15\times 10^{-4}}}
\providecommand{\MetricTGVThreeDReTenVDNOneTwoEightFPSixtyFourOmErr}{}\renewcommand{\MetricTGVThreeDReTenVDNOneTwoEightFPSixtyFourOmErr}{\ensuremath{4.96\times 10^{-4}}}
\providecommand{\MetricTGVThreeDReTenVDNOneTwoEightFPSixtyFourSweeps}{}\renewcommand{\MetricTGVThreeDReTenVDNOneTwoEightFPSixtyFourSweeps}{\ensuremath{18.2}}
\providecommand{\MetricTGVThreeDReTenVDNOneTwoEightFPSixtyFourWall}{}\renewcommand{\MetricTGVThreeDReTenVDNOneTwoEightFPSixtyFourWall}{\ensuremath{45.51}}

\providecommand{\MetricTGVThreeDReTenVDNTwoFiveSixFPSixtyFourDissErr}{}\renewcommand{\MetricTGVThreeDReTenVDNTwoFiveSixFPSixtyFourDissErr}{\ensuremath{2.86\times 10^{-4}}}
\providecommand{\MetricTGVThreeDReTenVDNTwoFiveSixFPSixtyFourEkErr}{}\renewcommand{\MetricTGVThreeDReTenVDNTwoFiveSixFPSixtyFourEkErr}{\ensuremath{2.40\times 10^{-4}}}
\providecommand{\MetricTGVThreeDReTenVDNTwoFiveSixFPSixtyFourOmErr}{}\renewcommand{\MetricTGVThreeDReTenVDNTwoFiveSixFPSixtyFourOmErr}{\ensuremath{2.43\times 10^{-4}}}
\providecommand{\MetricTGVThreeDReTenVDNTwoFiveSixFPSixtyFourSweeps}{}\renewcommand{\MetricTGVThreeDReTenVDNTwoFiveSixFPSixtyFourSweeps}{\ensuremath{25.0}}
\providecommand{\MetricTGVThreeDReTenVDNTwoFiveSixFPSixtyFourWall}{}\renewcommand{\MetricTGVThreeDReTenVDNTwoFiveSixFPSixtyFourWall}{\ensuremath{1051.03}}

\providecommand{\MetricTGVThreeDReTenVDNTwoFiveSixFPThirtyTwoDissErr}{}\renewcommand{\MetricTGVThreeDReTenVDNTwoFiveSixFPThirtyTwoDissErr}{\ensuremath{0.703}}
\providecommand{\MetricTGVThreeDReTenVDNTwoFiveSixFPThirtyTwoEkErr}{}\renewcommand{\MetricTGVThreeDReTenVDNTwoFiveSixFPThirtyTwoEkErr}{\ensuremath{1.466}}
\providecommand{\MetricTGVThreeDReTenVDNTwoFiveSixFPThirtyTwoOmErr}{}\renewcommand{\MetricTGVThreeDReTenVDNTwoFiveSixFPThirtyTwoOmErr}{\ensuremath{1.478}}
\providecommand{\MetricTGVThreeDReTenVDNTwoFiveSixFPThirtyTwoSweeps}{}\renewcommand{\MetricTGVThreeDReTenVDNTwoFiveSixFPThirtyTwoSweeps}{\ensuremath{2.0}}
\providecommand{\MetricTGVThreeDReTenVDNTwoFiveSixFPThirtyTwoWall}{}\renewcommand{\MetricTGVThreeDReTenVDNTwoFiveSixFPThirtyTwoWall}{\ensuremath{37.08}}

\providecommand{\MetricTGVThreeDReThreeKVDNTwoFiveSixFPSixtyFourDissErr}{}\renewcommand{\MetricTGVThreeDReThreeKVDNTwoFiveSixFPSixtyFourDissErr}{\emph{Blow-up}}
\providecommand{\MetricTGVThreeDReThreeKVDNTwoFiveSixFPSixtyFourEkErr}{}\renewcommand{\MetricTGVThreeDReThreeKVDNTwoFiveSixFPSixtyFourEkErr}{\emph{Blow-up}}
\providecommand{\MetricTGVThreeDReThreeKVDNTwoFiveSixFPSixtyFourOmErr}{}\renewcommand{\MetricTGVThreeDReThreeKVDNTwoFiveSixFPSixtyFourOmErr}{\emph{Blow-up}}
\providecommand{\MetricTGVThreeDReThreeKVDNTwoFiveSixFPSixtyFourSweeps}{}\renewcommand{\MetricTGVThreeDReThreeKVDNTwoFiveSixFPSixtyFourSweeps}{\emph{Blow-up}}
\providecommand{\MetricTGVThreeDReThreeKVDNTwoFiveSixFPSixtyFourWall}{}\renewcommand{\MetricTGVThreeDReThreeKVDNTwoFiveSixFPSixtyFourWall}{\emph{Blow-up}}

\providecommand{\MetricTGVThreeDStiffBGKNOneTwoEightWall}{}\renewcommand{\MetricTGVThreeDStiffBGKNOneTwoEightWall}{\ensuremath{1.26}}
\providecommand{\MetricTGVThreeDStiffBGKNTwoFiveSixEquivWall}{}\renewcommand{\MetricTGVThreeDStiffBGKNTwoFiveSixEquivWall}{\ensuremath{37.13}}

\providecommand{\MetricTGVThreeDStiffVDNOneTwoEightMomDrift}{}\renewcommand{\MetricTGVThreeDStiffVDNOneTwoEightMomDrift}{\ensuremath{5.10\times 10^{-18}}}

\providecommand{\MetricTGVThreeDStiffVDNOneTwoEightSweeps}{}\renewcommand{\MetricTGVThreeDStiffVDNOneTwoEightSweeps}{\ensuremath{136.4}}
\providecommand{\MetricTGVThreeDStiffVDNOneTwoEightWall}{}\renewcommand{\MetricTGVThreeDStiffVDNOneTwoEightWall}{\ensuremath{18.08}}

\providecommand{\MetricTGVThreeDStiffVDNTwoFiveSixMomDrift}{}\renewcommand{\MetricTGVThreeDStiffVDNTwoFiveSixMomDrift}{\ensuremath{4.48\times 10^{-18}}}

\providecommand{\MetricTGVThreeDStiffVDNTwoFiveSixSweeps}{}\renewcommand{\MetricTGVThreeDStiffVDNTwoFiveSixSweeps}{\ensuremath{232.4}}
\providecommand{\MetricTGVThreeDStiffVDNTwoFiveSixWall}{}\renewcommand{\MetricTGVThreeDStiffVDNTwoFiveSixWall}{\ensuremath{553.58}}

\providecommand{\MetricVarViscTwoDNTwoFiveSixOpErr}{}\renewcommand{\MetricVarViscTwoDNTwoFiveSixOpErr}{\ensuremath{3.94\times 10^{-4}}}
\providecommand{\MetricVarViscTwoDNTwoFiveSixOpOrder}{}\renewcommand{\MetricVarViscTwoDNTwoFiveSixOpOrder}{\ensuremath{2.00}}
\providecommand{\MetricVarViscTwoDNTwoFiveSixPDEEu}{}\renewcommand{\MetricVarViscTwoDNTwoFiveSixPDEEu}{\ensuremath{8.01\times 10^{-5}}}
\providecommand{\MetricVarViscTwoDNTwoFiveSixPDEOrder}{}\renewcommand{\MetricVarViscTwoDNTwoFiveSixPDEOrder}{\ensuremath{1.95}}

\providecommand{\MetricWomConvNOneTwoEightAlphaHalfEu}{}\renewcommand{\MetricWomConvNOneTwoEightAlphaHalfEu}{\ensuremath{1.89\times 10^{-6}}}

\providecommand{\MetricWomConvNOneTwoEightThomasSweeps}{}\renewcommand{\MetricWomConvNOneTwoEightThomasSweeps}{\ensuremath{3.0}}
\providecommand{\MetricWomConvNOneTwoEightUzawaEu}{}\renewcommand{\MetricWomConvNOneTwoEightUzawaEu}{\ensuremath{1.89\times 10^{-6}}}
\providecommand{\MetricWomConvNOneTwoEightUzawaSweeps}{}\renewcommand{\MetricWomConvNOneTwoEightUzawaSweeps}{\ensuremath{5560.9}}

\providecommand{\MetricWomConvOrderAlphaFifty}{}\renewcommand{\MetricWomConvOrderAlphaFifty}{\ensuremath{2.00}}
\providecommand{\MetricWomConvOrderAlphaHalf}{}\renewcommand{\MetricWomConvOrderAlphaHalf}{\ensuremath{2.00}}
\providecommand{\MetricWomConvOrderAlphaTenTau}{}\renewcommand{\MetricWomConvOrderAlphaTenTau}{\ensuremath{2.00}}
\providecommand{\MetricWomFifteenACDTSEuFifty}{}\renewcommand{\MetricWomFifteenACDTSEuFifty}{\ensuremath{1.46\times 10^{-3}}}
\providecommand{\MetricWomFifteenACDTSEuHalf}{}\renewcommand{\MetricWomFifteenACDTSEuHalf}{\ensuremath{9.64\times 10^{-7}}}

\providecommand{\MetricWomFifteenBGKEuFifty}{}\renewcommand{\MetricWomFifteenBGKEuFifty}{\ensuremath{5.55\times 10^{-3}}}
\providecommand{\MetricWomFifteenBGKEuHalf}{}\renewcommand{\MetricWomFifteenBGKEuHalf}{\ensuremath{1.40\times 10^{-6}}}

\providecommand{\MetricWomFifteenBGKWall}{}\renewcommand{\MetricWomFifteenBGKWall}{\ensuremath{95.72} / \ensuremath{0.40}}
\providecommand{\MetricWomFifteenCMEuFifty}{}\renewcommand{\MetricWomFifteenCMEuFifty}{\ensuremath{5.39\times 10^{-3}}}
\providecommand{\MetricWomFifteenCMEuHalf}{}\renewcommand{\MetricWomFifteenCMEuHalf}{\ensuremath{1.40\times 10^{-6}}}

\providecommand{\MetricWomFifteenCMWall}{}\renewcommand{\MetricWomFifteenCMWall}{\ensuremath{103.12} / \ensuremath{0.44}}
\providecommand{\MetricWomFifteenCumEuFifty}{}\renewcommand{\MetricWomFifteenCumEuFifty}{\ensuremath{5.39\times 10^{-3}}}
\providecommand{\MetricWomFifteenCumEuHalf}{}\renewcommand{\MetricWomFifteenCumEuHalf}{\ensuremath{1.40\times 10^{-6}}}

\providecommand{\MetricWomFifteenCumWall}{}\renewcommand{\MetricWomFifteenCumWall}{\ensuremath{108.16} / \ensuremath{0.46}}
\providecommand{\MetricWomFifteenDTSTRTEuFifty}{}\renewcommand{\MetricWomFifteenDTSTRTEuFifty}{\emph{Blow-up}}
\providecommand{\MetricWomFifteenDTSTRTEuHalf}{}\renewcommand{\MetricWomFifteenDTSTRTEuHalf}{\ensuremath{1.84\times 10^{-6}}}

\providecommand{\MetricWomFifteenDTSTRTWall}{}\renewcommand{\MetricWomFifteenDTSTRTWall}{\ensuremath{2.34} / \emph{Blow-up}}
\providecommand{\MetricWomFifteenELBMEuFifty}{}\renewcommand{\MetricWomFifteenELBMEuFifty}{\ensuremath{5.57\times 10^{-3}}}
\providecommand{\MetricWomFifteenELBMEuHalf}{}\renewcommand{\MetricWomFifteenELBMEuHalf}{\ensuremath{1.40\times 10^{-6}}}

\providecommand{\MetricWomFifteenELBMWall}{}\renewcommand{\MetricWomFifteenELBMWall}{\ensuremath{287.99} / \ensuremath{3.97}}
\providecommand{\MetricWomFifteenHRREuFifty}{}\renewcommand{\MetricWomFifteenHRREuFifty}{\ensuremath{6.93\times 10^{-3}}}
\providecommand{\MetricWomFifteenHRREuHalf}{}\renewcommand{\MetricWomFifteenHRREuHalf}{\ensuremath{1.40\times 10^{-6}}}

\providecommand{\MetricWomFifteenHRRWall}{}\renewcommand{\MetricWomFifteenHRRWall}{\ensuremath{144.92} / \ensuremath{0.61}}
\providecommand{\MetricWomFifteenKimMoinEuFifty}{}\renewcommand{\MetricWomFifteenKimMoinEuFifty}{\ensuremath{1.46\times 10^{-3}}}
\providecommand{\MetricWomFifteenKimMoinEuHalf}{}\renewcommand{\MetricWomFifteenKimMoinEuHalf}{\ensuremath{9.64\times 10^{-7}}}

\providecommand{\MetricWomFifteenMRTEuFifty}{}\renewcommand{\MetricWomFifteenMRTEuFifty}{\ensuremath{1.73\times 10^{-3}}}
\providecommand{\MetricWomFifteenMRTEuHalf}{}\renewcommand{\MetricWomFifteenMRTEuHalf}{\ensuremath{3.89\times 10^{-7}}}

\providecommand{\MetricWomFifteenMRTWall}{}\renewcommand{\MetricWomFifteenMRTWall}{\ensuremath{122.24} / \ensuremath{0.53}}
\providecommand{\MetricWomFifteenPIMPLEEuFifty}{}\renewcommand{\MetricWomFifteenPIMPLEEuFifty}{\ensuremath{1.46\times 10^{-3}}}
\providecommand{\MetricWomFifteenPIMPLEEuHalf}{}\renewcommand{\MetricWomFifteenPIMPLEEuHalf}{\ensuremath{9.64\times 10^{-7}}}

\providecommand{\MetricWomFifteenRREuFifty}{}\renewcommand{\MetricWomFifteenRREuFifty}{\ensuremath{5.39\times 10^{-3}}}
\providecommand{\MetricWomFifteenRREuHalf}{}\renewcommand{\MetricWomFifteenRREuHalf}{\ensuremath{1.40\times 10^{-6}}}

\providecommand{\MetricWomFifteenRRWall}{}\renewcommand{\MetricWomFifteenRRWall}{\ensuremath{105.90} / \ensuremath{0.45}}
\providecommand{\MetricWomFifteenSIMPLECEuFifty}{}\renewcommand{\MetricWomFifteenSIMPLECEuFifty}{\ensuremath{1.46\times 10^{-3}}}
\providecommand{\MetricWomFifteenSIMPLECEuHalf}{}\renewcommand{\MetricWomFifteenSIMPLECEuHalf}{\ensuremath{9.64\times 10^{-7}}}

\providecommand{\MetricWomFifteenTRTEuFifty}{}\renewcommand{\MetricWomFifteenTRTEuFifty}{\ensuremath{2.48\times 10^{-3}}}
\providecommand{\MetricWomFifteenTRTEuHalf}{}\renewcommand{\MetricWomFifteenTRTEuHalf}{\ensuremath{1.40\times 10^{-6}}}

\providecommand{\MetricWomFifteenTRTWall}{}\renewcommand{\MetricWomFifteenTRTWall}{\ensuremath{99.96} / \ensuremath{0.41}}
\providecommand{\MetricWomFifteenVDThomasEuFifty}{}\renewcommand{\MetricWomFifteenVDThomasEuFifty}{\ensuremath{5.23\times 10^{-3}}}
\providecommand{\MetricWomFifteenVDThomasEuHalf}{}\renewcommand{\MetricWomFifteenVDThomasEuHalf}{\ensuremath{1.85\times 10^{-6}}}

\providecommand{\MetricWomFifteenVDThomasWall}{}\renewcommand{\MetricWomFifteenVDThomasWall}{\ensuremath{0.46} / \ensuremath{0.72}}
\providecommand{\MetricWomFifteenWENOEuFifty}{}\renewcommand{\MetricWomFifteenWENOEuFifty}{\ensuremath{9.76\times 10^{-5}}}
\providecommand{\MetricWomFifteenWENOEuHalf}{}\renewcommand{\MetricWomFifteenWENOEuHalf}{\ensuremath{2.26\times 10^{-12}}}

}{}

\begin{document}

\begin{frontmatter}

\title{A Viscosity-Decoupled Uzawa Lattice Boltzmann Method for Incompressible Flows Across Reynolds Numbers}

\author[inst1,inst2,inst3]{Yuan Yu\texorpdfstring{\,\orcidlink{0000-0001-5125-2492}}{}}

\author[inst1]{Yaolong Yu\texorpdfstring{\,\orcidlink{0009-0002-9599-4353}}{}}

\author[inst1]{Yanhui Ding\texorpdfstring{\,\orcidlink{0009-0008-0420-6854}\corref{cor1}}{}}
\ead{yanhuidanth@163.com}

\cortext[cor1]{Corresponding author.}

\affiliation[inst1]{organization={School of Mathematics and Computational Science, Xiangtan University},
            city={Xiangtan},
            postcode={411105},
            country={China}}

\affiliation[inst2]{organization={National Center for Applied Mathematics in Hunan},
            city={Xiangtan},
            postcode={411105},
            country={China}}

\affiliation[inst3]{organization={Hunan Key Laboratory for Computation and Simulation in Science and Engineering, Xiangtan University},
            city={Xiangtan},
            postcode={411105},
            country={China}}

\begin{abstract}
Conventional lattice Boltzmann methods (LBMs) tie viscosity to collision relaxation, which induces diffusive stiffness at low Reynolds numbers, non-hydrodynamic instabilities at high Reynolds numbers, and passband breakdown in variable-relaxation dual-time schemes, whereas continuum incompressible solvers require global pressure-Poisson solves every step. We propose a Viscosity-Decoupled Uzawa Lattice Boltzmann Method (\VDULBM{}) in D2Q9 and D3Q19 that permanently fixes collision at unit relaxation under smooth harmonic scaling and clears pre-collision non-conserved kinetic memory in one sweep. Constant, heterogeneous, and Smagorinsky eddy viscosities enter through a parameter-free post-streaming bond-symmetric momentum-flux lift that conserves discrete momentum pairwise without perturbing continuity. Variable-step BDF1/BDF2 marching is accelerated by a matrix-free local polynomial Uzawa smoother on the $(d+1)$ conserved moments, paired with link-wise Bouzidi curved walls and shifted ghost open boundaries. Across $\mathrm{Re}\in[10^{-4},10^6]$, \VDULBM{} achieves smooth low-Mach second-order space--time accuracy, accelerates unsteady DFG~2D-2 shedding per cycle by \VDLBMDFGOpenFOAMRatio{}--$15.85\times$ over matched 20-core OpenFOAM-v2412, and lowers $256^3$ $\mathrm{Re}=1600$ resolved enstrophy error by $2.25\times$ versus hybrid recursive regularized LBM. On an NVIDIA A100 GPU, the fused D3Q19 solver attains $80.9\%\text{--}85.5\%$ of measured copy bandwidth in multi-pass port traffic and cuts bounded-relaxation diffusive step cost at high viscosity contrasts.
\end{abstract}


\end{frontmatter}

\section{Introduction}
\label{sec:intro}

Incompressible viscous flows across microfluidic devices, pulsatile biological conduits, separated wakes, and turbulent mixing layers span ten orders of magnitude in Reynolds number ($\ReNum \in \VDLBMReSpan$) and often involve wide spatial variations in effective viscosity~\citep{Womersley1955,Brachet1983,BrownMinion1995,SchaferTurek1996}. Simulating such flows with a single numerical method places competing demands on time-step stability, small-scale accuracy, and parallel locality. In continuum computational fluid dynamics, fractional-step projection schemes~\citep{Chorin1968,KimMoin1985} and colocated pressure-correction algorithms such as SIMPLE and PIMPLE~\citep{PatankarSpalding1972,RhieChow1983,Issa1986,Weller1998} discretize the macroscopic velocity and pressure directly and thus avoid non-hydrodynamic kinetic modes altogether. Nevertheless, enforcing the divergence-free constraint requires assembling and solving a global sparse elliptic pressure-Poisson equation at every time step or outer corrector. On stretched body-fitted meshes at moderate Courant numbers or on large three-dimensional grids, repeated Krylov and algebraic-multigrid reductions become a primary scalability bottleneck.

Explicit lattice Boltzmann methods (LBMs)~\citep{Qian1992,HeLuo1997,ChenDoolen1998,Krueger2017,Lallemand2021} circumvent global elliptic solves by lifting advection--diffusion onto a finite discrete-velocity set where streaming is exact along lattice links and nonlinearity resides locally in collision. In standard LBMs, however, kinematic viscosity is tied directly to symmetric collision relaxation. In creeping flows or regions of high local viscosity, keeping relaxation times bounded forces explicit steps to scale diffusively as $\dtphys \propto h^2/\numax$, whereas allowing large relaxation rates introduces relaxation-dependent boundary slip unless specialized multi-relaxation closures are tuned~\citep{Ginzburg2008,dHumieresGinzburg2009}. At the opposite extreme of high Reynolds numbers, symmetric relaxation approaches its inviscid stability boundary from above; non-hydrodynamic kinetic modes are then weakly damped and prone to dispersive oscillations or blow-up~\citep{LallemandLuo2000,Dellar2003,Coreixas2019}. Advanced collision models, including regularized and hybrid recursive regularized~\citep{LattChopard2006,Coreixas2017,Jacob2018}, non-orthogonal central-moment~\citep{DeRosis2016EPL,DeRosis2017PRE}, cumulant~\citep{Geier2015}, and entropic~\citep{AnsumaliKarlin2002} formulations, markedly extend high-$\ReNum$ stability by filtering or relaxing higher-order moments independently. Even so, higher-order moment filtering can introduce numerical hyper-viscosity that attenuates marginally resolved vortex stretching, while explicit time integration remains bound by acoustic and diffusive step limits when viscosity varies across several decades.

Pseudo-time dual-time-stepping LBMs~\citep{MerkleAthavale1987,GuoZhaoShi2004,Gsell2020,JiLuo2026} decouple the outer physical time step from lattice streaming through inner steady-state iterations, typically accelerated by geometric multigrid hierarchies. Recently, \citet{Ding2026MSCLBM} introduced a macroscopic shadow residual correction (\MSCLBM{}) showing that when symmetric and antisymmetric relaxation times are both fixed at unity, pre-collision non-conserved perturbations vanish in a single sweep and periodic inner iterations can be accelerated on the $(d+1)$ conserved moments in Fourier space. Yet extending dual-time LBMs across Reynolds numbers by varying collision relaxation away from unity re-introduces non-conserved mode memory. As shown in Section~\ref{subsec:sec2_vartau_failure}, this coupling triggers two-sided Floquet--Bloch passband instabilities and deceptive transient residual dips prior to divergence. From a complementary perspective, explicit unit-relaxation schemes such as Inamuro's Lattice Kinetic Scheme (\LKS{}~\citep{Inamuro2002}) and Asinari's Link-Wise Artificial Compressibility Method (\LWACM{}~\citep{Asinari2012}) demonstrated that fixing relaxation at unity allows physical viscosity to be adjusted outside standard BGK collision, though only under explicit acoustic stepping and non-conservative stress finite differences. Whether unit-relaxation collision can be retained globally inside an implicit dual-time framework while preserving exact discrete momentum conservation for heterogeneous viscosity and local, matrix-free inner preconditioning on general domains has remained open.

In this work, we address this question by developing a unified Viscosity-Decoupled Uzawa Lattice Boltzmann Method (\VDULBM{}) on standard D2Q9 and D3Q19 lattices. A smooth harmonic lattice scaling maps all flow regimes into bounded lattice viscosities and keeps the lattice pseudo-step time-invariant across variable-step BDF1/BDF2 outer marching. By permanently fixing both relaxation rates at unity, local collision becomes a memoryless map slaved solely to the $(d+1)$ conserved moments, which annihilates arbitrary pre-collision non-conserved kinetic perturbations in a single inner sweep. Physical diffusion spanning constant viscosity, multi-decade spatially varying viscosity, and parameter-free discrete Smagorinsky subgrid eddy viscosity is restored after streaming through a bond-symmetric momentum-flux lift across lattice links. Unlike pointwise scaling, symmetric bond averaging enforces discrete action--reaction balance: it conserves global momentum to machine precision ($\le \VDLBMMMSMomentumCons$ in manufactured-solution tests) without perturbing discrete continuity, and recovers the incompressible variable-viscosity Navier--Stokes equations at second order in space and time under smooth low-Mach scaling. Furthermore, Sylvester's determinant identity reduces the full $Q\times Q$ preconditioned inner amplification spectrum to an exact $(d+1)\times(d+1)$ conserved-moment symbol. Approximating the resulting acoustic--viscous inverse via a two-step regularized Uzawa smoother and a separable binomial Nyquist filter yields a compact, matrix-free inner iteration free of global linear solves; both its convective operating window and its strong-elliptic sweep scaling are established analytically.

Because unit-relaxation collision carries zero pre-collision non-equilibrium stress memory, solid-wall and open-channel boundaries require no non-equilibrium stress reconstruction. Planar walls and off-grid curved obstacles share a single parameter-free link-wise Bouzidi reflection operator paired with rest-population mass neutralization on closed domains and uncalibrated Eulerian hydrodynamic force evaluation, while open channels employ symmetric half-cell shifted ghost inlets and outlets that eliminate diagonal-link mass and diffusion defects. Using this single fixed formulation without empirical stabilization switches, we benchmark \VDULBM{} against representative kinetic and continuum solvers across two- and three-dimensional regimes. The test suite progresses from creeping-to-high-$\ReNum$ Taylor--Green vortices, pulsatile Womersley channels, and under-resolved double shear layers to lid-driven cavity corner cascades, supercritical Hopf bifurcation, cylinder wakes compared on matched 20-core bare-metal Linux hardware against OpenFOAM-v2412, four-decade viscosity contrasts, and $256^3$ D3Q19 turbulence ($\ReNum \in \VDLBMGPUReSet$) and diffusion-stiff crossovers on an NVIDIA A100 GPU.

The remainder of this paper is organized as follows. Section~\ref{sec:formulation} introduces the harmonic lattice scaling, unit-relaxation dual-time sweep, bond-symmetric viscous flux, and asymptotic consistency theorems. Section~\ref{sec:boundaries} formulates the unified solid-wall and open-boundary closures. Section~\ref{sec:spectral} establishes the spectral necessity of permanent unit relaxation, proves the Sylvester moment reduction, and constructs the matrix-free local polynomial Uzawa smoother and unified algorithm. Sections~\ref{sec:sec5_fundamental} and~\ref{sec:sec6_complex_3d} present two- and three-dimensional numerical validations and parallel performance analyses. Section~\ref{sec:sec7_conclusions} summarizes the main findings and methodological scope, and \ref{app:archival} archives detailed Taylor-expansion proofs, constructive spectral counterexamples, and multi-solver benchmark ledgers.

\section{Viscosity-decoupled dual-time kinetic formulation}
\label{sec:formulation}

We consider the incompressible Navier--Stokes equations with possibly heterogeneous or shear-dependent kinematic viscosity $\nueff(\bm{x},t) > 0$ on $\Omega \subset \R^d$ ($d \in \{2,3\}$),
\begin{subequations}
\label{eq:target_ns}
\begin{align}
  \nabla \cdot \bm{u}_{\mathrm{phys}} &= 0, \label{eq:target_ns_cont}\\
  \partial_t \bm{u}_{\mathrm{phys}} + \nabla \cdot \bigl(\bm{u}_{\mathrm{phys}} \otimes \bm{u}_{\mathrm{phys}}\bigr) + \frac{1}{\rho_0}\nabla p
  &= \nabla \cdot \Bigl[\nueff(\bm{x},t)\bigl(\nabla\bm{u}_{\mathrm{phys}} + (\nabla\bm{u}_{\mathrm{phys}})^{\mathsf{T}}\bigr)\Bigr] + \bm{b}_{\mathrm{phys}}, \label{eq:target_ns_mom}
\end{align}
\end{subequations}
where $\bm{u}_{\mathrm{phys}}(\bm{x},t) \in \R^d$ denotes velocity, $p(\bm{x},t)$ pressure, $\rho_0 > 0$ reference density, and $\bm{b}_{\mathrm{phys}}(\bm{x},t)$ body-force acceleration.

On a uniform Cartesian lattice of spacing $h$, let $(d, Q) \in \{(2, 9), (3, 19)\}$ denote the standard D2Q9 ($w_0 = 4/9$, $w_{1\text{--}4} = 1/9$, $w_{5\text{--}8} = 1/36$) and D3Q19 ($w_0 = 1/3$, $w_{\mathrm{axis}} = 1/18$, $w_{\mathrm{diag}} = 1/36$) velocity sets $\{\bm{c}_k\}_{k=0}^{Q-1} \subset \{-1,0,1\}^d$ with rest velocity $\bm{c}_0 = \bm{0}$, link involution $\bar{k} = \opp(k)$ ($\bm{c}_{\bar{k}} = -\bm{c}_k$), lattice sound speed $c_s^2 = 1/3$, and second- and fourth-order isotropy
\begin{equation}
\label{eq:lattice_isotropy}
\begin{aligned}
  \sum_{k=0}^{Q-1} w_k &= 1, \qquad
  \sum_{k=0}^{Q-1} w_k c_{k\alpha} c_{k\beta} = c_s^2 \delta_{\alpha\beta}, \\
  \sum_{k=0}^{Q-1} w_k c_{k\alpha} c_{k\beta} c_{k\gamma} c_{k\theta}
  &= c_s^4 \bigl(\delta_{\alpha\beta}\delta_{\gamma\theta} + \delta_{\alpha\gamma}\delta_{\beta\theta} + \delta_{\alpha\theta}\delta_{\beta\gamma}\bigr),
\end{aligned}
\end{equation}
while odd moments vanish by parity ($w_{\bar{k}} = w_k$). We define the conserved-moment projector $P : \R^Q \to \R^{d+1}$ and linear lift operators $E : \R^{d+1} \to \R^Q$, $E_u : \R^d \to \R^Q$ by
\begin{gather}
\label{eq:P_E_Eu_defs}
  P\bm{f} := \begin{pmatrix} P_0\bm{f} \\ P_u\bm{f} \end{pmatrix}
  = \begin{pmatrix} \sum_{k=0}^{Q-1} f_k \\[3pt] \sum_{k=0}^{Q-1} \bm{c}_k f_k \end{pmatrix}, \\
  (E\,\delta\bm{m})_k := w_k\!\left(\delta\hat{p} + \frac{\bm{c}_k \cdot \delta\bm{u}}{c_s^2}\right), \qquad
  (E_u\,\bm{v})_k := \frac{w_k}{c_s^2}(\bm{c}_k \cdot \bm{v}), \nonumber
\end{gather}
which satisfy the exact algebraic orthogonality relations
\begin{equation}
\label{eq:PE_identities}
  P E = I_{d+1}, \qquad
  P E_u(\bm{v}) = \begin{pmatrix} 0 \\ \bm{v} \end{pmatrix}, \qquad
  \sum_{k=0}^{Q-1} \bm{c}_k \otimes \bm{c}_k \,(E_u\bm{v})_k = \bm{0}.
\end{equation}
Following \citet{HeLuo1997} and \citet{Guo2002}, the dimensionless incompressible equilibrium $f_k^{\mathrm{eq}}(\hat{p}, \bm{u})$ and pseudo-time source distribution $\Phi_k(\bm{u}, \bm{s})$ for $\bm{s} = (s_p, \bm{s}_u)^{\mathsf{T}} \in \R^{d+1}$ are
\begin{subequations}
\label{eq:feq_Phi_defs}
\begin{align}
  f_k^{\mathrm{eq}}(\hat{p}, \bm{u})
  &= w_k\!\left[\hat{p} + \frac{\bm{c}_k \cdot \bm{u}}{c_s^2} + \frac{(\bm{c}_k \cdot \bm{u})^2}{2 c_s^4} - \frac{|\bm{u}|^2}{2 c_s^2}\right], \label{eq:feq_def}\\
  \Phi_k(\bm{u}, \bm{s})
  &= w_k\!\left[s_p + \frac{\bm{c}_k \cdot \bm{s}_u}{c_s^2} + \frac{(\bm{c}_k \cdot \bm{u})(\bm{c}_k \cdot \bm{s}_u)}{c_s^4} - \frac{\bm{u} \cdot \bm{s}_u}{c_s^2}\right], \label{eq:Phi_def}
\end{align}
\end{subequations}
so that $P\bm{f}^{\mathrm{eq}} = (\hat{p}, \bm{u})^{\mathsf{T}}$, $P\bm{\Phi} = (s_p, \bm{s}_u)^{\mathsf{T}}$, $\sum_k c_{k\alpha}c_{k\beta} f_k^{\mathrm{eq}} = c_s^2\hat{p}\,\delta_{\alpha\beta} + u_\alpha u_\beta$, and $\sum_k c_{k\alpha}c_{k\beta}\Phi_k = c_s^2 s_p\,\delta_{\alpha\beta} + u_\alpha s_{u\beta} + u_\beta s_{u\alpha}$. For a fixed lattice pseudo-step $\dtlat > 0$ and lattice speed $\clat := h / \dtlat$, dimensionless variables map to physical units via
\begin{equation}
\label{eq:unit_conversion}
  \bm{u} = \frac{\bm{u}_{\mathrm{phys}}}{\clat}, \qquad
  \hat{p} = \frac{p - p_0}{\rho_0 c_s^2 \clat^2} \quad \bigl(p_{\mathrm{lu}} = c_s^2 \hat{p}\bigr), \qquad
  \bm{b}_{\mathrm{lu}} = \frac{\dtlat}{\clat}\bm{b}_{\mathrm{phys}} = \frac{\dtlat^2}{h}\bm{b}_{\mathrm{phys}}.
\end{equation}

\subsection{Harmonic lattice scaling, deadbeat unit-relaxation sweep, and bond-symmetric viscous flux}
\label{subsec:sec2_harmonic_scaling}
\label{subsec:sec2_deadbeat_bdf2}
\label{subsec:sec3_varvisc}
\label{subsec:sec3_smagorinsky}

The unit conversion~\eqref{eq:unit_conversion} leaves $\dtlat$ as a free scale parameter whose selection must reconcile two constraints: variable-step BDF2 integration requires $\dtlat$ to remain time-invariant so that prior macroscopic histories share a common velocity scale $\clat$, whereas viscosity decoupling requires the dimensionless lattice viscosity $\nulu(\bm{x},t) = \nueff(\bm{x},t)\dtlat/h^2$ not to exceed the unit-relaxation kinetic viscosity $\nu_{\mathrm{unit}} = 1/6$. Given $(h, U_0, \umax)$ with $\umax = 0.05$ ($\MaNum := U_0/(c_s\clat) = \sqrt{3}\,\ulu < \MaNum_0 := \sqrt{3}\,\umax \approx 0.0866$) and base-viscosity ceiling $\numax \ge \sup_{\bm{x},t}\nu_0(\bm{x},t)$, we define the \emph{global $C^\infty$ harmonic lattice pseudo-step} by
\begin{equation}
\label{eq:harmonic_scaling}
  \dtlat := \frac{h^2}{6\numax + U_0 h / \umax}, \qquad
  \clat := \frac{h}{\dtlat} = \frac{6\numax}{h} + \frac{U_0}{\umax},
\end{equation}
which maps all flow regimes smoothly into bounded lattice parameters:
\begin{gather}
\label{eq:nulu_beta_bounds}
  \nulu(\bm{x},t) = \frac{\nueff(\bm{x},t)}{6\numax + U_0 h / \umax} \in \left(0, \frac{1}{6}\right], \\
  \beta(\bm{x},t) := 1 - 6\nulu(\bm{x},t) \in [0, 1), \qquad
  \ulu = \frac{U_0}{\clat} \in (0, \umax). \nonumber
\end{gather}
For uniform viscosity $\nueff \equiv \nu_0 = \numax$ with cell Reynolds number $\Reh := U_0 h / \nu_0$, Eq.~\eqref{eq:nulu_beta_bounds} yields the branch-free rational transition
\begin{equation}
\label{eq:rational_Reh_law}
  \nulu(\Reh) = \frac{1/6}{1 + \Reh/(6\umax)}, \quad
  \beta(\Reh) = \frac{\Reh/(6\umax)}{1 + \Reh/(6\umax)}, \quad
  \ulu(\Reh) = \frac{\umax\,\Reh/(6\umax)}{1 + \Reh/(6\umax)}.
\end{equation}
This rational law bridges the Stokes ($\Reh \to 0^+$: $\nulu \to \tfrac{1}{6}^-, \beta \to 0^+$) and high-Reynolds ($\Reh \to \infty$: $\nulu \to 0^+, \beta \to 1^-$) limits smoothly without switching collision relaxation. When discrete Smagorinsky large-eddy simulation (LES) is enabled ($C_s > 0$)~\citep{Smagorinsky1963,Hou1996} with discrete central-difference strain $\widehat{\bm{S}}_{\mathrm{phys}}(\bm{u}_{\mathrm{phys}}^*) := \tfrac{\clat}{h}\widehat{\bm{S}}_{\mathrm{lu}}(\bm{u}^*)$, substituting $\clat = h/\dtlat$ into $\nu_t(\bm{x}) = (C_s h)^2|\widehat{\bm{S}}_{\mathrm{phys}}(\bm{u}_{\mathrm{phys}}^*)|$ cancels $(h, \dtlat)$ under the kinetic ceiling $\nueff(\bm{x}) = \min\bigl(\tfrac{h^2}{6\dtlat},\, \nu_0(\bm{x}) + \nu_t(\bm{x})\bigr)$ and gives
\begin{gather}
\label{eq:smag_scale_invariance}
  \nu_{t,\mathrm{lu}}(\bm{x}) = (C_s h)^2\bigl|\widehat{\bm{S}}_{\mathrm{phys}}(\bm{u}_{\mathrm{phys}}^*)\bigr|\frac{\dtlat}{h^2} \equiv C_s^2\bigl|\widehat{\bm{S}}_{\mathrm{lu}}(\bm{u}^*)\bigr|, \\
\label{eq:smag_nulu_beta}
  \nulu(\bm{x}) = \min\!\left(\frac{1}{6},\; \nu_0(\bm{x})\frac{\dtlat}{h^2} + C_s^2\bigl|\widehat{\bm{S}}_{\mathrm{lu}}(\bm{u}^*)\bigr|\right), \qquad
  \beta(\bm{x}) = 1 - 6\nulu(\bm{x}) \in [0, 1).
\end{gather}
Here $\widehat{S}_{\mathrm{lu},\alpha\gamma}(\bm{u}^*) := \tfrac{1}{2}[(\NablaC u_\gamma^*)_\alpha + (\NablaC u_\alpha^*)_\gamma]$ and $|\widehat{\bm{S}}_{\mathrm{lu}}| := \sqrt{2\widehat{S}_{\mathrm{lu},\alpha\gamma}\widehat{S}_{\mathrm{lu},\alpha\gamma}}$ with $(\NablaC\phi)_\alpha(\bm{x}) := \tfrac{1}{2}[\phi(\bm{x}+\bm{e}_\alpha h) - \phi(\bm{x}-\bm{e}_\alpha h)]$ are evaluated at the frozen second-order Adams--Bashforth predictor
\begin{equation}
\label{eq:ab2_predictor}
  \bm{u}^{*,n+1} = (1 + \omega_n)\bm{u}^n - \omega_n\bm{u}^{n-1} \quad (n \ge 1), \qquad
  \bm{u}^{*,1} = \bm{u}^0 \quad (n = 0).
\end{equation}
Pairwise momentum conservation (Proposition~\ref{prop:varvisc_properties}(2)) holds unconditionally for any bounded $\beta(\bm{x}) \in [0,1)$, whereas non-smoothness of $|\widehat{\bm{S}}_{\mathrm{lu}}|$ at zero strain or ceiling activation places LES outside the $C^3$-viscosity assumption of Theorem~\ref{thm:hydro_consistency}.

With the time-invariant pseudo-step $\dtlat$ and effective lattice viscosity $\nulu(\bm{x},t)$ fixed, we discretize physical time on levels $0 = t_0 < t_1 < \dots < t_{n+1}$ with outer step $\Delta t_{\mathrm{phys},n+1} = \Delta t_{\mathrm{lu},n+1}\dtlat$ and step ratio $\omega_n := \Delta t_{\mathrm{lu},n+1}/\Delta t_{\mathrm{lu},n}$. Starting from $\bm{f}^0 = \bm{f}^{\mathrm{eq}}(\hat{p}^0, \bm{u}^0)$, step $1$ ($n=0$) uses BDF1 and steps $n \ge 1$ use variable-step BDF2 with coefficients
\begin{gather}
  a_q = \frac{1}{\Delta t_{\mathrm{lu},1}}, \quad \bm{h}_q = \frac{\bm{m}^0}{\Delta t_{\mathrm{lu},1}} \qquad (n = 0), \label{eq:bdf1_startup}\\
  a_q = \frac{1 + 2\omega_n}{1 + \omega_n}\frac{1}{\Delta t_{\mathrm{lu},n+1}}, \quad
  \bm{h}_q = \frac{(1 + \omega_n)^2 \bm{m}^n - \omega_n^2 \bm{m}^{n-1}}{(1 + \omega_n)\Delta t_{\mathrm{lu},n+1}} \qquad (n \ge 1), \label{eq:var_bdf2_coeffs}\\
  d_a := 1 + \frac{a_q}{2}, \qquad
  c_a := 1 - \frac{a_q}{2} \qquad (n \ge 0). \notag
\end{gather}
Given body-force moment $\bm{g}^{n+1} = (0, \bm{b}_{\mathrm{lu}}^{n+1})^{\mathsf{T}}$, each inner sweep evaluates the deadbeat $\tau^+ = \tau^- \equiv 1$ fixed-point map $T : \mathcal{X}_h \to \mathcal{X}_h$ on $\mathcal{X}_h := (\R^Q)^{\Omega_f}$ in four stages:
\begin{subequations}
\label{eq:inner_map_T}
\begin{align}
  \bm{m}(\bm{f}) &= \frac{P\bm{f} + \frac{1}{2}(\bm{h}_q + \bm{g}^{n+1})}{d_a}, &
  \bm{s}(\bm{f}) &= \bm{h}_q + \bm{g}^{n+1} - a_q \bm{m}(\bm{f}), &&\text{(moments)} \label{eq:inner_T_step1}\\
  f_k^\star(\bm{x}) &= f_k^{\mathrm{eq}}(\hat{p}, \bm{u}) + \frac{1}{2}\Phi_k(\bm{u}, \bm{s}),
  &&&&\text{($\tau^\pm \equiv 1$)} \label{eq:inner_T_step2}\\
  \bm{f}_{\mathrm{str}} &= \mathcal{S}_\Gamma \bm{f}^\star,
  &&&&\text{(streaming)} \label{eq:inner_T_step3}\\
  T(\bm{f})(\bm{x}) &= \bm{f}_{\mathrm{str}}(\bm{x}) + E_u\!\bigl(\DuVD(\bm{x})\bigr).
  &&&&\text{(viscous lift)} \label{eq:inner_T_step4}
\end{align}
\end{subequations}
In stage~\eqref{eq:inner_T_step4}, the post-collision momentum $\tilde{\bm{u}} := \bm{u} + \tfrac{1}{2}\bm{s}_u = P_u\bm{f}^\star$ drives the parameter-free \emph{bond-symmetric post-streaming momentum-flux lift} across interior 1-ring links,
\begin{gather}
\label{eq:DuVD_bulk}
  \Delta u_\alpha^{\mathrm{VD}}(\bm{x})
  := \sum_{k=1}^{Q-1} \bar{\beta}_k(\bm{x})\,\frac{w_k}{c_s^2} c_{k\alpha} c_{k\gamma}
     \bigl(\tilde{u}_\gamma(\bm{x}) - \tilde{u}_\gamma(\bm{x} - \bm{c}_k h)\bigr), \\
  \bar{\beta}_k(\bm{x}) := \frac{\beta(\bm{x}) + \beta(\bm{x} - \bm{c}_k h)}{2}, \nonumber
\end{gather}
whose boundary extension is constructed in Section~\ref{sec:boundaries}. Naive pointwise scaling $\beta(\bm{x})[(I_d - P_u\mathcal{S}E_u)\tilde{\bm{u}}](\bm{x})$ assigns asymmetric bond weights that violate discrete action--reaction balance and drop the viscosity-gradient stress $(\nabla\nueff)\cdot(\nabla\bm{u} + (\nabla\bm{u})^{\mathsf{T}})$. By contrast, the symmetric bond average $\bar{\beta}_k(\bm{x})$ preserves both pairwise conservation and variable-viscosity stress consistency; analytical elimination of pre-streaming and scalar-Laplacian alternatives is recorded in \ref{app:counterexamples}. Once inner sweeps converge to $\bm{f}^{n+1} = T(\bm{f}^{(\ell_\star)})$, outer conserved moments are committed via the exact half-source transformation
\begin{equation}
\label{eq:macroscopic_commit}
  \bm{m}^{n+1} = \begin{pmatrix}\hat{p}^{n+1} \\ \bm{u}^{n+1}\end{pmatrix}
  := \frac{P\bm{f}^{n+1} + \frac{1}{2}\bigl(\bm{h}_q + \bm{g}^{n+1}\bigr)}{d_a}.
\end{equation}

\subsection{Single-sweep ghost-mode annihilation and hydrodynamic consistency}
\label{subsec:sec2_consistency}

We establish the algebraic structure and asymptotic consistency of the bulk sweep~\eqref{eq:inner_map_T} in three steps: Theorem~\ref{thm:kerP_annihilation} proves single-sweep annihilation of pre-collision non-conserved modes, Proposition~\ref{prop:varvisc_properties} verifies pairwise momentum conservation and variable-viscosity stress recovery for $\DuVD$, and Theorem~\ref{thm:hydro_consistency} combines both results to obtain second-order low-Mach Navier--Stokes consistency.

\begin{theorem}[Single-Sweep Pre-Collision $\kerP$ Annihilation]
\label{thm:kerP_annihilation}
Let $\Pi_{\kerP} := I_Q - E P$ induce $\R^Q = \ranE \oplus \kerP$, where $\dim\kerP = Q - d - 1 \in \{6, 15\}$. For any $\bm{f} \in \mathcal{X}_h$ and arbitrary perturbation $\bm{\eta} \in \mathcal{X}_h$ satisfying $P\bm{\eta}(\bm{x}) \equiv \bm{0}$:
\begin{enumerate}
  \item $\bm{m}(\bm{f} + \bm{\eta}) = \bm{m}(\bm{f})$, $\bm{s}(\bm{f} + \bm{\eta}) = \bm{s}(\bm{f})$, $\bm{f}^\star(\bm{f} + \bm{\eta}) = \bm{f}^\star(\bm{f})$, and $T(\bm{f} + \bm{\eta}) = T(\bm{f})$ hold \emph{identically} after a single sweep;
  \item Around a source-free uniform rest state ($\bm{u}_0 = \bm{0}$, $\bm{s}_0 = \bm{0}$), $\Pi_{\kerP}\,\delta\bm{f}^\star \equiv \bm{0}$, while nonlinearly $\Pi_{\kerP}\bm{f}^\star = O(|\bm{u}|^2 + |\bm{u}||\bm{s}_u|)$ is purely slaved to instantaneous conserved moments with zero kinetic memory across sweeps.
\end{enumerate}
\end{theorem}

\begin{proof}
Since $P E = I_{d+1}$ and $P(\bm{f} + \bm{\eta}) = P\bm{f}$ pointwise, Eqs.~\eqref{eq:inner_T_step1}--\eqref{eq:inner_T_step2} produce identical $(\bm{m}, \bm{s}, \bm{f}^\star, \tilde{\bm{u}})$ because $(1 - 1/\tau^\pm)\bm{f}^{\mathrm{neq}} \equiv \bm{0}$ at $\tau^\pm \equiv 1$; hence $T(\bm{f}+\bm{\eta}) = T(\bm{f})$ in one sweep. Linearizing Eq.~\eqref{eq:feq_Phi_defs} at $\bm{u}_0 = \bm{0}$ and $\bm{s}_0 = \bm{0}$ places $\delta\bm{f}^\star \in \ranE$, so $\Pi_{\kerP}\,\delta\bm{f}^\star \equiv \bm{0}$.
\end{proof}

While Theorem~\ref{thm:kerP_annihilation} guarantees memoryless kinetic projection in collision, physical diffusion for $\beta > 0$ is governed by the post-streaming lift $\DuVD$, whose conservation and stress-divergence properties follow from undirected bond symmetry.

\begin{proposition}[Conservation and Variable-Viscosity Consistency of $\DuVD$]
\label{prop:varvisc_properties}
In both D2Q9 and D3Q19, the bond-symmetric operator~\eqref{eq:DuVD_bulk} satisfies:
\begin{enumerate}
  \item \emph{Exact uniform-viscosity collapse:} When $\beta(\bm{x}) \equiv \beta_0$, $\DuVD(\bm{x}) \equiv \beta_0(I_d - P_u\mathcal{S}E_u)\tilde{\bm{u}}(\bm{x})$, which reduces to standard unit-relaxation collision--streaming as $\beta_0 \to 0^+$ and $\nulu \to 1/6$.
  \item \emph{Exact pairwise discrete momentum conservation:} On any periodic lattice $\Omega_h$,
  \begin{equation}
  \label{eq:pairwise_mom_cons}
    \sum_{\bm{x} \in \Omega_h} \DuVD(\bm{x}) \equiv \bm{0}
  \end{equation}
  holds for arbitrary $\nulu(\bm{x},t)$, whereas pointwise scaling incurs global momentum drift $\tfrac{h^2}{6}\sum_{\bm{x}}(\nabla\beta)\cdot(\nabla\tilde{\bm{u}} + (\nabla\tilde{\bm{u}})^{\mathsf{T}} + (\nabla\cdot\tilde{\bm{u}})\bm{I}) + O(|\Omega_h|h^4)$.
  \item \emph{Second-order variable-viscosity tensor-divergence consistency:} On arbitrary smooth fields $\tilde{\bm{u}}_{\mathrm{phys}} = \clat\tilde{\bm{u}} \in C^4$ and $\nueff(\bm{x},t) \in C^3$ with uniformly bounded spatial derivatives,
  \begin{equation}
  \label{eq:varvisc_stress_recovery}
  \begin{aligned}
    &\frac{\clat}{\dtlat}\Bigl[\bigl(P_u\mathcal{S}E_u\tilde{\bm{u}} - \tilde{\bm{u}}\bigr)_\alpha + \Delta u_\alpha^{\mathrm{VD}}\Bigr] \\
    &\quad = \partial_\gamma\!\Bigl[\nueff(\bm{x},t)\bigl(\partial_\gamma\tilde{u}_{\mathrm{phys},\alpha} + \partial_\alpha\tilde{u}_{\mathrm{phys},\gamma} + (\nabla\cdot\tilde{\bm{u}}_{\mathrm{phys}})\delta_{\alpha\gamma}\bigr)\Bigr] + O(h^2),
  \end{aligned}
  \end{equation}
  which reduces under $\|\nabla\cdot\tilde{\bm{u}}_{\mathrm{phys}}\|_{C^1} = O(\varepsilon_{\mathrm{div}})$ to the incompressible symmetric stress divergence $\partial_\gamma[\nueff(\partial_\gamma\tilde{u}_{\mathrm{phys},\alpha} + \partial_\alpha\tilde{u}_{\mathrm{phys},\gamma})] + O(h^2, \|\nueff\|_{C^1}\varepsilon_{\mathrm{div}})$.
\end{enumerate}
\end{proposition}

\begin{proof}
(1)~Direct substitution via Eq.~\eqref{eq:lattice_isotropy}. (2)~Pairing directed bond $(\bm{x}, k)$ to $\bm{y} = \bm{x} - \bm{c}_k h$ with its reverse $(\bm{y}, \bar{k})$ gives $\bar{\beta}_{\bar{k}}(\bm{y}) = \bar{\beta}_k(\bm{x})$, $w_{\bar{k}} = w_k$, $c_{\bar{k}\alpha}c_{\bar{k}\gamma} = c_{k\alpha}c_{k\gamma}$, and opposite velocity difference $\tilde{u}_\gamma(\bm{y}) - \tilde{u}_\gamma(\bm{x})$, so every undirected bond pair cancels identically. (3)~Rewriting the combined diffusion link-by-link with $1 - \bar{\beta}_k = 3[\nueff(\bm{x})+\nueff(\bm{x}-\bm{c}_k h)]\dtlat/h^2$ cancels $\dtlat$ upon multiplication by $\clat/\dtlat$; Taylor expansion to $O(h^2)$ using odd-moment parity and fourth-order isotropy~\eqref{eq:lattice_isotropy} yields $(\delta_{\alpha\gamma}\delta_{\delta\theta} + \delta_{\alpha\delta}\delta_{\gamma\theta} + \delta_{\alpha\theta}\delta_{\gamma\delta})\partial_\theta(\nueff\partial_\delta\tilde{u}_{\mathrm{phys},\gamma}) + O(h^2)$, which gives Eq.~\eqref{eq:varvisc_stress_recovery}.
\end{proof}

Combining the single-sweep moment slaving of Theorem~\ref{thm:kerP_annihilation} with the stress-divergence expansion~\eqref{eq:varvisc_stress_recovery} yields the macroscopic consistency of converged outer steps.

\begin{theorem}[Asymptotic Second-Order Hydrodynamic Consistency]
\label{thm:hydro_consistency}
Consider a fixed physical problem on a periodic domain, or at interior lattice sites possessing complete 1-ring stencils, with scales $(U_0, L, \rho_0, \numax > 0)$ and smooth physical fields $(\bm{u}_{\mathrm{phys}}, p, \nueff, \bm{b}_{\mathrm{phys}})$ on $[0, T]$ possessing uniformly bounded derivatives through sixth order in space and third order in time, including $\|\nueff\|_{C^3} < \infty$. Assume pressure scales as $p - p_0 = O(\rho_0 U_0^2 + \rho_0\numax U_0 / L)$ with analogous bounds on $\partial_t^m p$ for $m \le 3$, so that $\hat{p} = O(\MaNum^2 + \MaNum\,h)$ as $(h, \dtphys) \to (0,0)$, where implicit $O(\cdot)$ constants depend on the fixed physical parameters and field derivative norms. Let $\bm{f}^{n+1} = T(\bm{f}^{n+1})$ be a converged fixed point of Eq.~\eqref{eq:inner_map_T} under harmonic scaling~\eqref{eq:harmonic_scaling}, bond-symmetric viscous lift~\eqref{eq:DuVD_bulk}, and macroscopic commit~\eqref{eq:macroscopic_commit}, with $\dtphys := \max_m \Delta t_{\mathrm{phys},m+1}$ under bounded step ratios $\sup_n\omega_n \le \bar{\omega}$. Then:
\begin{enumerate}
  \item At every interior point at step $n+1 \ge 2$, $(\hat{p}^{n+1}, \bm{u}^{n+1})$ satisfy the incompressible variable-viscosity Navier--Stokes equations~\eqref{eq:target_ns} asymptotically:
  \begin{subequations}
  \label{eq:consistency_bounds}
  \begin{gather}
    \nabla \cdot \bm{u}_{\mathrm{phys}}^{n+1}
    = O\!\left(\MaNum^2,\, \dtphys^2,\, h^2\right), \label{eq:cont_trunc}\\
    \begin{aligned}
      &\partial_t\bm{u}_{\mathrm{phys}}^{n+1}
       + \nabla\cdot\bigl(\bm{u}_{\mathrm{phys}}^{n+1}\otimes\bm{u}_{\mathrm{phys}}^{n+1}\bigr)
       + \frac{1}{\rho_0}\nabla p^{n+1} \\
      &\quad = \nabla\cdot\!\Bigl[\nueff\bigl(\nabla\bm{u}_{\mathrm{phys}}^{n+1} + (\nabla\bm{u}_{\mathrm{phys}}^{n+1})^{\mathsf{T}}\bigr)\Bigr]
       + \bm{b}_{\mathrm{phys}}^{n+1} + O(\dtphys^2, h^2, \MaNum h, \MaNum^2),
    \end{aligned} \label{eq:mom_trunc}
  \end{gather}
  \end{subequations}
  where $O(\MaNum\,h) = O(h^2)$ under convective low-Mach scaling $\MaNum = O(h)$ as well as at fixed $(U_0, \numax)$ in the creeping regime $\Reh \ll 1$ where $\MaNum = O(U_0 h/\numax) = O(h)$;
  \item Provided the semi-discrete spatial discretization is zero-stable on $[0, T]$, executing step $1$ with BDF1~\eqref{eq:bdf1_startup} and steps $n \ge 1$ with variable-step BDF2~\eqref{eq:var_bdf2_coeffs} under $\sup_n\omega_n \le \bar{\omega} < 1 + \sqrt{2}$~\citep{HairerWanner1996} (such as $\omega_n \in \{1/2, 1, 2\}$) preserves global $O(\dtphys^2)$ temporal convergence at any fixed horizon $T > 0$.
\end{enumerate}
\end{theorem}

\begin{proof}
Expanding $f_k(\bm{x}) = f_k^\star(\bm{x} - \bm{c}_k h) + (E_u\DuVD)_k(\bm{x})$ through $O(h^5)$ and applying $P_0(\cdot)$ cancels the spurious $\tfrac{h^2}{2}c_s^2\Delta\hat{p}$ acoustic pressure-Laplacian via the $O(h^3)$ $P_u(\cdot)$ balance and $P_0 E_u \equiv 0$. Meanwhile, Proposition~\ref{prop:varvisc_properties}(3) bounds the viscous divergence residual~\eqref{eq:varvisc_stress_recovery} at $O(\dtlat\numax + \dtlat h^2) = O(h^2)$ via $\dtlat\numax \le \tfrac{1}{6}h^2$, which yields Eq.~\eqref{eq:cont_trunc} and $\|\nabla\cdot\tilde{\bm{u}}_{\mathrm{phys}}\|_{C^1} = O(\MaNum^2, \dtphys^2, h^2)$. Applying $P_u(\cdot)$ through $O(h^5)$ then combines $\Pi_{\alpha\beta}^\star$ with unit-relaxation third-moment diffusion and Proposition~\ref{prop:varvisc_properties}(3), while the fourth-moment remainder satisfies $-\tfrac{h^2}{6}\partial_\beta\partial_\gamma\partial_\delta(\clat^2 R_{\alpha\beta\gamma\delta}^\star) = O(h^2)$. Summing the variable-step BDF2 local defects $O(\Delta t_{\mathrm{phys},n+1}\dtphys^2)$ under Grigorieff--Dahlquist zero-stability~\citep{HairerWanner1996} establishes Eq.~\eqref{eq:mom_trunc}; full algebra is given in \ref{app:hydro_proof}.
\end{proof}

\begin{remark}[Finite-Grid Translational Shear Shift and Outer-Step Stability]
\label{rem:finite_grid_galilean}
While Theorem~\ref{thm:kerP_annihilation} annihilates pre-collision $\kerP$ memory at every inner sweep, converging $T(\bm{f}^{n+1}) = \bm{f}^{n+1}$ solves the discrete outer step rather than enforcing unconditional outer-time $L_2$ contractivity. For a divergence-free transverse shear wave on uniform lattice translation $U_{\mathrm{lu}}$ with $H(k) := 1 - 2\nulu(1 - \cos k) - i U_{\mathrm{lu}}\sin k$, the quadratic equilibrium~\eqref{eq:feq_def} yields the exact finite-wavenumber symbol $\tfrac{2(H - 1)}{H + 1} = -i U_{\mathrm{lu}} k - (\nulu - \tfrac{1}{2}U_{\mathrm{lu}}^2)k^2 + O(k^3)$~\citep{HeLuo1997}. In physical units, the effective kinematic-viscosity shift $-\tfrac{1}{2}U_{\mathrm{phys}}^2\dtlat = O(\MaNum\,h)$ enters the $O(\MaNum\,h)$ viscous remainder of Eq.~\eqref{eq:mom_trunc} and vanishes under grid and Mach refinement. On a fixed coarse grid with $\nulu < \tfrac{1}{2}U_{\mathrm{lu}}^2$, however, transverse shear can grow slowly unless $\nulu > \tfrac{1}{2}U_{\mathrm{lu}}^2$ or cubic Mach corrections~\citep{Dellar2003} are included; for example, at $k = 2\pi/128$, $\nulu = 10^{-4}$, $U_{\mathrm{lu}} = 0.04997$, and $\dtlu = 4$, the 100-step amplitude ratio is $1.00103$ with single-step $|r_{\mathrm{BDF2}}| = 1.000011$. Likewise, at rest the longitudinal acoustic semi-discrete symbol obeys $\operatorname{Re}\mu_{\mathrm{ac}}(k) = (\tfrac{1}{12} - \tfrac{3}{2}\nulu)k^2 + O(k^4)$, which is strictly dissipative across the Brillouin zone for $\nulu \ge 1/18$. When $\nulu < 1/18$, low-wavenumber acoustic waves can grow slightly at small $\dtlu$ (for instance, a 100-step ratio of $1.01957$ and single-step $|r_{\mathrm{BDF2}}| = 1.00020$ at $k = 2\pi/128$, $\nulu = 10^{-4}$, and $\dtlu = 1$) until BDF2 fourth-order numerical damping $-\tfrac{1}{36}\dtlu^4 k^4$ dominates for $\dtlu^3 \gtrsim 3(1 - 18\nulu)/k_{\min}^2$. On $N = 128$ at $\nulu = 10^{-4}$, the discrete amplification crosses neutrality at $\dtlu \approx 11.26$ and remains contractive for $\dtlu \ge 11.3$.
\end{remark}

\section{Unified solid-wall and open-boundary closures}
\label{sec:boundaries}

Having established the bulk formulation and consistency of $T$ in Section~\ref{sec:formulation}, we now close $\mathcal{S}_\Gamma$ and $\DuVD$ along solid walls and open-channel boundaries. Because deadbeat collision at $\tau^+ = \tau^- \equiv 1$ carries zero pre-collision non-equilibrium stress memory (Theorem~\ref{thm:kerP_annihilation}), neither boundary class requires non-equilibrium stress reconstruction: impermeable walls close through a single link-wise reflection operator, and open inlets and outlets close through half-cell shifted ghost states.

\subsection{Unified planar and curved solid-wall closure and Eulerian hydrodynamic force}
\label{subsec:sec3_walls_force}

On a fluid mesh $\Omega_f$ bounded by impermeable walls moving tangentially with $\bm{u}_w(\bm{x}_w,t)\cdot\bm{n} = 0$, let $\mathcal{L}_\Gamma$ denote the cut links $(\bm{x}_f, i)$ leaving $\Omega_f$ with wall distance fraction $q_i := \|\bm{x}_f - \bm{x}_w\|/\|\bm{c}_i h\| \in (0, 1]$, where $q_i \equiv 1/2$ corresponds to planar lattice walls. With interior backward neighbor $\bm{x}_{ff} := \bm{x}_f - \bm{c}_i h$, reflected populations along $\bar{i} = \opp(i)$ follow the linear Bouzidi--Firdaouss--Lallemand rule~\citep{Bouzidi2001}
\begin{equation}
\label{eq:bouzidi_ibb}
  f_{\bar{i}}^{\mathrm{BFL}}(\bm{x}_f) :=
  \begin{cases}
    2 q_i f_i^\star(\bm{x}_f) + (1 - 2 q_i) f_i^\star(\bm{x}_{ff}) - \tfrac{2 w_i}{c_s^2}(\bm{c}_i \cdot \bm{u}_w),
      & 0 < q_i < \tfrac{1}{2},\; \bm{x}_{ff} \in \Omega_f, \\[4pt]
    f_i^\star(\bm{x}_f) - \tfrac{2 w_i}{c_s^2}(\bm{c}_i \cdot \bm{u}_w),
      & 0 < q_i < \tfrac{1}{2},\; \bm{x}_{ff} \notin \Omega_f, \\[4pt]
    \tfrac{1}{2 q_i} f_i^\star(\bm{x}_f) + \tfrac{2 q_i - 1}{2 q_i} f_{\bar{i}}^\star(\bm{x}_f) - \tfrac{w_i}{q_i c_s^2}(\bm{c}_i \cdot \bm{u}_w),
      & \tfrac{1}{2} \le q_i \le 1.
  \end{cases}
\end{equation}
On closed impermeable domains satisfying $\oint_{\partial\Omega}\bm{u}_w\cdot\bm{n}\,ds = 0$, curved cut-link quadrature induces a pre-correction global mass defect $\delta M_\Gamma(\bm{f}^\star) := \sum_{(\bm{x}_f, i) \in \mathcal{L}_\Gamma}\bigl(f_{\bar{i}}^{\mathrm{BFL}}(\bm{x}_f) - f_i^\star(\bm{x}_f)\bigr)$, which we neutralize uniformly across zero-velocity rest populations ($\bm{c}_0 = \bm{0}$) via
\begin{equation}
\label{eq:rest_mass_neutrality}
  f_0^{\mathrm{str}}(\bm{x}) =
  \begin{cases}
    f_0^\star(\bm{x}) - \dfrac{\delta M_\Gamma(\bm{f}^\star)}{|\Omega_f|}, & \text{closed impermeable domains}, \\[6pt]
    f_0^\star(\bm{x}), & \text{open inlet--outlet domains}.
  \end{cases}
\end{equation}
This rest-state adjustment leaves momentum fluxes, $\DuVD$, and wall forces strictly invariant; the single-sweep post-correction operator mass defect $|\delta M_{\mathrm{post}}| := \bigl|\sum_{\bm{x}\in\Omega_f}(P_0 T(\bm{f})(\bm{x}) - P_0\bm{f}^\star(\bm{x}))\bigr|$ is bounded to machine precision ($\le \VDLBMCurvedMassCons$) and vanishes identically when $q_i \equiv 1/2$.

Applying the same reflection operator $\SBFL$ to $(E_u\tilde{\bm{u}})$ closes the post-streaming viscous lift across interior, planar, and curved boundary cells through
\begin{equation}
\label{eq:DuVD_unified_wall}
\begin{aligned}
  \Delta u_\alpha^{\mathrm{VD}}(\bm{x}_f)
  &= \sum_{\bar{i}:\,\bm{x}_f-\bm{c}_{\bar{i}}h \in \Omega_f}
     \bar{\beta}_{\bar{i}}(\bm{x}_f)\frac{w_{\bar{i}}}{c_s^2}c_{\bar{i}\alpha}c_{\bar{i}\gamma}
     \bigl(\tilde{u}_\gamma(\bm{x}_f) - \tilde{u}_\gamma(\bm{x}_f - \bm{c}_{\bar{i}}h)\bigr) \\
  &\quad + \beta(\bm{x}_f)\sum_{i:\,(\bm{x}_f,i)\in\mathcal{L}_\Gamma}
     c_{\bar{i}\alpha}\Bigl[(E_u\tilde{\bm{u}})_{\bar{i}}(\bm{x}_f) - \SBFL(E_u\tilde{\bm{u}}; \bm{u}_w)_{\bar{i}}(\bm{x}_f)\Bigr].
\end{aligned}
\end{equation}
For planar walls ($q_i \equiv 1/2$), Eq.~\eqref{eq:DuVD_unified_wall} reduces to the corner-safe same-node reflection $2\beta(\bm{x}_f)\sum_{i\in\mathcal{L}_\Gamma}\allowbreak\tfrac{w_i}{c_s^2}c_{i\alpha}c_{i\gamma}\allowbreak(\tilde{u}_\gamma(\bm{x}_f) - u_{w,\gamma})$ and avoids the $O(1-6\nulu)$ wall-shear jump caused by setting $\beta=0$ at boundary cells (\ref{app:counterexamples}). For curved walls with interior neighbor $\bm{x}_{ff} \in \Omega_f$, exact linear extrapolation $\SBFL(E_u\bm{v}; \bm{u}_w)_{\bar{i}}(\bm{x}_f) \equiv\allowbreak (E_u\bm{v})_{\bar{i}}(\bm{x}_f + \bm{c}_i h)$ on any affine shear field $\bm{v}(\bm{x}) = \bm{u}_w + \bm{G}(\bm{x}-\bm{x}_w)$ eliminates $O(h)$ wall-slip errors across all uniform $\beta \in [0, 1)$, whereas narrow-gap degenerate links with $q_i < 1/2$ and $\bm{x}_{ff} \notin \Omega_f$ revert locally to $O(h)$ half-way reflection. Summing link momentum exchange across $\mathcal{L}_\Gamma$ then yields the uncalibrated Eulerian hydrodynamic force and torque
\begin{subequations}
\label{eq:vdlbm_eulerian_force}
\begin{align}
  \bm{F}_{\mathrm{obs,lu}}^{n+1}
  &= \sum_{(\bm{x}_f, i) \in \mathcal{L}_\Gamma} \bm{c}_i \Bigl\{
     f_i^\star(\bm{x}_f) + f_{\bar{i}}^{\mathrm{BFL}}(\bm{x}_f)
     - \beta(\bm{x}_f)\Bigl[(E_u\tilde{\bm{u}})_i(\bm{x}_f) + \SBFL(E_u\tilde{\bm{u}}; \bm{u}_w)_{\bar{i}}(\bm{x}_f)\Bigr]
     \Bigr\}, \label{eq:F_obs_lu}\\
  \bm{F}_{\mathrm{obs,phys}}^{n+1}
  &= \rho_0 \clat^2 h^{d-1}\bm{F}_{\mathrm{obs,lu}}^{n+1}, \qquad
  \bm{T}_{\mathrm{obs,phys}}^{n+1}
  = \rho_0 \clat^2 h^{d-1}\sum_{(\bm{x}_f, i) \in \mathcal{L}_\Gamma}
    (\bm{x}_w - \bm{x}_c) \times \bm{f}_{\mathrm{link},(\bm{x}_f,i)}^{n+1}, \label{eq:F_T_obs_phys}
\end{align}
\end{subequations}
where $\bm{f}_{\mathrm{link},(\bm{x}_f,i)}^{n+1}$ denotes the $(\bm{x}_f,i)$ link summand of Eq.~\eqref{eq:F_obs_lu} and $\bm{x}_c$ is the obstacle torque center.

\subsection{Symmetric shifted ghost velocity inlets and pressure outlets}
\label{subsec:sec3_open_bc}

Turning from closed walls to open channels $[0,L]\times[0,H]$ with cell centers $x_i = (i + \tfrac{1}{2})h$ ($i=0,\dots,N_x-1$), we consider prescribed inflow $\bm{u}_{\mathrm{in}}(y)$ at $x=0$ and zero-pressure outflow $\hat{p}=0$, $\partial_x\bm{u}=\bm{0}$ at $x=L$~\citep{SchaferTurek1996}. Single-node Ladd velocity bounce-back~\citep{Ladd1994} and outlet anti-bounce-back~\citep{Ginzburg2008,Lou2013} omit off-row diagonal neighbors $y_{j\mp 1}$, which introduces an $O(\beta\,\Delta_{yy}^h u_{\mathrm{in}})$ inlet mass source and backward vertical outlet diffusion $(\nulu - 1/12)\Delta_{yy}^h\tilde{u}_x$ whenever $\nulu < 1/12$. Because $\tau^\pm \equiv 1$ requires no Zou--He non-equilibrium closure~\citep{ZouHe1997}, we instead evaluate $\bm{f}^\star$ directly at half-cell shifted ghost columns $i = -1$ ($x_{-1}=-h/2$) and $i = N_x$ ($x_{N_x}=L+h/2$) via
\begin{subequations}
\label{eq:ghost_inlet_outlet}
\begin{align}
  \text{Inlet ($i = -1$):}\quad
  &\begin{aligned}[t]
    \hat{p}_{-1,j} &= 2\hat{p}_{0,j} - \hat{p}_{1,j}, &
    \bm{u}_{-1,j} &= 2\bm{u}_{\mathrm{in}}(y_j) - \bm{u}_{0,j}, \\
    s_{p,-1,j} &= 2 s_{p,0,j} - s_{p,1,j}, &
    \bm{s}_{u,-1,j} &= -\bm{s}_{u,0,j},
  \end{aligned} \label{eq:ghost_inlet}\\[3pt]
  \text{Outlet ($i = N_x$):}\quad
  &\begin{aligned}[t]
    \hat{p}_{N_x,j} &= -\hat{p}_{N_x-1,j}, &
    \bm{u}_{N_x,j} &= \bm{u}_{N_x-1,j}, \\
    s_{p,N_x,j} &= -s_{p,N_x-1,j}, &
    \bm{s}_{u,N_x,j} &= \bm{s}_{u,N_x-1,j}.
  \end{aligned} \label{eq:ghost_outlet}
\end{align}
\end{subequations}
This shifted-ghost closure eliminates both diagonal-link defects and supplies consistent ghost velocities $\tilde{\bm{u}}_{-1,j} = 2\bm{u}_{\mathrm{in}}(y_j) - \tilde{\bm{u}}_{0,j}$ and $\tilde{\bm{u}}_{N_x,j} = \tilde{\bm{u}}_{N_x-1,j}$ together with zero-gradient ghost viscosity $\beta_{-1,j}=\beta_{0,j}$ and $\beta_{N_x,j}=\beta_{N_x-1,j}$ to $\DuVD$, while solid-wall links take precedence at $j \in \{0, N_y-1\}$. Discretely, Eq.~\eqref{eq:ghost_outlet} enforces the same second-order zero-pressure, zero-normal-gradient outflow as finite-volume pressure-correction schemes (Section~\ref{sec:sec6_2_curved_dfg}), whereas Eq.~\eqref{eq:ghost_inlet} imposes linear pressure extrapolation $\delta_{xx}^h\hat{p}_{0,j}=0$ that is exact for Poiseuille inflow, unlike homogeneous zero-normal-gradient inlet pressure closures. With the single-sweep operator $T$ now completely specified across bulk, wall, and open-boundary nodes, we turn to its inner convergence acceleration.

\section{Sylvester moment reduction and matrix-free local Uzawa smoothing}
\label{sec:spectral}

Although deadbeat unit relaxation ($\tau^\pm \equiv 1$) clears all pre-collision ghost modes in one sweep (Theorem~\ref{thm:kerP_annihilation}), unaccelerated Picard iteration $\bm{f}^{(\ell+1)} = T(\bm{f}^{(\ell)})$ stalls on spatially uniform modes as $a_q \to 0^+$ and diverges via off-diagonal acoustic coupling at low $\nulu$. With both bulk and boundary operators of $T$ closed in Sections~\ref{sec:formulation} and~\ref{sec:boundaries}, we now establish why permanent unit relaxation is spectrally necessary under moment preconditioning, prove an exact $Q\times Q \leftrightarrow (d+1)\times(d+1)$ Sylvester spectrum reduction, and construct the matrix-free local polynomial Uzawa smoother of \VDULBM{}.

\subsection{Spectral necessity of permanent unit relaxation and Sylvester moment reduction}
\label{subsec:sec4_spectrum}
\label{subsec:sec2_vartau_failure}

Consider residual-form preconditioned fixed-point updates on $\Omega_f$ of the form
\begin{equation}
\label{eq:precond_update_general}
  \bm{f}^{(\ell+1)} = T\bigl(\bm{f}^{(\ell)}\bigr) + \mathcal{K}\Bigl(T\bigl(\bm{f}^{(\ell)}\bigr) - \bm{f}^{(\ell)}\Bigr)
  = \bm{f}^{(\ell)} + \bigl(I + \mathcal{K}\bigr)\Bigl(T\bigl(\bm{f}^{(\ell)}\bigr) - \bm{f}^{(\ell)}\Bigr),
\end{equation}
where $\mathcal{K} := E K_m P$ lifts a $(d+1)$-component conserved-moment operator $K_m$ back to $\R^Q$.

\begin{proposition}[Nonsingular Fixed-Point Equivalence and Full-Residual Acceptance]
\label{prop:fixed_point_equivalence}
For any linear operator $\mathcal{K} = E K_m P$ on $\mathcal{X}_h = (\R^Q)^{\Omega_f}$:
\begin{enumerate}
  \item If $I + K_m$ is nonsingular on $(\R^{d+1})^{\Omega_f}$, then $(I + E K_m P)^{-1} = I - E(I + K_m)^{-1} K_m P$ exists on $\mathcal{X}_h$ since $P E = I_{d+1}$, so Eq.~\eqref{eq:precond_update_general} and $T$ share the \emph{exact same fixed-point set}, even when $K_m$ uses simplified constant-coefficient or homogeneous boundary stencils.
  \item On a periodic domain around a source-free uniform base state, a uniform conserved-moment error $\delta\bm{f} = E\,\delta\bm{m}$ contracts under Picard at $|c_a|/d_a \to 1^-$ as $a_q \to 0^+$, while its residual-to-error ratio shrinks as $\|P(T(\bm{f})-\bm{f})\|/\|P\delta\bm{f}\| = 1 - c_a/d_a = a_q/d_a = O(a_q)$. Across all domains, inner sweeps terminate only when the full $Q$-component pre-commit residual satisfies
  \begin{equation}
  \label{eq:full_residual_gate}
  \begin{aligned}
    R_\ell &:= \sqrt{\frac{1}{Q|\Omega_f|}\sum_{\bm{x}\in\Omega_f}\sum_{k=0}^{Q-1}\Bigl(T\bigl(\bm{f}^{(\ell)}\bigr)_k(\bm{x}) - f_k^{(\ell)}(\bm{x})\Bigr)^2} \\
    &< \varepsilon_{\mathrm{abs}} := \varepsilon_{\mathrm{rel}}\min(1, a_q)\,\ulu
    \qquad (\ell + 1 \ge 2),
  \end{aligned}
  \end{equation}
  whereupon the solver commits $\bm{f}^{n+1} = T(\bm{f}^{(\ell)})$. Unconverged iterates with $R_\ell$ non-finite, $R_\ell > 10^4\ulu$, or $\ell + 1 = K_{\max}$ are never committed: fixed-step tests abort, whereas adaptive runs halve $\Delta t_{\mathrm{lu},n+1}$ and retry from $t_n$.
\end{enumerate}
\end{proposition}

\begin{proof}
Direct multiplication via $P E = I_{d+1}$ gives $(I + E K_m P)(I - E(I + K_m)^{-1} K_m P) = I$, so $(I + \mathcal{K})(T(\bm{f}) - \bm{f}) = \bm{0} \iff T(\bm{f}) = \bm{f}$; at $\bm{k} = \bm{0}$, $P(T(\bm{f}) - \bm{f}) = (c_a/d_a - 1)P\delta\bm{f}$ gives $1 - c_a/d_a = a_q/d_a$.
\end{proof}

In steady-state runs, outer convergence is monitored via the zero-source residual $R_{\mathrm{steady}} := \|T_{\bm{h}_q=\bm{0},a_q=0}(\bm{f}) - \bm{f}\|_{\mathrm{rms}}$. Although Proposition~\ref{prop:fixed_point_equivalence} guarantees fixed-point equivalence for any nonsingular $I + K_m$, whether Eq.~\eqref{eq:precond_update_general} contracts across Reynolds numbers depends critically on collision relaxation. Exact Floquet--Bloch analysis explains why varying $\tau^\pm(\nulu)$ along $\Lambda = (\tau^+ - \tfrac{1}{2})(\tau^- - \tfrac{1}{2}) = 1/4$, rather than freezing $\tau^\pm \equiv 1$ as in periodic \MSCLBM{}~\citep{Ding2026MSCLBM} and \VDULBM{}, causes spectral breakdown away from $\nulu = 1/6$ (Fig.~\ref{fig:fig01_spectral_foundation}a,b). With $\omega^+ = 2/(1+6\nulu)$ and $\omega^- = 12\nulu/(1+6\nulu)$, linearizing a variable-$\tau$ dual-time sweep around rest with low-pass Fourier shadow $H(\bm{k}) = \bigl(\begin{smallmatrix} a_q & i\bm{k}^{\mathsf{T}} \\ i c_s^2\bm{k} & b(\bm{k})I_d \end{smallmatrix}\bigr)$, passband cutoff $\chi(\bm{k}) = \mathbf{1}_{\{\|\bm{k}\|_\infty \le \pi/2\}}$, and momentum diagonal $b(\bm{k}) = a_q + \tfrac{1}{6}|\bm{k}|^2$ for $H_{\mathrm{unit}}$ or $a_q + \nulu|\bm{k}|^2$ for $H_\nu$ yields
\begin{gather}
\label{eq:msc_amplification_G}
  \widehat{G}_{\mathrm{var}\text{-}\tau}(\bm{k}; \nulu)
  = \widehat{T}_{\mathrm{var}\text{-}\tau}(\bm{k}; \nulu)
  + \chi(\bm{k})\,E H(\bm{k})^{-1} P \Bigl(\widehat{T}_{\mathrm{var}\text{-}\tau}(\bm{k}; \nulu) - I_Q\Bigr), \\
  \bigl|1 - \omega^\pm(\nulu)\bigr| = \frac{|6\nulu - 1|}{6\nulu + 1}. \nonumber
\end{gather}
Whenever $\nulu \neq 1/6$, nonzero ghost retention $|1 - \omega^\pm| \in (0,1)$ creates two failure boundaries, evaluated at $N=128$ and $\dtlu = 38.9$. First, for $\nulu > 1/6$ where $\tau^- \to \tfrac{1}{2}^+$, odd-parity ghost over-relaxation ($1-\omega^- \to -1^+$) destabilizes the passband above $\nu_{\mathrm{crit}}^+ = 0.6357$ under $H_{\mathrm{unit}}$ and $0.3916$ under $H_\nu$. Second, for $\nulu < 1/6$ where $\tau^+ \to \tfrac{1}{2}^+$, retained ghost modes couple destructively with $H(\bm{k})^{-1}$ at the passband corner $\bm{k} = (\pi/2,\pi/2)$ and diverge below $\nu_{\mathrm{crit}}^- = 0.0974$ ($\rho = 1.0495$ at $\nulu = 0.09$ under $H_{\mathrm{unit}}$) and $\nu_{\mathrm{crit}}^- = 0.1272$ ($\rho = 1.6071$ under $H_\nu$). Moreover, when initialized from smooth equilibrium $\bm{f}^{(0)} = \bm{f}^{\mathrm{eq}}(\bm{m}^0)$, rapid damping of the fundamental mode $(1,1)$ produces a deceptive transient U-shaped residual dip down to $R_{\min}/\ulu = 1.13\times 10^{-9}$ at sweep $\ell=27$ under $H_\nu$ before IEEE-754 roundoff on mode $(32,32)$ triggers divergence (Fig.~\ref{fig:fig01_spectral_foundation}b).

Having seen why $\tau^\pm$ cannot be varied away from unity under moment preconditioning, we next show that permanently locking $\tau^\pm \equiv 1$ collapses the $Q\times Q$ inner amplification problem onto the $(d+1)$ conserved moments across all $\nulu \in (0, 1/6]$.

\begin{theorem}[Exact $Q\times Q \leftrightarrow (d+1)\times(d+1)$ Sylvester Spectrum Reduction]
\label{thm:sylvester_spectrum}
Linearizing Eq.~\eqref{eq:inner_map_T} around a source-free uniform rest state $(\hat{p}_0, \bm{u}_0 = \bm{0}, \bm{s}_0 = \bm{0})$ with frozen history $\bm{h}_q$, zero body force, and uniform $\beta \in [0, 1)$ on a $d$-dimensional periodic lattice ($d \in \{2,3\}$), and defining
\begin{equation}
\label{eq:Shat_moment_blocks}
  \widehat{S}(\bm{k}) := P\,\diag\!\bigl(e^{-i\bm{c}_k\cdot\bm{k}}\bigr) E
  = \begin{pmatrix}
      \widehat{s}_{00}(\bm{k}) & \widehat{s}_{0u}(\bm{k}) \\
      \widehat{s}_{u0}(\bm{k}) & \widehat{s}_{uu}(\bm{k})
    \end{pmatrix} \in \C^{(d+1)\times(d+1)},
\end{equation}
we obtain:
\begin{enumerate}
  \item The Fourier symbol of $T$ factors through $P$ as $\widehat{\mathcal{T}}_{\mathrm{lin}}(\bm{k}) = \mathcal{U}(\bm{k}) P$, with reduced moment propagation matrix
  \begin{equation}
  \label{eq:A_moment_matrix}
    A(\bm{k}) := P\,\mathcal{U}(\bm{k})
    = \frac{c_a}{d_a}\begin{pmatrix}
        \widehat{s}_{00}(\bm{k}) & \widehat{s}_{0u}(\bm{k}) \\[3pt]
        \widehat{s}_{u0}(\bm{k}) & \beta I_d + (1 - \beta)\widehat{s}_{uu}(\bm{k})
      \end{pmatrix}.
  \end{equation}
  \item For any lifted preconditioner $\widehat{\mathcal{K}}(\bm{k}) = E K(\bm{k}) P$, the full $Q\times Q$ amplification symbol $\widehat{\mathcal{G}}_{\mathrm{full}}(\bm{k}) := \widehat{\mathcal{T}}_{\mathrm{lin}}(\bm{k}) + \widehat{\mathcal{K}}(\bm{k})(\widehat{\mathcal{T}}_{\mathrm{lin}}(\bm{k}) - I_Q)$ obeys
  \begin{gather}
  \label{eq:sylvester_charpoly}
    \det\!\Bigl(\lambda I_Q - \widehat{\mathcal{G}}_{\mathrm{full}}(\bm{k})\Bigr)
    = \lambda^{Q - (d+1)} \det\!\Bigl(\lambda I_{d+1} - B(\bm{k})\Bigr), \\
    B(\bm{k}) := A(\bm{k}) + K(\bm{k})\bigl(A(\bm{k}) - I_{d+1}\bigr), \nonumber
  \end{gather}
  so all $Q - (d+1)$ non-conserved kinetic modes ($6$ in D2Q9, $15$ in D3Q19) possess identically zero eigenvalues across $[-\pi, \pi]^d$, and $\rho(\widehat{\mathcal{G}}_{\mathrm{full}}(\bm{k})) \equiv \rho(B(\bm{k}))$.
\end{enumerate}
\end{theorem}

\begin{proof}
At rest ($\bm{u}_0 = \bm{0}$, $\bm{s}_0 = \bm{0}$), Eq.~\eqref{eq:inner_T_step1} gives $\delta\bm{m} + \tfrac{1}{2}\delta\bm{s} = \tfrac{c_a}{d_a}P\delta\bm{f}$, so Eqs.~\eqref{eq:inner_T_step2}--\eqref{eq:inner_T_step4} and Proposition~\ref{prop:varvisc_properties}(1) yield $\widehat{\mathcal{T}}_{\mathrm{lin}}(\bm{k}) = \tfrac{c_a}{d_a}[\widehat{S}_{\mathrm{str}}(\bm{k}) E + \beta E_u(I_d - \widehat{s}_{uu}(\bm{k}))\Pi_u]P =: \mathcal{U}(\bm{k})P$. Applying $P E_u(\bm{v}) = (0, \bm{v})^{\mathsf{T}}$ yields Eq.~\eqref{eq:A_moment_matrix}, and Sylvester's identity $\det(\lambda I_Q - X P) = \lambda^{Q-(d+1)}\det(\lambda I_{d+1} - P X)$ applied to $X = \mathcal{U} + E K(A - I_{d+1})$ gives Eq.~\eqref{eq:sylvester_charpoly}.
\end{proof}

By Theorem~\ref{thm:sylvester_spectrum}, an exact periodic preconditioner would invert $I_{d+1} - A(\bm{k})$ in Fourier space. To build a local real-space counterpart applicable on general bounded domains without FFTs, we expand Eq.~\eqref{eq:Shat_moment_blocks} at low wavenumber to obtain $\widehat{s}_{00} = 1 - \tfrac{1}{6}|\bm{k}|^2 + O(|\bm{k}|^4)$, $\widehat{s}_{0u} = -i\bm{k}^{\mathsf{T}} + O(|\bm{k}|^3)$, $\widehat{s}_{u0} = -ic_s^2\bm{k} + O(|\bm{k}|^3)$, and $I_d - [\beta I_d + (1-\beta)\widehat{s}_{uu}] = \nulu(|\bm{k}|^2 I_d + 2\bm{k}\bm{k}^{\mathsf{T}}) + O(|\bm{k}|^4)$. Dropping longitudinal $2\nulu\bm{k}\bm{k}^{\mathsf{T}}$ coupling for scalar separability, setting $c_a^+ := \max(c_a, 0)$, and mapping $i\bm{k} \mapsto \NablaC$ and $|\bm{k}|^2 \mapsto -\DeltaAx$ with $-\DeltaAx\phi(\bm{x}) := \sum_{\alpha=1}^d[2\phi(\bm{x}) - \phi(\bm{x}+\bm{e}_\alpha h) - \phi(\bm{x}-\bm{e}_\alpha h)]$ yields the real-space moment surrogate $\Lsurr$ for $\bm{r}_m = P(T(\bm{f}) - \bm{f})$,
\begin{equation}
\label{eq:Lsurr_def}
  \Lsurr\begin{pmatrix}\delta\hat{p} \\ \delta\bm{u}\end{pmatrix}
  := \frac{1}{d_a}\begin{pmatrix}
       a_q\delta\hat{p} - \frac{1}{6}c_a^+\DeltaAx\delta\hat{p} + c_a\NablaC\cdot\delta\bm{u} \\[4pt]
       c_a c_s^2\NablaC\delta\hat{p} + \bigl(a_q - c_a^+\nulu(\bm{x})\DeltaAx\bigr)\delta\bm{u}
     \end{pmatrix}.
\end{equation}

\subsection{Matrix-free local polynomial Uzawa smoother, spectral envelope, and unified algorithm}
\label{subsec:sec4_uzawa}
\label{subsec:sec4_counterexamples}

To approximate $\Lsurr^{-1}$ matrix-free on general 2D and 3D domains without global PPE assembly, \VDULBM{} combines three local stages:
\begin{enumerate}
  \item \emph{Regularized single-sweep Schur preconditioner $\Ploc^{-1}$.} Replacing $-\DeltaAx$ by $2d$ with acoustic-shear damping $\gamma_u = 0.35$ maps $\bm{q} = (q_p, \bm{q}_u)^{\mathsf{T}}$ to $\bm{z} = (z_p, \bm{z}_u)^{\mathsf{T}} := \Ploc^{-1}\bm{q}$ via positive pointwise denominators and two $2d$-point axial passes~\citep{Uzawa1958,BenziGolubLiesen2005},
  \begin{gather}
  \label{eq:bs_ss_defs}
    b_s(\bm{x}) := a_q + \gamma_u c_a^+ + 2 d\,c_a^+\nulu(\bm{x}), \qquad
    s_s(\bm{x}) := a_q + \frac{d}{3}c_a^+ + \frac{d\,c_s^2 (c_a^+)^2}{b_s(\bm{x})}, \\
  \label{eq:Ploc_sweep}
    z_p(\bm{x}) = \frac{d_a q_p(\bm{x}) - \frac{c_a d_a}{b_s(\bm{x})}\bigl(\NablaC\cdot\bm{q}_u\bigr)(\bm{x})}{s_s(\bm{x})}, \qquad
    \bm{z}_u(\bm{x}) = \frac{d_a \bm{q}_u(\bm{x}) - c_a c_s^2\bigl(\NablaC z_p\bigr)(\bm{x})}{b_s(\bm{x})}.
  \end{gather}
  \item \emph{Two-step Richardson coupling ($\kLoc = 2$).} Starting from $\bm{m}_{\mathrm{tot}}^{(0)} = \bm{r}_m$, we iterate
  \begin{equation}
  \label{eq:ksw_richardson}
    \bm{m}_{\mathrm{tot}}^{(s)} = \bm{m}_{\mathrm{tot}}^{(s-1)}
    + \Ploc^{-1}\!\left(\bm{r}_m - \Lsurr\bm{m}_{\mathrm{tot}}^{(s-1)}\right), \qquad s = 1, \dots, \kLoc.
  \end{equation}
  \item \emph{Separable $3^d$-point binomial Nyquist filter $\Fd$.} Passing the moment increment through $d$ sequential 1D three-point averages $(\mathcal{B}_\alpha\phi)(\bm{x}) := \tfrac{1}{4}\phi(\bm{x}-\bm{e}_\alpha h) + \tfrac{1}{2}\phi(\bm{x}) + \tfrac{1}{4}\phi(\bm{x}+\bm{e}_\alpha h)$ suppresses grid-checkerboard modes where $(\widehat{\NablaC})_\alpha(\pm\pi) = 0$ via
  \begin{equation}
  \label{eq:binomial_filter}
    \Delta\bm{m}_{\mathrm{corr}} = \Fd\!\left(\bm{m}_{\mathrm{tot}}^{(\kLoc)} - \bm{r}_m\right), \qquad
    \Fhat(\bm{k}) = \prod_{\alpha=1}^d \cos^2\!\left(\frac{k_\alpha}{2}\right) \in [0, 1].
  \end{equation}
\end{enumerate}
The kinetic state then updates as $\bm{f}^{(\ell+1)} = T(\bm{f}^{(\ell)}) + E\,\Delta\bm{m}_{\mathrm{corr}}$. On closed curved domains we subtract the mean pressure increment $\langle\Delta\hat{p}_{\mathrm{corr}}\rangle$, and at non-fluid neighbors we apply homogeneous Neumann pressure and Dirichlet velocity stencils; by Proposition~\ref{prop:fixed_point_equivalence}(1), exact fixed points of $T$ remain unchanged whenever $I + K_m$ is nonsingular.

\begin{figure}[tbp]
  \centering
  \includegraphics[width=\textwidth]{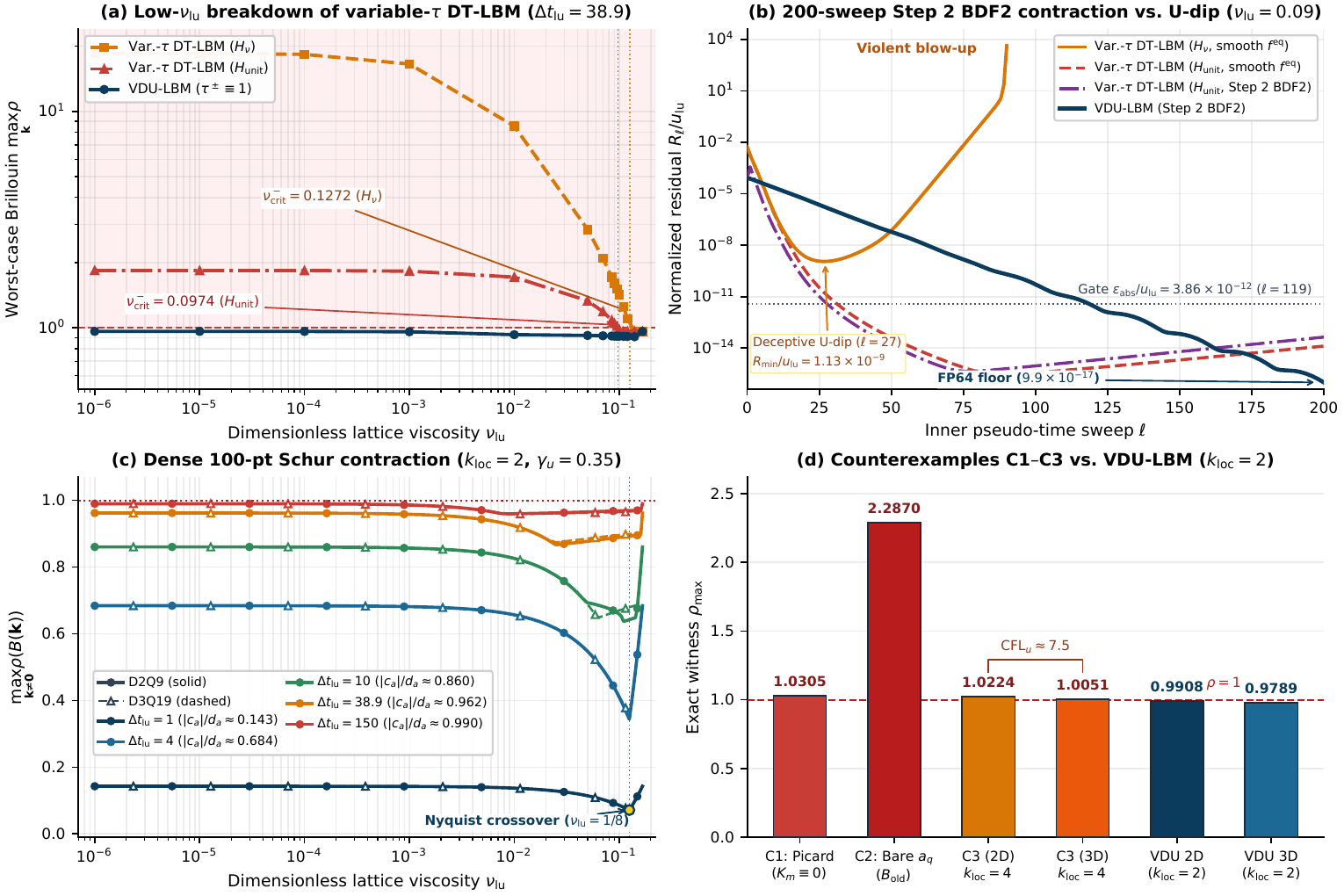}
  \caption{Spectral foundation of \VDULBM{} across D2Q9 and D3Q19. (a)~Worst-case Brillouin amplification $\max_{\bm{k}}\rho$ across $\nulu \in [10^{-6}, 1/6]$ ($\dtlu = 38.9$), contrasting low-viscosity passband breakdown of variable-$\tau$ dual-time LBM ($\nu_{\mathrm{crit}}^- = 0.0974$ under $H_{\mathrm{unit}}$, $0.1272$ under $H_\nu$) against viscosity-invariant \VDULBM{}. (b)~Inner sweep trajectories at $\nulu = 0.09$, contrasting the deceptive transient U-dip of variable-$\tau$ $H_\nu$ against actual BDF1-started Step~2 BDF2 contraction of \VDULBM{} (crossing $\varepsilon_{\mathrm{abs}}/\ulu = 3.86\times 10^{-12}$ at $\ell=119$ and reaching $9.9\times 10^{-17}$ at $\ell=200$). (c)~Dense 100-point discrete Brillouin scan of \VDULBM{} ($\kLoc = 2, \gamma_u = 0.35$) across $(\dtlu, \nulu)$ in 2D and 3D, resolving the axial-Nyquist mode crossover valley near $\nulu = 1/8$. (d)~Constructive spectral witnesses C1--C3 versus contractive baseline $\kLoc = 2$ (\ref{app:counterexamples}).}
  \label{fig:fig01_spectral_foundation}
\end{figure}

Figure~\ref{fig:fig01_spectral_foundation}c,d and three exact Fourier witnesses C1--C3 detailed in \ref{app:counterexamples} quantify the operating envelope of this local Uzawa smoother. We evaluate $\rho(B(\bm{k}))$ across $100$ discrete viscosities in $\nulu \in [10^{-6}, 1/6]$ ($70$ logarithmically spaced below $0.02$ and $30$ linearly spaced on $[0.02, 1/6]$) and $\dtlu \in \{1, 4, 10, 38.9, 150\}$ over symmetry-reduced $128^2$ D2Q9 ($2{,}144$ nonzero wavevectors) and $48^3$ D3Q19 ($2{,}924$ wavevectors) Brillouin grids $\mathcal{G}_d$. Across every sampled pair $(\dtlu, \nulu)$, the baseline parameters $\kLoc = 2$ and $\gamma_u = 0.35$ suppress both unpreconditioned acoustic instability (C1: $\rho = 1.0305$) and unfiltered Schur over-correction (C2: $\rho = 2.2870$), satisfying the uniform rest-state contraction bound
\begin{equation}
\label{eq:rest_contraction_envelope}
  \max_{\bm{k} \in \mathcal{G}_d \setminus \{\bm{0}\}} \rho\bigl(B(\bm{k})\bigr) \le \frac{|1 - a_q/2|}{1 + a_q/2} = \frac{|c_a|}{d_a} < 1.
\end{equation}
On the unfiltered axial Nyquist face $\bm{k} = (\pi, 0)$ where $\Fhat(\pi,0) = 0$, transverse shear contracts at $|\lambda_\perp| = \tfrac{|c_a|}{d_a}(1 - 4\nulu)$ whereas the longitudinal mode obeys $|\lambda_\parallel| = \tfrac{|c_a|}{d_a}|1 - 12\nulu|$; these two branches cross at $\nulu = 1/8 = 0.125$ ($0.3421$ at $\dtlu=4$, while low-wavenumber 3D shear reaches $0.3720$ at $(0,0,\pi/24)$) and return to $|c_a|/d_a$ at $\nulu = 1/6$, producing the valley seen in Fig.~\ref{fig:fig01_spectral_foundation}c. Under nonzero background advection, omitting $-i(\bm{u}_0\cdot\bm{k})$ from $\Lsurr$ causes fourth-degree $\kLoc = 4$ to overshoot at $\CFLu \approx 7.5$ (C3: $\rho = 1.0224$ in 2D and $1.0051$ in 3D), whereas $\kLoc = 2$ remains contractive at $0.9908$ and $0.9789$; we therefore fix $\kLoc = 2$ with $\CFLu \le 2$ in convection-active runs. Finally, in the strong-elliptic limit $a_q = o(\nulu k_{\min}^2)$ as $k_{\min} = 2\pi/N \to 0$, expanding Eq.~\eqref{eq:sylvester_charpoly} on the fundamental shear branch gives
\begin{equation}
\label{eq:elliptic_gap_scaling}
  1 - \rho\bigl(B(k_{\min}\bm{e}_1)\bigr) = \nulu\!\left(1 + \frac{2}{0.35 + 2d\,\nulu}\right) k_{\min}^2 + o(k_{\min}^2) = O(h^2) \quad \bigl(a_q = o(\nulu k_{\min}^2)\bigr).
\end{equation}
This spectral gap imposes an $\Omega(h^{-2})$ local sweep lower bound on the fundamental shear branch alongside $\Omega(a_q^{-1})$ unfiltered axial Nyquist modes as $\nulu \to 0^+$ or $\nulu \to 1/6^-$ when $a_q \ll 1$, unless paired with coarse-to-fine grid prolongation for steady states or nonlocal elliptic preconditioning.

Table~\ref{tab:stencil_complexity} and Algorithm~\ref{alg:vdlbm_master} summarize per-sweep locality, auxiliary memory footprint, and the complete boundary-closed execution loop. For comparison in the $a_q \to 0^+$ elliptic limit, we also list two nonlocal analytical references: a unidirectional 1D-channel wall-normal tridiagonal shear-block solver (channel Thomas reference, $O(N_y)$ cost) and a periodic Fourier symbol inverse ($2(d+1)$ FFTs, which reduces at $\nulu=1/6$ to periodic \MSCLBM{}~\citep{Ding2026MSCLBM}).

\begin{table}[tbp]
\centering
\caption{Computational stencil locality, auxiliary memory footprint, and global synchronization per inner sweep ($N_{\Omega} = |\Omega_f|$, baseline $\kLoc = 2$).}
\label{tab:stencil_complexity}
\scriptsize
\setlength{\tabcolsep}{4.0pt}
\renewcommand{\arraystretch}{1.14}
\begin{tabular}{l c c c p{0.34\textwidth}}
\toprule
\textbf{Solver / Preconditioner} & \textbf{$\ell_\infty$ Radius} & \textbf{Aux.\ Mem.} & \textbf{Global Solves} & \textbf{Global Reductions per Sweep} \\
\midrule
Unaccelerated Picard & $1$ cell & $O(Q N_\Omega)$ & None & $1$ ($R_\ell$); $2$ on closed curved walls \\
\textbf{\VDULBM{} (local Uzawa)} & $1 / 7 / 8$ cells\textsuperscript{a} & $O((d\!+\!1)N_\Omega)$ & \textbf{None} & \textbf{$1$ ($R_\ell$)} open/planar; $2$--$3$ closed curved \\
1D channel Thomas ref. & Nonlocal in $y$ & $O(N_y)$ & $O(N_y)$ Thomas & $1$ ($R_\ell$); $O(1)$-sweep 1D elliptic limit \\
Continuum PIMPLE / MAC & Global (PPE) & CSR + vecs & Sparse PPE & Repeated CG/AMG dot-product reductions \\
\bottomrule
\end{tabular}
\par\smallskip
\raggedright\scriptsize \textsuperscript{a}Local stencil radii: base sweep $T$ has $\ell_\infty$ radius $1$; corrector $K_m\bm{r}_m$ executes $3\kLoc + d$ 1D axial/binomial passes ($\ell_\infty$ radius $3\kLoc + 1 = 7$ at $\kLoc=2$); composed local update has $\ell_\infty$ radius $3\kLoc + 2 = 8$ (closed curved walls add scalar global reductions for $\delta M_\Gamma$ and $\langle\Delta\hat{p}_{\mathrm{corr}}\rangle$).
\end{table}

\begin{algorithm}[tbp]
\caption{Viscosity-Decoupled Uzawa Lattice Boltzmann Method (\VDULBM{})}
\label{alg:vdlbm_master}
\linespread{1.04}\selectfont\scriptsize
\begin{algorithmic}[1]
\Require Mesh $\Omega_f$ ($h$), scales $(U_0, \umax, \numax)$, Smagorinsky $C_s \ge 0$, $\kLoc = 2$, $\gamma_u = 0.35$, tolerance $\varepsilon_{\mathrm{rel}}$.
\State Fix harmonic pseudo-step $\dtlat \gets h^2 / (6\numax + U_0 h / \umax)$, $\clat \gets h / \dtlat$ \Comment{Eq.~\eqref{eq:harmonic_scaling}}
\State Initialize $\bm{m}^0 = (\hat{p}^0, \bm{u}^0)^{\mathsf{T}}$, $\bm{f}^0 \gets \bm{f}^{\mathrm{eq}}(\hat{p}^0, \bm{u}^0)$, $n \gets 0$.
\While{$t_n < T_{\mathrm{end}}$}
  \State Select outer step $\Delta t_{\mathrm{lu},n+1}$ ($\CFLu := \ulu\Delta t_{\mathrm{lu},n+1} \le 2$ on unsteady convection-resolved runs; $\Delta t_{\mathrm{lu},n+1} \le 2\Delta t_{\mathrm{lu},n}$ for $n \ge 1$).
  \State Compute BDF1/BDF2 $(a_q, \bm{h}_q, d_a, c_a)$ and AB2 predictor $\bm{u}^{*,n+1}$; freeze $\nulu, \beta, b_s, s_s$ \Comment{Eqs.~\eqref{eq:smag_nulu_beta}--\eqref{eq:var_bdf2_coeffs}, \eqref{eq:bs_ss_defs}}
  \State Set $\bm{f}^{(0)} \gets \bm{f}^n$ (or optional linear seed $(1+\omega_n)\bm{f}^n - \omega_n\bm{f}^{n-1}$ for $n \ge 1$ in the unsteady 2D runs of Sections~\ref{sec:sec6_1_ldc}--\ref{sec:sec6_3_varvisc_les}, reverting to $\bm{f}^n$ if unconverged), $\varepsilon_{\mathrm{abs}} \gets \varepsilon_{\mathrm{rel}}\min(1, a_q)\ulu$.
  \For{$\ell = 0, 1, \dots, K_{\max} - 1$}
    \State Reconstruct $(\bm{m}, \bm{s}, \tilde{\bm{u}})$ and collide $\bm{f}^\star \gets \bm{f}^{\mathrm{eq}}(\hat{p},\bm{u}) + \tfrac{1}{2}\bm{\Phi}(\bm{u},\bm{s})$ \Comment{Eq.~\eqref{eq:inner_map_T}, $\kerP$ cleared}
    \State Pull-stream $\bm{f}_{\mathrm{str}} \gets \mathcal{S}_\Gamma\bm{f}^\star$ via Section~\ref{sec:boundaries} (+ rest-mass neutrality~\eqref{eq:rest_mass_neutrality} if closed curved domain).
    \State Add bond-symmetric viscous lift $\bm{f}_T \gets \bm{f}_{\mathrm{str}} + E_u(\DuVD)$~\eqref{eq:DuVD_unified_wall} and compute $R_\ell \gets \|\bm{f}_T - \bm{f}^{(\ell)}\|_{\mathrm{rms}}$.
    \If{$R_\ell < \varepsilon_{\mathrm{abs}}$ \textbf{and} $\ell + 1 \ge 2$}
      \State Commit $\bm{f}^{n+1} \gets \bm{f}_T$, $\bm{m}^{n+1} \gets (P\bm{f}^{n+1} + \tfrac{1}{2}(\bm{h}_q + \bm{g}^{n+1}))/d_a$, apply pressure gauge, evaluate $\bm{F}_{\mathrm{obs,phys}}^{n+1}$~\eqref{eq:vdlbm_eulerian_force}; \textbf{break}.
    \ElsIf{$R_\ell$ non-finite \textbf{or} $R_\ell > 10^4\ulu$}
      \State \textbf{break}
    \EndIf
    \State Set $\bm{r}_m \gets P(\bm{f}_T - \bm{f}^{(\ell)})$, $\bm{m}_{\mathrm{tot}}^{(0)} \gets \bm{r}_m$; apply $\kLoc$ Uzawa passes~\eqref{eq:ksw_richardson} and binomial filter~\eqref{eq:binomial_filter}; set $\bm{f}^{(\ell+1)} \gets \bm{f}_T + E\,\Delta\bm{m}_{\mathrm{corr}}$.
  \EndFor
  \State If converged, advance $t_{n+1} \gets t_n + \Delta t_{\mathrm{lu},n+1}\dtlat$, $n \gets n + 1$; otherwise halve $\Delta t_{\mathrm{lu},n+1}$, retain accepted history $(\bm{m}^n,\bm{m}^{n-1})$ while resetting $\bm{f}^{(0)} \gets \bm{f}^n$, and retry.
\EndWhile
\end{algorithmic}
\end{algorithm}

\FloatBarrier
\section{Fundamental accuracy, kinetic robustness, and shear-layer resolution}
\label{sec:sec5_fundamental}

We validate the parameter-free \VDULBM{} formulation across three foundational periodic and channel suites against representative kinetic and continuum solvers. The comparison set comprises explicit BGK~\citep{Qian1992}, hybrid recursive regularized LBM with $\sigma=0.98$ (HRR)~\citep{Jacob2018}, cumulant LBM~\citep{Geier2015}, variable-$\tau$ dual-time LBM (DTS-TRT)~\citep{JiLuo2026}, colocated finite-volume PIMPLE with TVD van Leer MUSCL convection~\citep{Issa1986}, and staggered projection (Kim--Moin MAC)~\citep{KimMoin1985}, supplemented by fifth-order WENO5-RK3~\citep{JiangShu1996} on under-resolved shear layers. Complete \VDLBMLBMComparatorCount{}-kinetic and \VDLBMCFDComparatorCount{}-continuum archival tables are provided in \ref{app:all_solvers}. All tests enforce the low-Mach bound $\umax = 0.05$ ($\MaNum_0 \approx 0.0866$), diffusive and convective step protection $\Delta t_{\mathrm{expl}} \le \min\bigl(\umax h/U_0,\, h^2/(6\numax)\bigr)$ for explicit baselines, and the zero-mean pressure gauge $p_{\mathrm{phys}} = c_s^2\clat^2(\hat{p}-\langle\hat{p}\rangle)$. Pre-flight smooth convergence checks at $\ReNum=100$ confirm second-order spatial accuracy across all solver families ($p_x = \MetricPreflightVDLBMUzawaOrder$ for \VDULBM{}, $p_x \in \MetricPreflightPeerLBMOrderRange$ for peer LBMs, and $p_x \in \MetricPreflightCFDTwoOrderRange$ for second-order continuum solvers).

\subsection{Ten-decade Reynolds-number invariance in 2D Taylor--Green vortex flow}
\label{sec:sec5_1_accuracy}
\label{sec:sec5_1_1_tgv2d}

To isolate the rational regime transition~\eqref{eq:rational_Reh_law} and bulk space--time consistency (Theorem~\ref{thm:hydro_consistency}) from boundary closures, we first integrate the analytical 2D Taylor--Green vortex~\citep{TaylorGreen1937} on $\Omega = [0,1)^2$ with $U_0 = 1$ and $\nu_{\mathrm{phys}} = 1/\ReNum$,
\begin{equation}
  \begin{aligned}
    u_x^{\mathrm{ex}}(x,y,t) &= -\cos(2\pi x)\sin(2\pi y)\,e^{-8\pi^2 \nu_{\mathrm{phys}} t}, \quad
    u_y^{\mathrm{ex}}(x,y,t) = \sin(2\pi x)\cos(2\pi y)\,e^{-8\pi^2 \nu_{\mathrm{phys}} t}, \\
    p^{\mathrm{ex}}(x,y,t)   &= -\frac{1}{4}\bigl(\cos 4\pi x + \cos 4\pi y\bigr)\,e^{-16\pi^2 \nu_{\mathrm{phys}} t},
  \end{aligned}
  \label{eq:sec5_tgv2d_exact}
\end{equation}
over the regime-adaptive horizon $T_{\mathrm{phys}}(\ReNum) = 0.02/(4\pi^2/\ReNum + 1)$ with outer step $\dtphys = T_{\mathrm{phys}}/S$. We record relative $L_2$ errors $E_u = \|\bm{u}_h - \bm{u}^{\mathrm{ex}}\|_2 /\allowbreak \|\bm{u}^{\mathrm{ex}}\|_2$ and $E_p = \|(p_h - \langle p_h\rangle) - p^{\mathrm{ex}}\|_2 /\allowbreak \|p^{\mathrm{ex}}\|_2$, together with the effective numerical viscosity ratio $\nu_{\mathrm{num}}/\nu_{\mathrm{phys}} = -\ln\bigl(\|\bm{u}_h(T_{\mathrm{phys}})\|_2 /\allowbreak \|\bm{u}^{\mathrm{ex}}(0)\|_2\bigr) /\allowbreak (8\pi^2\nu_{\mathrm{phys}}T_{\mathrm{phys}})$.

\begin{figure}[tbp]
  \centering
  \includegraphics[width=\textwidth]{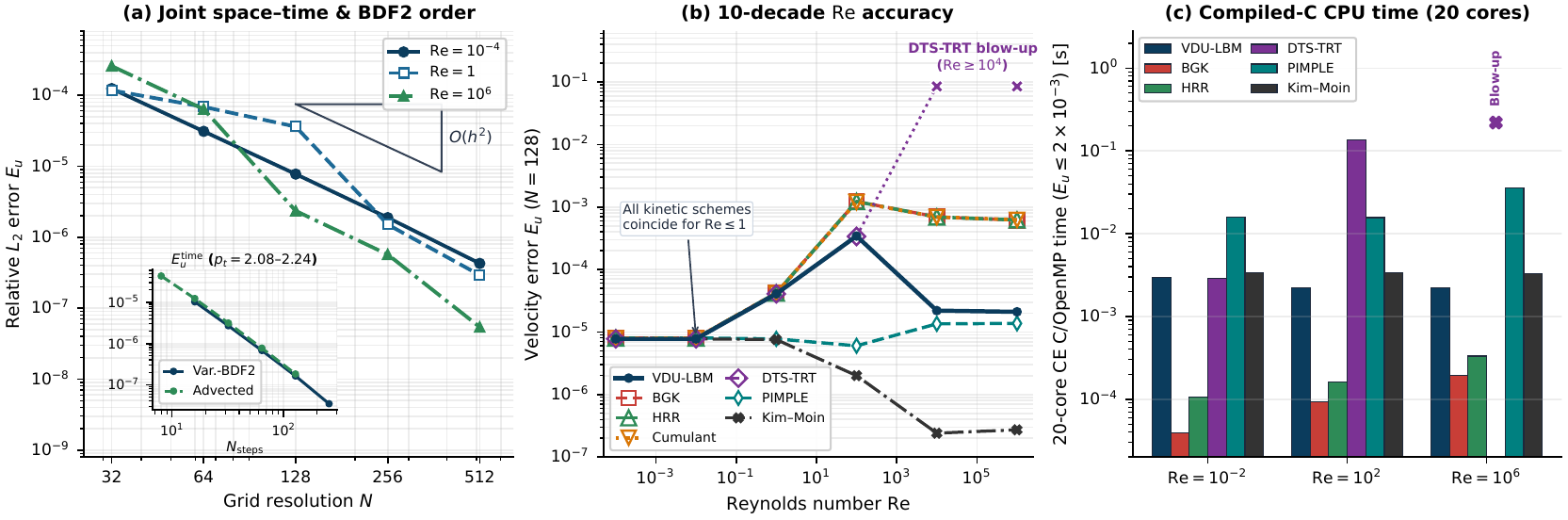}
  \caption{2D Taylor--Green vortex across $\ReNum \in [10^{-4}, 10^6]$ (archival multi-solver ledger in Table~\ref{tab:app_tgv2d_all16}). (a)~Joint space--time $O(h^2,\dtphys^2)$ convergence ($u_{\max,\mathrm{eff}} = 0.05\times 32/N$) across $\ReNum \in \{10^{-4}, 1, 10^6\}$, with Galilean-advected (fine-step self-reference $S_{\mathrm{ref}}=512$ on $N=64, \ReNum=100$) and variable-step BDF2 ($N_{\mathrm{steps}} = 2S$) temporal convergence inset ($p_t = \MetricTGVBDFAdvSOneTwoEightOrder\text{--}\MetricTGVBDFVarSOneTwoEightOrder$). (b)~Velocity error $E_u$ across ten $\ReNum$ decades ($N=128$, $S=64$, $T_{\mathrm{phys}} \le 0.02$), contrasting uniform \VDULBM{} accuracy against variable-$\tau$ DTS-TRT blow-up at $\ReNum \ge 10^4$ (all kinetic schemes coincide at $\ReNum \le 1$). (c)~Equal-accuracy ($E_u \le 2\times 10^{-3}$) bare-metal Linux C runtime across kinetic and continuum solvers at $\ReNum \in \{10^{-2}, 10^2, 10^6\}$.}
  \label{fig:fig02_tgv2d_all_re_efficiency}
\end{figure}

Figure~\ref{fig:fig02_tgv2d_all_re_efficiency} and Table~\ref{tab:app_tgv2d_all16} summarize joint space--time convergence, ten-decade Reynolds-number invariance, and bare-metal CPU runtime across $\ReNum \in \VDLBMReSpan$. Under joint space--time refinement with $u_{\max,\mathrm{eff}} = 0.05\times 32/N$ so that $\MaNum = O(h)$ (Fig.~\ref{fig:fig02_tgv2d_all_re_efficiency}a), $E_u$ at $\ReNum=1$ decreases from \MetricTGVJointUzawaNThirtyTwoReOneEu{} on $32^2$ to \MetricTGVJointUzawaNFiveOneTwoReOneEu{} on $512^2$ at convergence order \MetricTGVJointUzawaNFiveOneTwoOrder{} within $\MetricTGVJointUzawaNThirtyTwoSweeps\text{--}\MetricTGVJointUzawaNFiveOneTwoSweeps$ local sweeps per step. Temporal refinement across $N_{\mathrm{steps}} = 2S = 16\text{--}256$ likewise confirms second-order BDF2 accuracy under stationary, Galilean-advected ($\bm{u}_0=(0.2,0.1)$ on $N=64$ at $\ReNum=100$, measured against a fine-step temporal self-reference $S_{\mathrm{ref}}=512$), and $0.5\times$-halving variable-step schedules, with measured orders $p_t = \MetricTGVBDFStatSOneTwoEightOrder$, $\MetricTGVBDFAdvSOneTwoEightOrder$, and $\MetricTGVBDFVarSOneTwoEightOrder$, respectively.

At fixed resolution $N=128$ and $S=64$ (Fig.~\ref{fig:fig02_tgv2d_all_re_efficiency}b), \VDULBM{} maintains $E_u \in [\MetricTGVSixteenVDUzawaEuReMin, \MetricTGVSixteenVDUzawaEuReHun]$ across $\ReNum \in [10^{-4}, 10^6]$; its effective viscosity ratio $\nu_{\mathrm{num}}/\nu_{\mathrm{phys}}$ equals \MetricTGVSixteenVDUzawaNuRatio{} at $\ReNum=10^2$, stays within $[0.9991, 1.0023]$ across $\ReNum \in [10^{-4}, 10^4]$, and shifts to $1.4126$ only at the nearly inviscid short-horizon endpoint $\ReNum=10^6$ where exact viscous decay $8\pi^2\nu_{\mathrm{phys}}T_{\mathrm{phys}} \approx 1.58\times 10^{-6}$ approaches machine roundoff. Consistent with the Floquet--Bloch passband breakdown in Section~\ref{subsec:sec2_vartau_failure} (Fig.~\ref{fig:fig01_spectral_foundation}a), variable-$\tau$ DTS-TRT diverges for $\ReNum \ge 10^4$ as $\tau^+ \to 1/2^+$, whereas explicit BGK, HRR, and Cumulant remain stable on this smooth mode but incur roughly $30\times$ larger convective error under acoustic stepping (\MetricTGVSixteenBGKEuReMax{} versus \MetricTGVSixteenVDUzawaEuReMax{} for \VDULBM{} at $\ReNum=10^6$). Under identical GCC~13.2 \text{-O3} compilation on a 20-core Intel Xeon Platinum~8358 allocation (OpenMP enabled for $N^2 \ge 4096$) at target error $E_u \le 2\times 10^{-3}$ over $T_{\mathrm{phys}} \le 0.02$ (Fig.~\ref{fig:fig02_tgv2d_all_re_efficiency}c), \VDULBM{} ($\MetricTGVCostFineVDUzawaReHighT\text{ s}$ on $16^2$ at $\ReNum=10^6$) outpaces compiled-C colocated PIMPLE ($3.61\times 10^{-2}\text{ s}$ on $32^2$, $16.3\times$) and staggered Kim--Moin ($3.30\times 10^{-3}\text{ s}$ on $16^2$, $1.49\times$). Single-sweep explicit BGK ($1.94\times 10^{-4}\text{ s}$) and HRR ($3.31\times 10^{-4}\text{ s}$) require fewer total updates on this short periodic horizon, whereas long diffusive horizons or stiff bounded-$\tau$ viscosity contrasts reverse the step-cost balance in Sections~\ref{sec:sec5_1_2_womersley}, \ref{sec:sec6_2_curved_dfg}, and~\ref{sec:sec6_4_3d_gpu}.

\FloatBarrier
\subsection{\texorpdfstring{Pulsatile Womersley channel ($\alpha \in [0.5, 50]$) and single-sweep $\kerP$ immunity}{Pulsatile Womersley channel (alpha in [0.5, 50]) and single-sweep ker P immunity}}
\label{sec:sec5_1_2_womersley}
\label{sec:sec5_2_kinetic_robustness}

\begin{figure}[tbp]
  \centering
  \includegraphics[width=\textwidth]{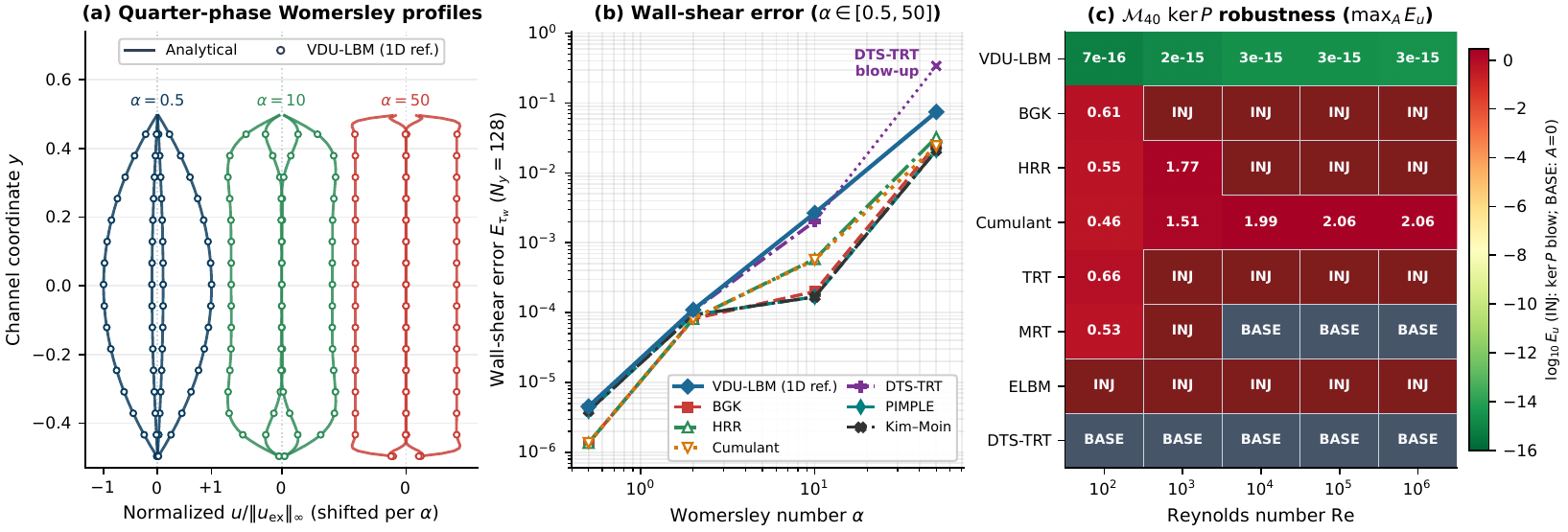}
  \caption{Pulsatile Womersley channel and kinetic $\kerP$ immunity (15-method ledger in Table~\ref{tab:app_wom_dsl_all15}). (a)~Quarter-phase velocity profiles across $\alpha \in \{0.5, 10, 50\}$ ($N_y=128$), resolving thin Stokes wall layers without slip. (b)~Wall-shear error $E_{\tau_w}(\alpha)$ across $\alpha \in [0.5, 50]$ ($N_y=128$), showing DTS-TRT blow-up at $\alpha = 50$. (c)~Synchronized $\mathcal{M}_{40}$ $\kerP$ perturbation heatmap across $\ReNum \in [10^2, 10^6]$, contrasting single-sweep machine-zero annihilation ($\le \MetricKerPVDUzawaMaxPert$) in \VDULBM{} against $\kerP$-injection blow-up (\text{INJ}) and unperturbed $A=0$ baseline failure (\text{BASE}) in variable-$\tau$ LBMs.}
  \label{fig:fig03_womersley_and_kerp}
\end{figure}

Moving from unbounded periodic decay to wall-bounded oscillatory shear, we test the planar link-wise closure~\eqref{eq:DuVD_unified_wall} ($q_i \equiv 1/2$) across Womersley numbers $\alpha = R\sqrt{2\pi / \nu_{\mathrm{phys}}} \in \VDLBMAlphaSpan$ together with direct $\kerP$ perturbation immunity. On $y \in [-0.5, 0.5]$ with half-width $R = 0.5$ and oscillatory body force $b_x(t) = 2\pi\cos(2\pi t)$, the analytical profile~\citep{Womersley1955,Loudon1998} reads
\begin{gather}
  \label{eq:sec5_womersley_exact}
  u_{\mathrm{ex}}(y, t) = \operatorname{Re}\!\left[\frac{1}{i}\left(1 - \frac{\cosh(\sqrt{i}\,\alpha\,y/R)}{\cosh(\sqrt{i}\,\alpha)}\right)e^{2\pi i t}\right], \\
  \left.\frac{\partial u_{\mathrm{ex}}}{\partial y}\right|_{y=-R} = \frac{1}{R}\operatorname{Re}\!\left[\frac{\sqrt{i}\,\alpha\tanh(\sqrt{i}\,\alpha)}{i}e^{2\pi i t}\right]. \nonumber
\end{gather}
Grid refinement across $N_y = 32 \to 1024$ using the 1D channel Thomas reference delivers second-order convergence across all $\alpha \in [0.5, 50]$ for both velocity ($p_u = \MetricWomConvOrderAlphaHalf$ and $\MetricWomConvOrderAlphaFifty$ at $\alpha = 0.5$ and $50$) and Eulerian wall-shear stress ($p_{\tau_w} = \MetricWomConvOrderAlphaTenTau$ at $\alpha = 10$, with finest-grid order within $1.998\text{--}2.001$ across all $\alpha$; Fig.~\ref{fig:fig03_womersley_and_kerp}a,b). Cross-checking purely local Uzawa smoothing at the quasi-steady diffusive extreme $\alpha = 0.5$ across $N_y \in \{32, 64, 128\}$ reproduces the exact 1D Thomas error ($E_u = \MetricWomConvNOneTwoEightUzawaEu$ versus $\MetricWomConvNOneTwoEightAlphaHalfEu$ at $N_y=128$) as inner sweeps grow from $1160.0 \to 2634.5 \to \MetricWomConvNOneTwoEightUzawaSweeps$ ($1256.24\text{ s}$ at $N_y=128$), which confirms the $\Omega(h^{-2})$ strong-elliptic sweep scaling predicted by Eq.~\eqref{eq:elliptic_gap_scaling}. By comparison, the 1D Thomas reference converges in $\MetricWomConvNOneTwoEightThomasSweeps$ sweeps ($0.43\text{--}0.46\text{ s}$), achieving a $207\text{--}623\times$ speedup over explicit BGK--ELBM baselines at $95.7\text{--}288.0\text{ s}$ (Table~\ref{tab:app_wom_dsl_all15}). At the thin-Stokes-layer extreme $\alpha = 50$, \VDULBM{} resolves the oscillatory wall layer without slip, whereas variable-$\tau$ DTS-TRT again diverges.

To test Theorem~\ref{thm:kerP_annihilation} directly at finite amplitude, we inject a 6-mode orthonormal basis $B \in \R^{6\times 9}$ of $\kerP$ satisfying $\|P B^{\mathsf{T}}\|_\infty = \MetricKerPBasisProjMax$ across the 20-cell $\mathcal{M}_{40}$ matrix spanning $\ReNum \in [10^2, 10^6]$ (Fig.~\ref{fig:fig03_womersley_and_kerp}c). A single sweep of \VDULBM{} clears all $\kerP$ modes to machine precision (\MetricKerPSingleCollisionAMin{}--\MetricKerPSingleCollisionAMax{}, with overall minimum $6.94\times 10^{-18}$) and remains stable across all $20/20$ cells with maximum macroscopic perturbation \MetricKerPVDUzawaMaxPert. By contrast, high-$\ReNum$ variable-$\tau$ baselines suffer either baseline failure already at zero injection (all $20$ DTS-TRT cells and $12$ MRT cells) or $\kerP$-triggered instability ($3$ additional MRT cells, high-$\ReNum$ BGK/TRT/HRR/ELBM blow-up, and $O(1)$ Cumulant drift of \MetricKerPCumMaxPert{} driven by nonlinear ghost--moment coupling).

\FloatBarrier
\subsection{Double shear layers: Brown--Minion dispersion, grid convergence, and physical LES}
\label{sec:sec5_3_dsl}

\begin{figure}[tbp]
  \centering
  \includegraphics[width=\textwidth]{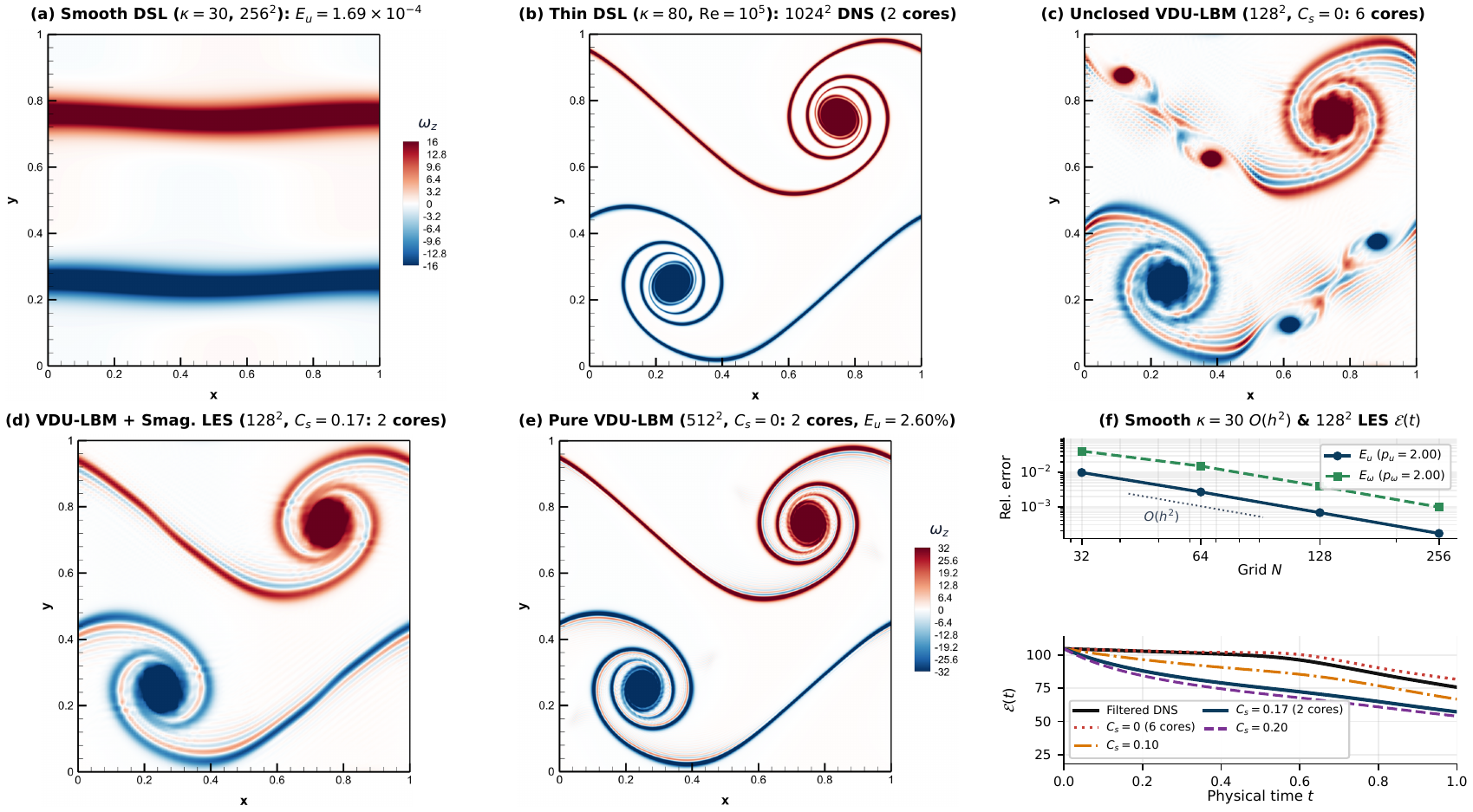}
  \caption{Double-shear-layer dispersion, physical LES closure, and grid convergence (15-method $128^2$ ledger in Table~\ref{tab:app_wom_dsl_all15}). (a)~Resolved smooth DSL ($\kappa=30, \ReNum=10^3, 256^2$, pure \VDULBM{} $C_s=0$, $E_u = 1.69\times 10^{-4}$). (b)~Thin DSL ($\kappa=80, \ReNum=10^5, t=1.0$): $1024^2$ pseudospectral DNS ($N_{\mathrm{cores}}=2$). (c)~Under-resolved $128^2$ pure \VDULBM{} ($C_s=0$): bounded energy with classical Brown--Minion central-dispersion braid vortices ($N_{\mathrm{cores}}=6$, while BGK/TRT/MRT/DTS-TRT/Kim--Moin blow up). (d)~Under-resolved $128^2$ \VDULBM{} + Smagorinsky LES ($C_s=0.17$): physical eddy viscosity restores two-vortex braids ($N_{\mathrm{cores}}=2, E_u=8.10\%$). (e)~Grid-refined $512^2$ pure \VDULBM{} ($C_s=0$): parameter-free convergence to the DNS spiral sheet ($N_{\mathrm{cores}}=2, E_u=2.60\%$). (f)~Smooth $\kappa=30$ $O(h^2)$ convergence (top) and $128^2$ resolved enstrophy evolution $\mathcal{E}(t)$ across $C_s$ (bottom).}
  \label{fig:fig04_dsl_vorticity_hero}
\end{figure}

Escalating from laminar analytical profiles to under-resolved nonlinear Kelvin--Helmholtz roll-up, periodic double shear layers~\citep{BrownMinion1995,MinionBrown1997} on $\Omega = [0,1)^2$ with $u_y(x,y,0) = 0.05\sin(2\pi(x + 0.25))$ and
\begin{equation}
  u_x(x,y,0) = \begin{cases}
    \tanh\bigl(\kappa(y - 0.25)\bigr), & y \le 0.5, \\
    \tanh\bigl(\kappa(0.75 - y)\bigr), & y > 0.5,
  \end{cases}
  \label{eq:sec5_dsl_kappa80}
\end{equation}
provide a canonical diagnostic separating kinetic ghost-mode blow-up, central dispersive roll-up, and numerical versus physical subgrid dissipation. On a resolved smooth layer ($\kappa = 30$, $\ReNum = 10^3$, $t = 0.2$), parameter-free \VDULBM{} ($C_s = 0$) exhibits second-order accuracy across $N = 32 \to 256$: $E_u$ decreases from $9.81\times 10^{-3}$ to $1.69\times 10^{-4}$ with orders $p_u = \MetricDSLThirtyNTwoFiveSixOrderU$ and $p_\omega = \MetricDSLThirtyNTwoFiveSixOrderOmega$ (Fig.~\ref{fig:fig04_dsl_vorticity_hero}a,f).

In the severe Brown--Minion thin-layer regime ($\kappa = 80$, $\ReNum \in [10^4, 10^6]$, $t = 1.0$), assessed against a $1024^2$ pseudospectral DNS reference with self-convergence $\le \MetricDSLDNSSelfConvReOneM$ (Fig.~\ref{fig:fig04_dsl_vorticity_hero}b), a coarse $128^2$ grid places fewer than two cells across the stretched vorticity braids. At this resolution, BGK, TRT, MRT, DTS-TRT, and central Kim--Moin diverge before $t = 1.0$ for $\ReNum \ge 10^5$ (Table~\ref{tab:app_wom_dsl_all15}). Because \VDULBM{} annihilates $\kerP$ at every inner sweep (Theorem~\ref{thm:kerP_annihilation}) while retaining a non-dissipative central viscous flux $\DuVD$, unclosed $128^2$ runs ($C_s = 0$) remain bounded throughout $t \in [0, 1]$ across all tested $\ReNum \in [10^4, 10^6]$. As expected for a centered second-order discretization, they form the classical Brown--Minion dispersive braid vortices ($N_{\mathrm{cores}} = 6$ and $E_u = \MetricDSLEightyVDPsiOneReHunKEu$ at $\ReNum = 10^5$), analogous to colocated PIMPLE ($4$ cores and $E_u = \MetricDSLEightyPIMPLEReHunKEu$; Fig.~\ref{fig:fig04_dsl_vorticity_hero}c).

Whereas explicit HRR ($\sigma = 0.98$) suppresses coarse-grid braids through built-in $O(h^3\text{--}h^4)$ numerical hyper-viscosity, \VDULBM{} resolves the thin sheet without artificial hyper-viscosity, either through physical subgrid closure at fixed coarse resolution or through parameter-free grid refinement. On the $128^2$ mesh, activating the scale-invariant Smagorinsky eddy viscosity ($C_s = 0.17$, Eq.~\eqref{eq:smag_nulu_beta}) eliminates spurious secondary vortices ($N_{\mathrm{cores}} = 2$, $E_u = 8.10\times 10^{-2}$, $23.4\text{ s}$) and regulates resolved enstrophy monotonically across $C_s \in [0.10, 0.20]$, while pairwise momentum remains conserved to $\le 1.08\times 10^{-16}$ over $t \in [0, 1]$ (Fig.~\ref{fig:fig04_dsl_vorticity_hero}d,f). Conversely, refining the mesh across $128^2 \to 256^2 \to 512^2$ with zero subgrid model ($C_s = 0$) removes the dispersive braid satellites monotonically: $N_{\mathrm{cores}}$ drops from $6 \to 4 \to 2$, $E_u$ falls from $16.2\% \to 4.95\% \to 2.60\%$, and the high-wavenumber spectral index $\mathcal{I}_{\mathrm{high}\text{-}k} = 0.180 \to 0.096 \to 0.022$ tracks the grid-restricted DNS values $0.241 \to 0.119 \to 0.016$ to recover the two-vortex spiral topology of the $1024^2$ spectral reference (Fig.~\ref{fig:fig04_dsl_vorticity_hero}e).

\FloatBarrier
\section{Complex geometries, heterogeneous viscosity, and three-dimensional GPU turbulence}
\label{sec:sec6_complex_3d}

Having verified regime-invariant second-order accuracy, single-sweep $\kerP$ immunity, and shear-layer convergence on periodic and planar channel domains, we now extend the assessment to complex boundary conditions, strongly heterogeneous viscosity fields, and three-dimensional turbulence. Specifically, Sections~\ref{sec:sec6_1_ldc}--\ref{sec:sec6_4_3d_gpu} examine corner-singular cavity flows, off-grid curved Couette and Sch\"afer--Turek cylinder wakes benchmarked against OpenFOAM-v2412, four-decade variable-viscosity flows in two and three dimensions, and $256^3$ D3Q19 turbulence and diffusion-stiff crossovers on an NVIDIA A100 GPU.

\subsection{Steady corner-eddy cascade and supercritical Hopf cavity flow across \texorpdfstring{$\mathrm{Re} \in [10^{-2}, 10^4]$}{Re in [1e-2, 1e4]}}
\label{sec:sec6_1_ldc}

\begin{figure}[tbp]
  \centering
  \includegraphics[width=\textwidth]{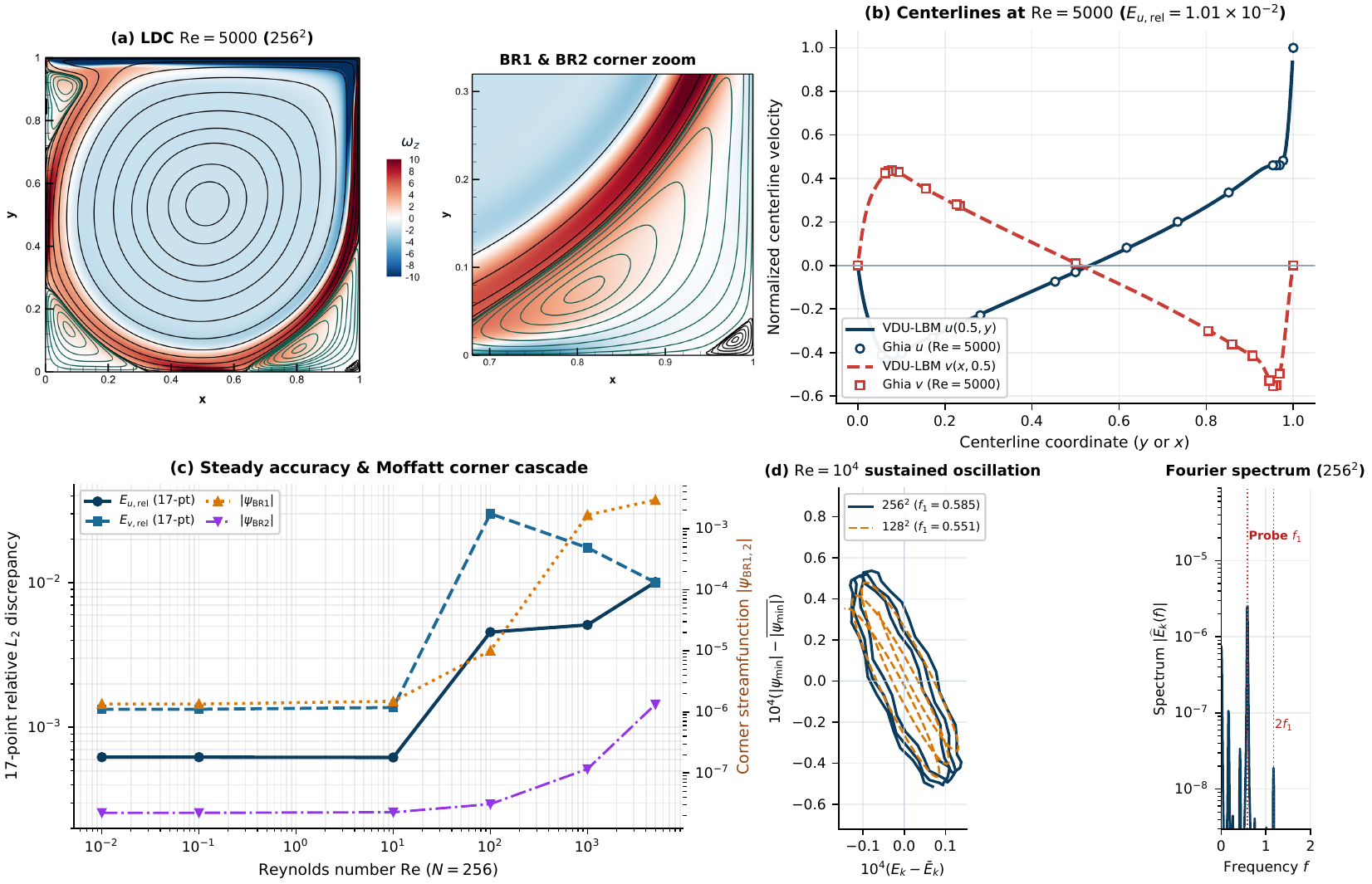}
  \caption{Lid-driven cavity across $\ReNum \in [10^{-2}, 10^4]$. (a)~Vorticity field and exact closed streamfunction isolines at $\ReNum=5000$ ($256^2$), paired with an external bottom-right cascade zoom resolving both primary corner eddy BR1 and secondary Moffatt eddy BR2 without occluding the main cavity. (b)~Centerline velocity profiles at $\ReNum=5000$ ($256^2$) versus \citet{Ghia1982}. (c)~17-point centerline discrepancy ($\ReNum \le 10$ against $512^2$ self-reference; $\ReNum \ge 100$ against \citet{Ghia1982}) and BR1/BR2 corner streamfunction amplitudes across $\ReNum \in [10^{-2}, 5000]$. (d)~Sustained supercritical oscillation at $\ReNum=10^4$ over $t \in [250, 350]$ ($256^2$ vs.\ $128^2$) and detrended kinetic-energy Fourier spectrum (probe frequency $f_1 = \MetricLDCHopfNTwoFiveSixFreq$ at $(0.8,0.8)$).}
  \label{fig:fig05_ldc_bifurcation_hero}
\end{figure}

To move from periodic and unidirectional shear to closed two-dimensional recirculation with corner singularities, we consider the lid-driven cavity in $\Omega = [0, 1]^2$ on cell-centered grids $N \in \{64, 128, 256, 512\}$ with stationary walls at $x=0,1$ and $y=0$ and a moving lid $\bm{u}_{\mathrm{lid}} = (U_0/\clat, 0)$ at $y=1$ ($U_0=1$). Planar walls use the corner-safe reflection~\eqref{eq:DuVD_unified_wall} at $q_i \equiv 1/2$, for which $\delta M_\Gamma \equiv 0$. Steady states across $\ReNum \in [10^{-2}, 5000]$ employ coarse-to-fine grid prolongation across $64 \to 128 \to 256 \to 512$ at $\dtlu \in [10, 40]$ and $\varepsilon_{\mathrm{rel}} = 10^{-10}$ until $R_{\mathrm{steady}}/\ulu < 10^{-9}$, which bypasses the single-grid $a_q \to 0$ sweep lower bound~\eqref{eq:elliptic_gap_scaling}. Beyond the supercritical Hopf threshold $\ReNum_c \in [8000, 8050]$~\citep{BruneauSaad2006}, time-accurate BDF2 integration at $\ReNum=10^4$ with $\CFLu \le 1$ is integrated past $T_{\mathrm{phys}} = 250$ spin-up and sampled over $t \in [250, 350]$; results across all six $\ReNum$ decades are summarized in Fig.~\ref{fig:fig05_ldc_bifurcation_hero} and Table~\ref{tab:ldc_consolidated}.

\begin{table}[tbp]
  \centering
  \caption{Quantitative 2D lid-driven cavity benchmark: steady primary/corner vortex extrema against \citet{Ghia1982}, \citet{BotellaPeyret1998}$^\mathrm{a}$, and \citet{BruneauSaad2006}$^\mathrm{b}$, alongside supercritical oscillation statistics over $t \in [250, 350]$ at $\ReNum=10^4$.}
  \label{tab:ldc_consolidated}
  \scriptsize
  \setlength{\tabcolsep}{2.8pt}
  \renewcommand{\arraystretch}{1.10}
  \begin{tabular}{cc ccccc}
    \toprule
    $\ReNum$ & Grid $N$ & $\mathrm{RelL2}_{17}(u)$ & $\psi_{\min}^{\mathrm{PV}}$ (Ref.) & $u_{\min}(0.5,y)$ (Ref.) & $v_{\min}(x,0.5)$ (Ref.) & $\psi_{\mathrm{BR1}}$ / Probe $f_1$ \\
    \midrule
    $100$  & $256$ & \MetricLDCReHunRelU  & \MetricLDCReHunPsiPV\ ($-0.10351^\mathrm{b}$) & \MetricLDCReHunUmin\ ($-0.21404^\mathrm{b}$) & \MetricLDCReHunVmin\ ($-0.25380^\mathrm{b}$) & $1.01\times 10^{-5}$ \\
    $1000$ & $512$ & \MetricLDCReOneKRelU & \MetricLDCReOneKPsiPV\ ($-0.11894^\mathrm{a}$) & \MetricLDCReOneKUmin\ ($-0.38857^\mathrm{a}$) & \MetricLDCReOneKVmin\ ($-0.52708^\mathrm{a}$) & \MetricLDCMoffattReOneKBRone \\
    $5000$ & $512$ & \MetricLDCReFiveKRelU & \MetricLDCReFiveKPsiPV\ ($-0.12197^\mathrm{b}$) & \MetricLDCReFiveKUmin\ ($-0.44733^\mathrm{b}$) & \MetricLDCReFiveKVmin\ ($-0.57428^\mathrm{b}$) & \MetricLDCMoffattReFiveKBRone \\
    \midrule
    $10^4$ (osc.) & $128$ & --- & $\langle\psi_{\min}\rangle_t = \MetricLDCHopfNOneTwoEightPsiMean$ & $\langle u_{\min}\rangle_t = \MetricLDCHopfNOneTwoEightUminMean$ & $\mathrm{std}_t^\ast(E_k) = \MetricLDCHopfNOneTwoEightEkStd$ & $f_1 = \MetricLDCHopfNOneTwoEightFreq$ \\
    $10^4$ (osc.) & $256$ & --- & $\langle\psi_{\min}\rangle_t = \MetricLDCHopfNTwoFiveSixPsiMean$ & $\langle u_{\min}\rangle_t = \MetricLDCHopfNTwoFiveSixUminMean$ & $\mathrm{std}_t^\ast(E_k) = \MetricLDCHopfNTwoFiveSixEkStd$ & $f_1 = \MetricLDCHopfNTwoFiveSixFreq$ \\
    \bottomrule
  \end{tabular}
  \par\smallskip
  \raggedright\scriptsize Notes: $^\mathrm{a}$Spectral Chebyshev reference~\citep{BotellaPeyret1998}; $^\mathrm{b}$Fine-grid multigrid reference ($1024^2/2048^2$)~\citep{BruneauSaad2006}; $^\ast$linearly detrended standard deviation over $t\in[250,350]$ ($f_1$ from detrended $u_x(0.8,0.8)$ probe; quarter-window mean $E_k$ drift $\approx 7.2\times 10^{-5}$, linear-fit drift $9.6\times 10^{-5}$, on $256^2$ over $[250,350]$). Creeping-flow $512^2$ self-convergence ($\ReNum \in \{10^{-2}, 10^{-1}, 10\}$: $\mathrm{RelL2}_{17}(u) = \MetricLDCReZeroPointZeroOneRelU\text{--}\MetricLDCReTenRelU$) and BR1/BR2 amplitudes are plotted in Fig.~\ref{fig:fig05_ldc_bifurcation_hero}c ($512^2$ Moffatt ratios at $\ReNum=1000$: $r_{\mathrm{BR1}}/r_{\mathrm{BR2}} = \MetricLDCMoffattReOneKSizeRatio$, $|\psi_{\mathrm{BR1}}/\psi_{\mathrm{BR2}}| = \MetricLDCMoffattReOneKPsiRatio$ vs.\ Stokes asymptotes $16.57$ and $3.63\times 10^4$~\citep{Moffatt1964}).
\end{table}

Exact streamfunction contours capture closed primary, TL1, BL1, BR1, and secondary Moffatt BR2 eddies without streamline breakage (Fig.~\ref{fig:fig05_ldc_bifurcation_hero}a). Centerline $u$-velocity profiles match the $512^2$ creeping-flow self-reference for $\ReNum \le 10$ and the 17-point \citet{Ghia1982} benchmark for $\ReNum \ge 100$ within $0.06\%\text{--}0.79\%$, while $v$-centerline profiles agree within $0.14\%$ for $\ReNum \le 10$ and within $1.1\%\text{--}3.0\%$ against the coarse $129^2$ \citet{Ghia1982} tabulation (Fig.~\ref{fig:fig05_ldc_bifurcation_hero}b,c). Primary vortex and velocity extrema match spectral and fine-grid multigrid references within $0.2\%\text{--}1.1\%$ for $\ReNum \le 1000$ and $1.6\%\text{--}3.1\%$ at $\ReNum=5000$ on $512^2$, and time-accurate marching at $\ReNum=10^4$ sustains a periodic supercritical limit cycle with fundamental probe frequency $f_1 = \MetricLDCHopfNTwoFiveSixFreq$ on $256^2$ (Fig.~\ref{fig:fig05_ldc_bifurcation_hero}d).

\FloatBarrier
\subsection{Off-grid curved Couette flow and Sch\"afer--Turek cylinder wakes vs.\ OpenFOAM}
\label{sec:sec6_2_curved_dfg}

\begin{figure}[tbp]
  \centering
  \includegraphics[width=\textwidth]{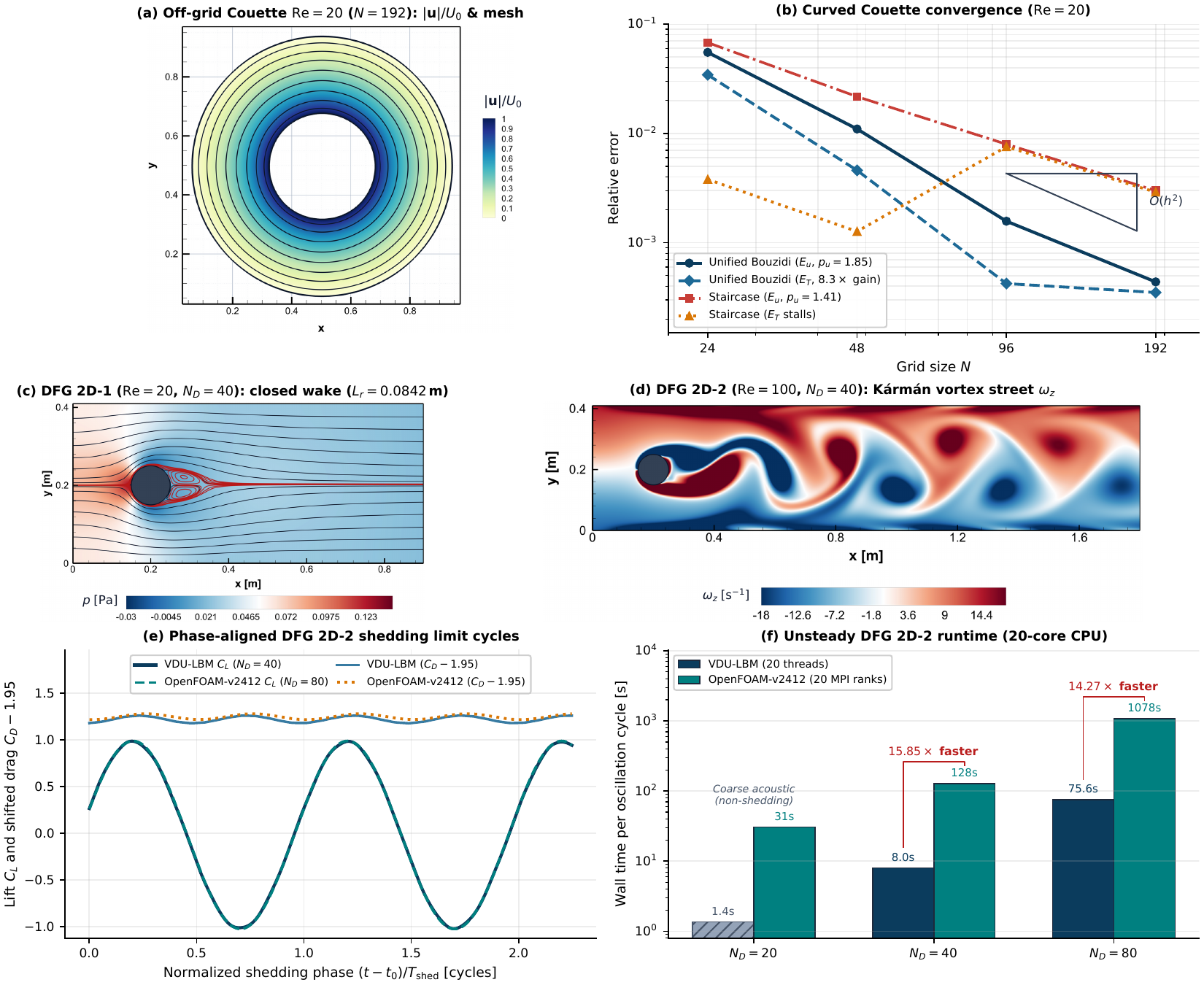}
  \caption{Off-grid curved boundaries and Sch\"afer--Turek cylinder wakes. (a)~Off-grid cylindrical Couette ($\ReNum=20, N=192$) velocity magnitude and circular streamlines across cut Cartesian cells. (b)~Couette velocity ($E_u$) and torque ($E_T$) convergence across $N \in \{24,\dots,192\}$ versus staircase bounce-back. (c)~Steady DFG~2D-1 ($\ReNum=20, N_D=40$) pressure field and closed twin recirculation bubble ($L_r = 0.0842\text{ m}$). (d)~Unsteady DFG~2D-2 ($\ReNum=100, N_D=40, t=14.82\text{ s}$) K\'arm\'an vortex street vorticity $\omega_z$. (e)~Phase-aligned periodic lift $C_L$ and shifted drag $C_D - 1.95$ over tail shedding cycles (\VDULBM{} at $N_D=40$ vs.\ OpenFOAM-v2412 at $N_D=80$). (f)~Matched 20-core per-cycle wall time on bare-metal Linux across $N_D \in \{20, 40, 80\}$ ($N_D=20$ hatched bar denotes coarse non-shedding acoustic mode).}
  \label{fig:fig06_curved_dfg_openfoam_hero}
\end{figure}

To verify curved-boundary accuracy without staircase artifacts, we first test off-grid cylindrical Couette flow at $\ReNum=20$ between a rotating inner cylinder ($R_1=0.18$, $U_0=1$) and a stationary outer cylinder ($R_2=0.44$) centered at $(0.503, 0.497)$ in $[0,1]^2$ across $N \in \{24, 48, 96, 192\}$ ($\dtlu = 8.0$, $u_{\max,\mathrm{eff}} = 0.05\times 24/N$, $\varepsilon_{\mathrm{rel}} = 10^{-11}$). As shown in Fig.~\ref{fig:fig06_curved_dfg_openfoam_hero}a,b, staircase bounce-back ($q_i\equiv 1/2$) degrades to first order ($p_u = \MetricCouetteStairOrderU$, torque error $E_T(192) = \MetricCouetteStairNOneNineTwoET$). By contrast, link-wise Bouzidi closure ($q_i\in(0,1]$) paired with rest-mass neutralization~\eqref{eq:rest_mass_neutrality} and Eulerian force evaluation~\eqref{eq:vdlbm_eulerian_force} produces smooth concentric velocity contours across cut Cartesian cells, restores second-order velocity convergence ($E_u(192) = \MetricCouetteBouzNOneNineTwoEu$, $p_u \in [1.85, 2.80]$), enforces machine-zero mass conservation ($|\delta M| \le \VDLBMCurvedMassCons$), and reduces cylinder torque error by \VDLBMCouetteTorqueGain{} to $E_T(192) = \MetricCouetteBouzNOneNineTwoET$.

Combining this curved-wall closure with the shifted ghost inlets and outlets~\eqref{eq:ghost_inlet_outlet}, we evaluate the official DFG~2D-1 ($\ReNum=20$, steady) and DFG~2D-2 ($\ReNum=100$, periodic vortex shedding) cylinder wakes~\citep{SchaferTurek1996,John2004} in $[0, 2.20]\times[0, 0.41]\text{ m}$ around a cylinder of diameter $D=0.10\text{ m}$ centered at $(0.20, 0.20)\text{ m}$ ($\nu = 10^{-3}\text{ m}^2/\text{s}$) across $N_D = D/h \in \{20, 40, 80\}$ ($440\times 82$ to $1760\times 328$; Table~\ref{tab:dfg_couette_openfoam}). Matched 20-core OpenFOAM-v2412 steady SIMPLE and unsteady PIMPLE baselines run on body-fitted O/H-grid meshes matched in cell count within $\pm 2\%$. These continuum runs employ second-order slope-limited linear-upwind convection, BDF2 time stepping at $\dtphys = 10^{-3}\text{ s}$, and up to 80 residual-controlled outer PIMPLE correctors with six inner geometric--algebraic multigrid pressure-correction solves per corrector to a $10^{-8}$ step tolerance. Both solvers execute natively on bare-metal Linux Intel Xeon Platinum~8358 cluster nodes using 20 core-pinned OpenMP threads for C \VDULBM{} and 20 core-bound OpenMPI~4.1.8 ranks on RAM tmpfs for OpenFOAM-v2412.\footnote{Extending the channel from $L=2.20\text{ m}$ to $L_{\mathrm{ext}}=2.60\text{ m}$ at $N_D=40$ shifts steady DFG~2D-1 observables ($C_D, C_L, \Delta p$) by $< 0.02\%$ (\MetricDFGSensRelCD, \MetricDFGSensRelCL, \MetricDFGSensRelDpOne) and unsteady DFG~2D-2 observables ($C_{D,\max}, C_{L,\max}, \StNum, \Delta p$) by $\le \VDLBMOutletSensBound$ (\MetricDFGSensRelCDmax, \MetricDFGSensRelCLmax, \MetricDFGSensRelSt, \MetricDFGSensRelDpTwo).}

\begin{table}[tbp]
  \centering
  \caption{Official Sch\"afer--Turek DFG~2D-1 ($\ReNum=20$) and DFG~2D-2 ($\ReNum=100$) cylinder wake benchmark comparing off-grid Cartesian \VDULBM{}, body-fitted OpenFOAM (20 physical cores), and reference intervals~\citep{SchaferTurek1996}.}
  \label{tab:dfg_couette_openfoam}
  \scriptsize
  \setlength{\tabcolsep}{2.9pt}
  \renewcommand{\arraystretch}{1.08}
  \begin{tabular}{lc cccccc}
    \toprule
    \multicolumn{8}{l}{\emph{Part A: Steady DFG~2D-1 ($\ReNum=20$, bare-metal Linux Xeon Platinum~8358, 20 cores)}} \\
    \midrule
    Solver & $N_D$ & Cells & $C_D$ & $C_L$ & $L_r$ [m] & $\Delta p$ [Pa] & 20-core $T_{\mathrm{wall}}$ [s] \\
    \midrule
    \VDULBM{} & $20$ & \MetricDFGOneVDNDTwentyCells & \MetricDFGOneVDNDTwentyCD & \MetricDFGOneVDNDTwentyCL & \MetricDFGOneVDNDTwentyLr & \MetricDFGOneVDNDTwentyDp & \MetricDFGOneVDNDTwentyWall \\
    \VDULBM{} & $40$ & \MetricDFGOneVDNDFortyCells & \MetricDFGOneVDNDFortyCD & \MetricDFGOneVDNDFortyCL & \MetricDFGOneVDNDFortyLr & \MetricDFGOneVDNDFortyDp & \MetricDFGOneVDNDFortyWall \\
    \VDULBM{} & $80$ & \MetricDFGOneVDNDEightyCells & \MetricDFGOneVDNDEightyCD & \MetricDFGOneVDNDEightyCL & \MetricDFGOneVDNDEightyLr & \MetricDFGOneVDNDEightyDp & \MetricDFGOneVDNDEightyWall \\
    OpenFOAM (SIMPLE) & $40$ & \MetricDFGOneOFNDFortyCells & \MetricDFGOneOFNDFortyCD & \MetricDFGOneOFNDFortyCL & \MetricDFGOneOFNDFortyLr & \MetricDFGOneOFNDFortyDp & \MetricDFGOneOFNDFortyWall \\
    OpenFOAM (SIMPLE) & $80$ & \MetricDFGOneOFNDEightyCells & \MetricDFGOneOFNDEightyCD & \MetricDFGOneOFNDEightyCL & \MetricDFGOneOFNDEightyLr & \MetricDFGOneOFNDEightyDp & \MetricDFGOneOFNDEightyWall \\
    \citet{SchaferTurek1996} & Ref. & --- & $[5.5700, 5.5900]$ & $[0.0104, 0.0110]$ & $[0.0842, 0.0852]$ & $[0.1172, 0.1176]$ & --- \\
    \midrule
    \multicolumn{8}{l}{\emph{Part B: Unsteady DFG~2D-2 ($\ReNum=100$, bare-metal Linux Xeon Platinum~8358, 20 cores, $\dtphys \le 10^{-3}\text{ s}$)}} \\
    \midrule
    Solver & $N_D$ & Cells & $C_{D,\max}$ & $C_{L,\max}$ & $\StNum$ & $\Delta p$ [Pa] & $T_{\mathrm{wall}}/T_{\mathrm{shed}}$ [s] \\
    \midrule
    \VDULBM{} & $20$ & \MetricDFGTwoVDNDTwentyCells & \MetricDFGTwoVDNDTwentyCDmax & \MetricDFGTwoVDNDTwentyCLmax & \MetricDFGTwoVDNDTwentySt & \MetricDFGTwoVDNDTwentyDp & \MetricDFGTwoVDNDTwentyWall$^\dagger$ \\
    \VDULBM{} & $40$ & \MetricDFGTwoVDNDFortyCells & \MetricDFGTwoVDNDFortyCDmax & \MetricDFGTwoVDNDFortyCLmax & \MetricDFGTwoVDNDFortySt & \MetricDFGTwoVDNDFortyDp & \MetricDFGTwoVDNDFortyWall \\
    \VDULBM{} & $80$ & \MetricDFGTwoVDNDEightyCells & \MetricDFGTwoVDNDEightyCDmax & \MetricDFGTwoVDNDEightyCLmax & \MetricDFGTwoVDNDEightySt & \MetricDFGTwoVDNDEightyDp & \MetricDFGTwoVDNDEightyWall \\
    OpenFOAM (PIMPLE) & $40$ & \MetricDFGTwoOFNDFortyCells & \MetricDFGTwoOFNDFortyCDmax & \MetricDFGTwoOFNDFortyCLmax & \MetricDFGTwoOFNDFortySt & \MetricDFGTwoOFNDFortyDp & \MetricDFGTwoOFNDFortyWall \\
    OpenFOAM (PIMPLE) & $80$ & \MetricDFGTwoOFNDEightyCells & \MetricDFGTwoOFNDEightyCDmax & \MetricDFGTwoOFNDEightyCLmax & \MetricDFGTwoOFNDEightySt & \MetricDFGTwoOFNDEightyDp & \MetricDFGTwoOFNDEightyWall \\
    \citet{SchaferTurek1996} & Ref. & --- & $[3.2200, 3.2400]$ & $[0.9900, 1.0100]$ & $[0.2950, 0.3050]$ & $[2.4600, 2.5000]$ & --- \\
    \bottomrule
  \end{tabular}
  \par\smallskip
  \raggedright\scriptsize $^\dagger$Coarse unstretched Cartesian $N_D=20$ exhibits a non-shedding acoustic grid oscillation ($\StNum=\MetricDFGTwoVDNDTwentySt$ vs.\ $\MetricDFGTwoOFNDTwentySt$ on body-fitted OpenFOAM $N_D=20$ at $\MetricDFGTwoOFNDTwentyWall\text{ s}$/cycle) that disappears upon resolving the boundary layer at $N_D=40, 80$. Part~B observables and periodic per-cycle runtimes are evaluated over the tail window of fixed-horizon runs ($t\in[0,30]\text{ s}$ for \VDULBM{}, $t\in[0,12]\text{ s}$ for OpenFOAM; OpenFOAM $N_D=40$ uses corrected lift-column re-evaluation on the $t\in[0,12]\text{ s}$ base run, \text{ExecutionTime} $= 4560.95\text{ s}$, \text{ClockTime} $= 4618\text{ s}$, with step timing traces archived).
\end{table}

Figure~\ref{fig:fig06_curved_dfg_openfoam_hero}c--f and Table~\ref{tab:dfg_couette_openfoam} clarify both wake accuracy and the steady-versus-unsteady runtime balance. At $N_D=80$, Cartesian \VDULBM{} matches body-fitted OpenFOAM closely in steady DFG~2D-1: $C_D$, $C_L$, and $L_r$ lie inside the \citet{SchaferTurek1996} reference intervals, and $\Delta p = \MetricDFGOneVDNDEightyDp\text{ Pa}$ falls within $0.11\%$ of $0.1176\text{ Pa}$. In unsteady DFG~2D-2, $N_D=80$ places all four observables inside the official \citet{SchaferTurek1996} intervals ($C_{D,\max}=\MetricDFGTwoVDNDEightyCDmax$, $C_{L,\max}=\MetricDFGTwoVDNDEightyCLmax$, $\StNum=\MetricDFGTwoVDNDEightySt$, $\Delta p = \MetricDFGTwoVDNDEightyDp\text{ Pa}$), while coarser $N_D=40$ approaches them within $0.4\%\text{--}2.2\%$ with 5-cycle $C_{L,\max}$ modulation of $0.019$ and $0.0043$ on the fixed integration horizons. On bare-metal Linux for steady DFG~2D-1, non-time-accurate SIMPLE relaxation and pseudo-time-marched \VDULBM{} reach near parity within $7.9\%$ ($\MetricDFGOneOFNDEightyWall\text{ s}$ versus $\MetricDFGOneVDNDEightyWall\text{ s}$ at $N_D=80$). In time-accurate unsteady DFG~2D-2 at $\dtphys = 10^{-3}\text{ s}$, however, thin near-wall O-grid cells drive OpenFOAM's local Courant number to $\mathrm{Co}_{\max} \approx 4.08\text{--}8.15$ and force $10.1\text{--}13.9$ outer PIMPLE iterations ($60.5\text{--}83.2$ global multigrid PPE solves) per physical step. Directly reflecting the elimination of global PPE solves noted in Section~\ref{sec:intro} and Table~\ref{tab:stencil_complexity}, matrix-free \VDULBM{} completes each shedding cycle $15.85\times$ faster at $N_D=40$ ($\MetricDFGTwoVDNDFortyWall\text{ s}$ versus $\MetricDFGTwoOFNDFortyWall\text{ s}$) and \VDLBMDFGOpenFOAMRatio{} faster at $N_D=80$ ($\MetricDFGTwoVDNDEightyWall\text{ s}$ versus $\MetricDFGTwoOFNDEightyWall\text{ s}$).

\FloatBarrier
\subsection{Spatially variable viscosity MMS and four-decade stratified shear}
\label{sec:sec6_3_varvisc_les}

\begin{figure}[tbp]
  \centering
  \includegraphics[width=\textwidth]{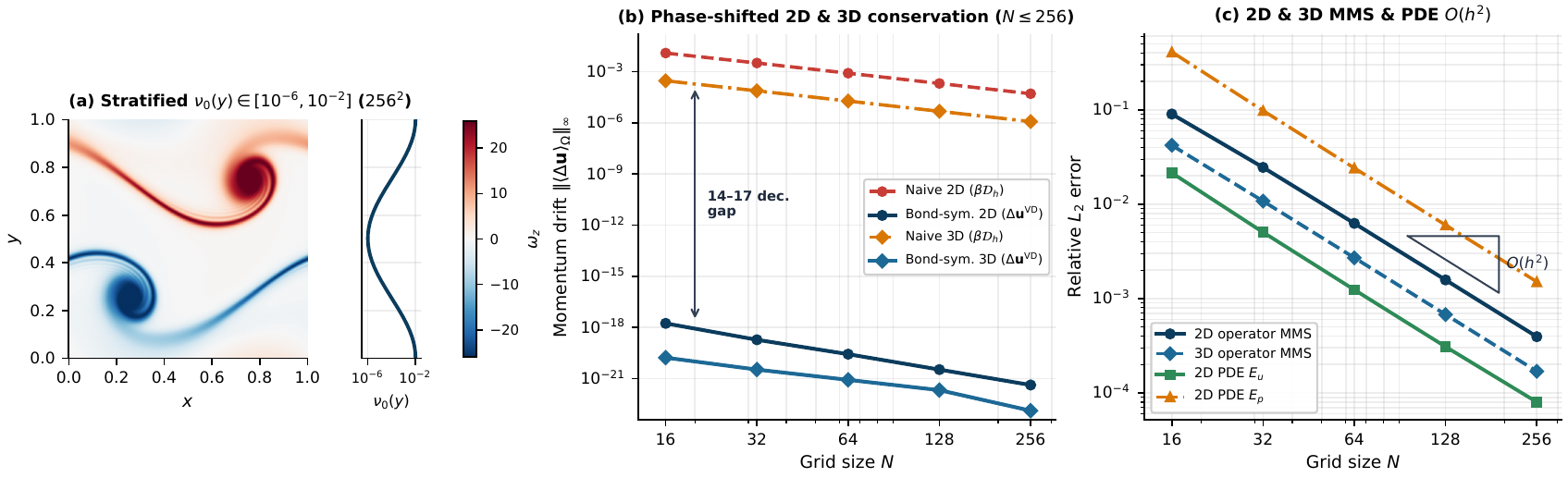}
  \caption{Bond-symmetric spatially variable viscosity across 2D and 3D ($N \in \{16, 32, 64, 128, 256\}$). (a)~Instantaneous vorticity $\omega_z$ at $t=1.0$ alongside the four-decade stratified viscosity profile $\nu_0(y) \in [10^{-6}, 10^{-2}]$ ($256^2$). (b)~Pairwise global momentum drift $\|\langle\Delta\bm{u}\rangle_\Omega\|_\infty$ across $N=16\text{--}256$ on phase-shifted 2D/3D viscosity fields, showing a 14-to-17-decade conservation gap between bond-symmetric $\DuVD$ ($\le \VDLBMMMSMomentumCons$) and naive node scaling. (c)~Second-order 2D/3D variable-viscosity operator MMS ($\nu\in[0.02,0.10], [0.04,0.10]$) and 2D solved-PDE ($E_u, E_p$) convergence through $N=256$.}
  \label{fig:fig07_varvisc_stratified_hero}
\end{figure}

Whereas Sections~\ref{sec:sec5_1_accuracy}--\ref{sec:sec6_2_curved_dfg} prescribed a spatially uniform base viscosity, Proposition~\ref{prop:varvisc_properties}(2)--(3) guarantees pairwise momentum conservation and second-order stress-divergence recovery for heterogeneous $\nueff(\bm{x})$. To verify both properties across $N \in \{16, 32, 64, 128, 256\}$, we examine smooth manufactured fields $\nu(\bm{x}) = 0.06 + 0.04\sin x\cos 2y \in [0.02, 0.10]$ in 2D and $\nu(\bm{x}) = 0.07 + 0.03\sin x\cos y\sin z \in [0.04, 0.10]$ in 3D, together with a 2D four-decade stratified shear layer $\nu_0(y) = 10^{-2}\cdot 10^{-4\sin^2(\pi y)} \in [10^{-6}, 10^{-2}]$ evolved to $T_{\mathrm{end}} = 1.0$. Global momentum defects in Fig.~\ref{fig:fig07_varvisc_stratified_hero}b are evaluated on phase-shifted profiles $\nu(x+0.37,\, y-0.23)$ in 2D and $\nu(x+0.37,\, y-0.23,\, z+0.19)$ in 3D so as to break accidental trigonometric parity cancellation of the naive operator. Across all five grid levels in Fig.~\ref{fig:fig07_varvisc_stratified_hero}b,c, bond-symmetric $\DuVD$ maintains a 14-to-17-decade global momentum conservation advantage over naive node-local scaling ($\le \VDLBMMMSMomentumCons{}$ in 2D and $1.32\times 10^{-23}$ at $256^3$ in 3D, versus $5.03\times 10^{-5}$ and $1.17\times 10^{-6}$ for naive $\beta\mathcal{D}_h$). Simultaneously, \VDULBM{} recovers second-order convergence through $N=256$ in 2D operator MMS ($E = \MetricVarViscTwoDNTwoFiveSixOpErr$, order $\MetricVarViscTwoDNTwoFiveSixOpOrder$), 3D operator MMS ($E = 1.70\times 10^{-4}$, order $2.00$), and 2D steady solved-PDE MMS ($E_u = \MetricVarViscTwoDNTwoFiveSixPDEEu$, order $\MetricVarViscTwoDNTwoFiveSixPDEOrder$). On the four-decade stratified shear layer (Fig.~\ref{fig:fig07_varvisc_stratified_hero}a), \VDULBM{} integrates stably in $400$ convective steps with $\langle K_{\mathrm{sweep}}\rangle = \MetricStratVDNTwoFiveSixSweeps$ on $256^2$ and trajectory momentum drift $\MetricStratVDNOneTwoEightMomDrift$ on $128^2$ and $\MetricStratVDNTwoFiveSixMomDrift$ on $256^2$, free from the near-singular $\tau_{\min} = 0.5000192$ stiffness encountered by explicit BGK.

\FloatBarrier
\subsection{Unified 3D D3Q19 GPU turbulence, \texorpdfstring{$\mathrm{Re}=3000$}{Re=3000} LES, and diffusion-stiff crossover}
\label{sec:sec6_4_3d_gpu}

\begin{figure}[tbp]
  \centering
  \includegraphics[width=\textwidth]{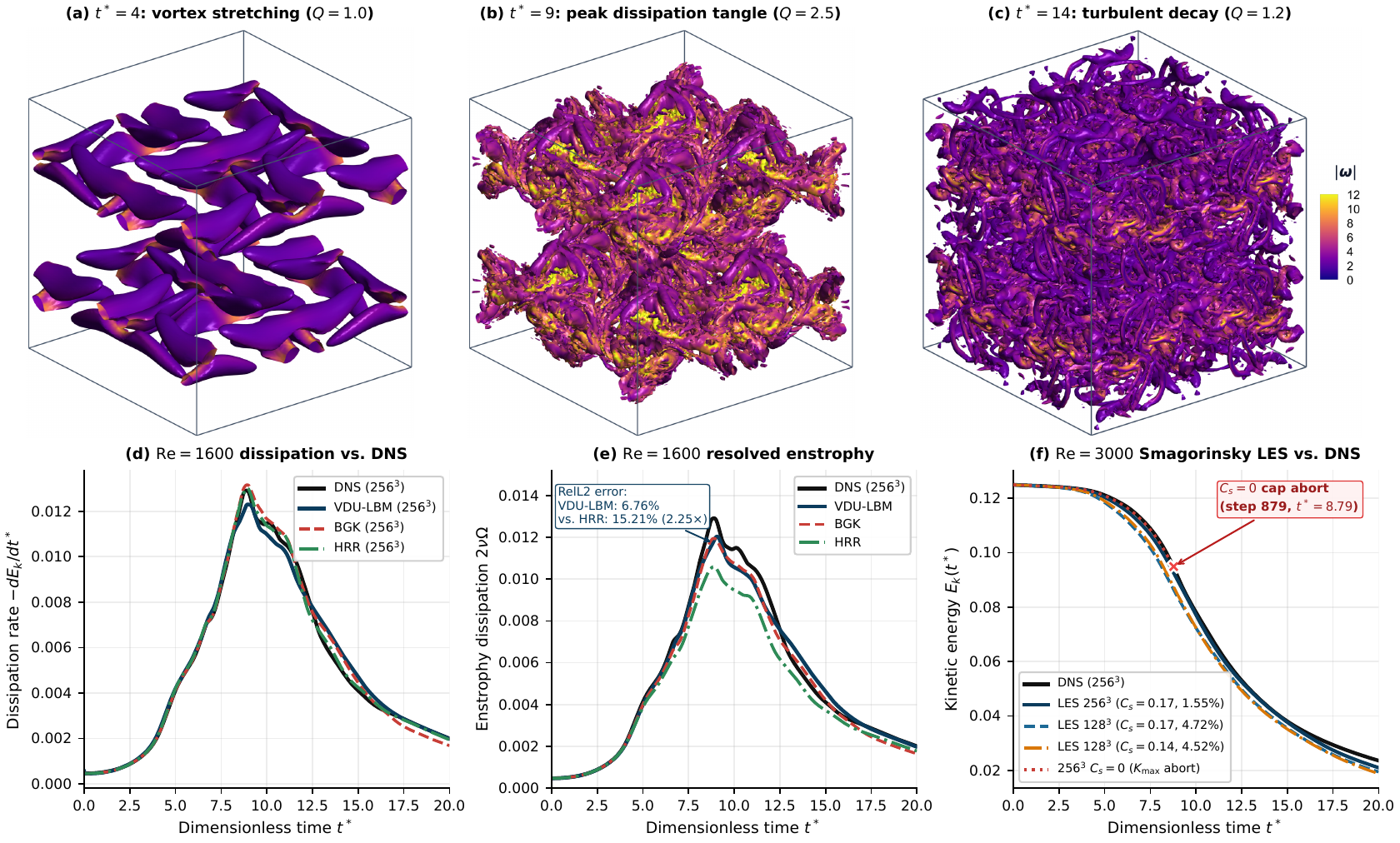}
  \caption{Three-dimensional D3Q19 Taylor--Green turbulence against $256^3$ FP64 spectral DNS. (a--c)~Phong-shaded $Q$-criterion vortex isosurfaces colored by $|\bm{\omega}|$ ($128^3$ rendering subsample of the $256^3$ \VDULBM{} run at $\ReNum=1600$): (a)~$t^*=4$ vortex-sheet stretching ($Q=1.0$), (b)~$t^*=9$ peak turbulent tangle ($Q=2.5$), and (c)~$t^*=14$ fine-scale decay ($Q=1.2$). (d)~$\ReNum=1600$ kinetic-energy dissipation rate $-dE_k/dt^*$. (e)~$\ReNum=1600$ resolved enstrophy dissipation $2\nu\Omega$, where \VDULBM{} ($6.76\%$) matches BGK ($6.49\%$) and achieves $2.25\times$ lower error than HRR ($15.21\%$). (f)~$\ReNum=3000$ Smagorinsky LES on $(128^3, C_s\in\{0.14,0.17\})$ and $(256^3, C_s=0.17)$ versus unclosed $C_s=0$ inner-cap abort at $t^*=8.79$.}
  \label{fig:fig08_tgv3d_turbulence_hero}
\end{figure}

Finally, we extend the evaluation to three dimensions on D3Q19 lattices over $\Omega = [0, 2\pi)^3$ from the Taylor--Green initial state $\bm{u}_0(\bm{x}) = U_0(\sin x\cos y\cos z,\, -\cos x\sin y\cos z,\, 0)$ and $p_0(\bm{x}) = \tfrac{\rho_0 U_0^2}{16}(\cos 2x + \cos 2y)(\cos 2z + 2)$ ($U_0 = L_0 = 1$). Our fused FP64/FP32 CUDA implementation stores $62$ base scalars per cell ($7.75\text{ GiB}$ at $256^3$ FP64) with verified CPU--GPU pointwise equivalence $\MetricGPUCPUEquivMaxDiff < 10^{-13}$. All GPU timings execute on an NVIDIA A100-SXM4-80GB ($B_{\mathrm{peak}} = 2039.04\text{ GB/s}$, measured device-to-device copy ceiling $B_{\mathrm{D2D}} = \MetricRoofAOneHundredMeasuredDTwoDBW\text{ GB/s}$) under per-section zero-foreign-process GPU monitoring on the main suite and pre-run snapshot checks on the supplement runs. Accuracy across $\ReNum \in \{10, 100, 1600, 3000\}$ over $t^* \in [0, 20]$ is measured under AIAA normalization~\citep{DeBonis2013,Brachet1983,vanRees2011} against a $2/3$-dealiased $256^3$ FP64 cuFFT spectral DNS reference (computed on a Quadro~P4000 and 20-step cross-verified on the A100, resolving peak dissipation $\MetricTGVThreeDDNSPeakDissipReSixteenHun$ at $t^*=\MetricTGVThreeDDNSPeakTimeReSixteenHun$ with an $8.95\%$ $128^3\to 256^3$ self-difference in $2\nu\Omega$ at $\ReNum=1600$). At $\ReNum=1600$ on $256^3$ FP64 (Fig.~\ref{fig:fig08_tgv3d_turbulence_hero}a--e and Table~\ref{tab:tgv3d_gpu_summary}, Part~A), $Q$-criterion isosurfaces capture vortex-sheet roll-up, peak tubular tangle, and fine-scale turbulent decay. On resolved enstrophy dissipation $2\nu\Omega$, parameter-free \VDULBM{} ($\MetricTGVThreeDReSixteenHunVDNTwoFiveSixFPSixtyFourOmErr$) matches explicit BGK ($\MetricTGVThreeDReSixteenHunBGKNTwoFiveSixFPSixtyFourOmErr$) and achieves $2.25\times$ lower error than HRR ($\MetricTGVThreeDReSixteenHunHRRNTwoFiveSixFPSixtyFourOmErr$), whereas on total kinetic-energy decay $-dE_k/dt^*$ HRR yields $2.56\%$ compared with $4.50\%$ for BGK and $5.71\%$ for \VDULBM{}.

\begin{figure}[tbp]
  \centering
  \includegraphics[width=\textwidth]{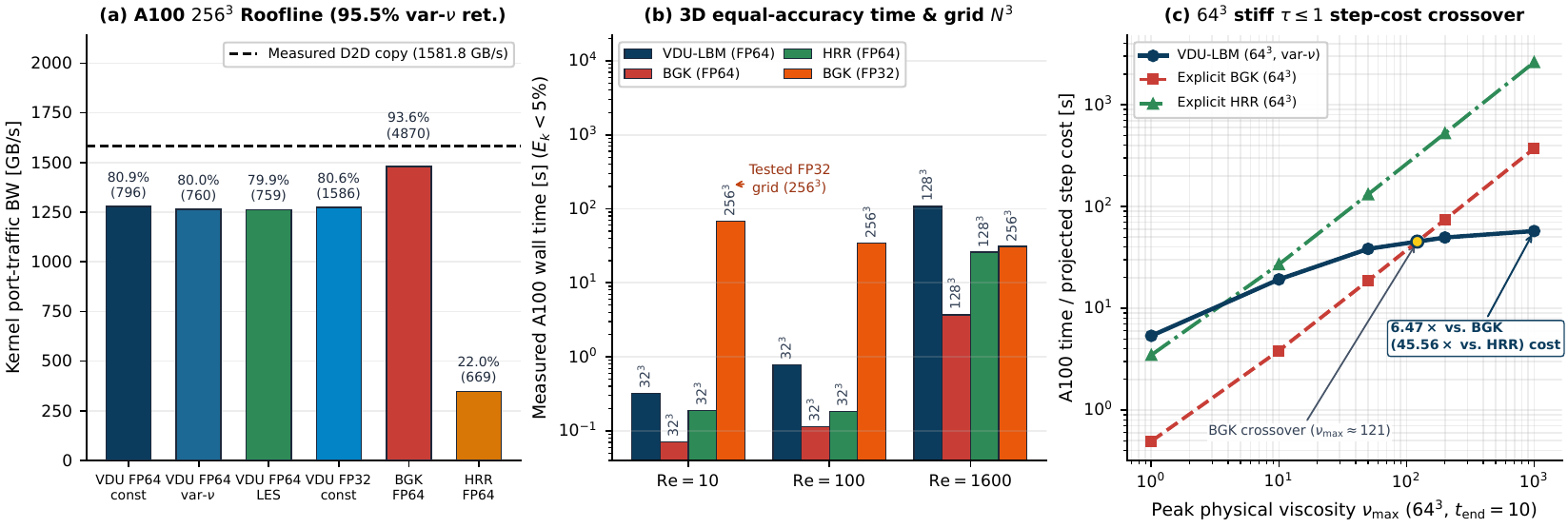}
  \caption{A100 GPU Roofline and 3D multi-regime performance. (a)~$256^3$ kernel port-traffic bandwidth ($201$ scalars/cell for constant-$\nu$ \VDULBM{} [$197$ deduplicated, $38$ nominal state], $208$ for variable-$\nu$/LES, $38$ for explicit BGK, and $65$ for HRR) versus measured A100 D2D copy ceiling ($1581.8\text{ GB/s}$), reaching $80.9\%$ in constant-$\nu$ FP64 ($95.5\%$ variable-$\nu$ and $95.4\%$ LES kernel-rate retention). (b)~Equal-accuracy ($\mathrm{RelL2}(E_k)<5\%$) A100 runtime and grid $N^3$ across $\ReNum \in \{10, 100, 1600\}$ (FP32 BGK tested at $256^3$). (c)~Uniform-$64^3$ ($t_{\mathrm{end}}=10$, $20\times$ stratification) bounded-$\tau$ ($\tau\le 1$) diffusive step-cost crossover across $\numax \in \{1, 10, 50, 200, 1000\}$: explicit BGK is faster for $\numax \le 50$, crosses interpolated parity near $\numax \approx 121$ ($\approx 3.6$ vs.\ HRR; both crossovers bracketed by full-trajectory runs, with BGK at $\numax=1000$ and HRR at $\numax\ge 50$ scaled from live $64^3$ kernel rates), and incurs a $6.47\times$ ($45.56\times$ for HRR) higher kernel-rate-projected diffusive step cost at $\numax=1000$.}
  \label{fig:fig09_gpu_roofline_stiff3d}
\end{figure}

Because the local Uzawa smoother replaces global Krylov inner-product reductions with regular 1D stencil passes (Table~\ref{tab:stencil_complexity}), GPU execution remains strongly bandwidth-efficient. Under its $201$-scalar multi-pass kernel port-traffic model, constant-viscosity \VDULBM{} attains $80.9\%$ of measured A100 D2D copy bandwidth at $256^3$ and $85.5\%$ at $64^3$ ($79.3\%$ under the deduplicated $197$-scalar count and $15.3\%$ in nominal $38$-scalar state traffic, versus $38$-scalar single-pass BGK at $93.6\%$ and $65$-scalar HRR at $22.0\%$). Enabling $208$-scalar 18-bond variable viscosity and Smagorinsky LES retains $95.4\%\text{--}95.5\%$ of that kernel throughput (Fig.~\ref{fig:fig09_gpu_roofline_stiff3d}a). On moderate-viscosity periodic cubes ($\ReNum \in [10, 1600]$ or stratified $\nu_0(z) \in [10^{-4}, 1]$ in Table~\ref{tab:app_tgv3d_full}), single-sweep explicit BGK remains faster in raw wall time even though \VDULBM{} takes $58\text{--}101\times$ fewer outer steps (Fig.~\ref{fig:fig09_gpu_roofline_stiff3d}b).

At the more demanding Reynolds number $\ReNum=3000$ (Fig.~\ref{fig:fig08_tgv3d_turbulence_hero}f), fixed-step unclosed $256^3$ \VDULBM{} ($C_s=0$) reaches the $K_{\max}=200$ inner iteration cap near peak vortex stretching at step $879/2000$ ($t^* = 8.79$). Activating the discrete Smagorinsky closure~\eqref{eq:smag_nulu_beta} across $(128^3, C_s=0.14)$, $(128^3, C_s=0.17)$, and $(256^3, C_s=0.17)$ integrates stably through $t^*=20$ without step halvings, yielding $\mathrm{RelL2}(E_k) = \MetricTGVThreeDLESNTwoFiveSixCsSeventeenEkErr$ ($\mathrm{RelL2}(2\nu\Omega) = \MetricTGVThreeDLESNTwoFiveSixCsSeventeenOmErr$ against unfiltered DNS) on $256^3$ and $\MetricTGVThreeDLESNOneTwoEightCsSeventeenEkErr$ ($C_s=0.17$) or $\MetricTGVThreeDLESNOneTwoEightCsFourteenEkErr$ ($C_s=0.14$) on $128^3$.

To test the high-viscosity $\dtphys \propto h^2/\numax$ bottleneck highlighted in Section~\ref{sec:intro}, we prescribe the heterogeneous viscosity $\nu_0(z) = \numax\cdot 20^{-\sin^2(z/2)}$ across $\numax \in [1, 1000]$ with oscillatory body forcing $\bm{b}_{\mathrm{phys}}(\bm{x},t) = 3\nu_0(z)\bm{u}_0(\bm{x})\cos(2\pi t/t_{\mathrm{end}})$ over $t \in [0, 10]$ in $20$ outer steps ($\dtphys = 0.5$, nominal convective Courant ratio $U_0\dtphys/h = 2.55$ on $32^3$ and $5.09$ on $64^3$). This sweep spans creeping regimes for $\numax \ge 50$ ($\ReNum_{\mathrm{loc}} \le 0.4$, $\ulu \le 3.25\times 10^{-4} \ll \nulu$, $\Reh \le 0.04$, matching exact oscillatory Stokes $E_k(10) = 0.14422$ to $0.05\%$ at $\numax=1000$) as well as transitional regimes $\ReNum_{\mathrm{loc}} \in [0.1, 20]$ at $\numax \in \{1, 10\}$ on $64^3$, and is compared against unforced uniform-$\numax$ bounded-relaxation ($\tau \le 1$, $\nulu \le 1/6$) explicit step budgets $N_{\mathrm{steps,exp}} = \lceil t_{\mathrm{end}}/\dtlat(\numax)\rceil$ of BGK and HRR (Fig.~\ref{fig:fig09_gpu_roofline_stiff3d}c and Table~\ref{tab:tgv3d_gpu_summary}, Part~B). As $\numax$ increases on $64^3$, explicit diffusive step budgets grow linearly up to $6.23\times 10^6$ whereas 20-step forced \VDULBM{} inner sweeps saturate. The step cost crosses log-log interpolated parity with HRR near $\numax \approx 3.6$ and with BGK near $\numax \approx 121$, yielding $1.72\times$ ($10.62\times$ vs.\ HRR) direct-timed ($32^3, \numax=400$) and $6.47\times$ ($45.56\times$ vs.\ HRR) kernel-rate-projected ($64^3, \numax = 1000$) step-cost reductions relative to uniform-$\numax$ ($\tau\le 1$) explicit diffusive budgets while conserving momentum to $1.66\times 10^{-16}$.

\begin{table}[tbp]
  \centering
  \caption{Headline 3D D3Q19 accuracy, NVIDIA A100-SXM4-80GB Roofline ($B_{\mathrm{D2D}}=\MetricRoofAOneHundredMeasuredDTwoDBW\text{ GB/s}$), and diffusion-stiff bounded-$\tau$ GPU step-cost comparison (additional multi-grid rows in Table~\ref{tab:app_tgv3d_full}).}
  \label{tab:tgv3d_gpu_summary}
  \scriptsize
  \setlength{\tabcolsep}{2.0pt}
  \renewcommand{\arraystretch}{1.08}
  \begin{tabular}{ll cccccc}
    \toprule
    \multicolumn{8}{l}{\emph{Part A: 3D Taylor--Green turbulence ($t^*\in[0,20]$ vs.\ $256^3$ DNS) and $256^3$ A100 kernel port-traffic Roofline}} \\
    \midrule
    Regime & Solver & Grid & $\mathrm{RelL2}(E_k)$ & $\mathrm{RelL2}(2\nu\Omega)$ & $\langle K_{\mathrm{sweep}}\rangle$ & MLUPS (\% $B_{\mathrm{D2D}}$) & $T_{\mathrm{wall}}$ [s] \\
    \midrule
    $\ReNum=1600$ & \textbf{\VDULBM{}} ($C_s=0$) & $256^3$ & \MetricTGVThreeDReSixteenHunVDNTwoFiveSixFPSixtyFourEkErr & \textbf{\MetricTGVThreeDReSixteenHunVDNTwoFiveSixFPSixtyFourOmErr} & \MetricTGVThreeDReSixteenHunVDNTwoFiveSixFPSixtyFourSweeps & \MetricRoofVDNTwoFiveSixFPSixtyFourSweepMLUPS\ (\MetricRoofVDNTwoFiveSixFPSixtyFourPctDTwoD) & \MetricTGVThreeDReSixteenHunVDNTwoFiveSixFPSixtyFourWall \\
    $\ReNum=1600$ & BGK & $256^3$ & \MetricTGVThreeDReSixteenHunBGKNTwoFiveSixFPSixtyFourEkErr & \MetricTGVThreeDReSixteenHunBGKNTwoFiveSixFPSixtyFourOmErr & $1.0$ & \MetricRoofBGKNTwoFiveSixFPSixtyFourSweepMLUPS\ (\MetricRoofBGKNTwoFiveSixFPSixtyFourPctDTwoD) & \MetricTGVThreeDReSixteenHunBGKNTwoFiveSixFPSixtyFourWall \\
    $\ReNum=1600$ & HRR & $256^3$ & \MetricTGVThreeDReSixteenHunHRRNTwoFiveSixFPSixtyFourEkErr & \MetricTGVThreeDReSixteenHunHRRNTwoFiveSixFPSixtyFourOmErr & $1.0$ & \MetricRoofHRRNTwoFiveSixFPSixtyFourSweepMLUPS\ (\MetricRoofHRRNTwoFiveSixFPSixtyFourPctDTwoD) & \MetricTGVThreeDReSixteenHunHRRNTwoFiveSixFPSixtyFourWall \\
    $\ReNum=3000$ & \textbf{\VDULBM{}} ($C_s=0.17$) & $128^3$ & \MetricTGVThreeDLESNOneTwoEightCsSeventeenEkErr & \MetricTGVThreeDLESNOneTwoEightCsSeventeenOmErr & \MetricTGVThreeDLESNOneTwoEightCsSeventeenSweeps & $793.1^\mathrm{b}$ ($95.4\%^\mathrm{a}$) & \MetricTGVThreeDLESNOneTwoEightCsSeventeenWall \\
    $\ReNum=3000$ & \textbf{\VDULBM{}} ($C_s=0.17$) & $256^3$ & \textbf{\MetricTGVThreeDLESNTwoFiveSixCsSeventeenEkErr} & \MetricTGVThreeDLESNTwoFiveSixCsSeventeenOmErr & \MetricTGVThreeDLESNTwoFiveSixCsSeventeenSweeps & \MetricRoofVDLESNTwoFiveSixFPSixtyFourSweepMLUPS\ (\MetricRoofVDLESNTwoFiveSixFPSixtyFourRetention$^\mathrm{a}$) & \MetricTGVThreeDLESNTwoFiveSixCsSeventeenWall \\
    Var.-$\nu$ & \textbf{\VDULBM{}} (18-bond) & $256^3$ & --- & --- & --- & \MetricRoofVDVarViscNTwoFiveSixFPSixtyFourSweepMLUPS\ (\MetricRoofVDVarViscNTwoFiveSixFPSixtyFourRetention$^\mathrm{a}$) & --- \\
    \midrule
    \multicolumn{8}{l}{\emph{Part B: 3D stiff forced variable-$\nu(z)$ \VDULBM{} vs.\ bounded-$\tau$ ($\tau\le 1$) explicit step cost ($t_{\mathrm{end}}=10$, 20 steps)}} \\
    \midrule
    $\nu_0(z)$ span & Grid & Sweeps & Mom.\ drift & \VDULBM{} [s] & BGK [s] & HRR [s] & Ratio (BGK/HRR) \\
    \midrule
    $[10, 200]$  & $32^3$ & $37{,}365$ & $3.51\times 10^{-16}$ & $3.62$ & $3.24$ & $19.96$ & $0.90\times$/$\mathbf{5.51\times}$ \\
    $[20, 400]$  & $32^3$ & $40{,}228$ & $1.52\times 10^{-16}$ & $\mathbf{3.75}$ & $6.45$ & $39.87$ & $\mathbf{1.72\times}$/$\mathbf{10.62\times}$ \\
    $[50, 1000]$ & $64^3$ & $167{,}288$ & $1.66\times 10^{-16}$ & $\mathbf{57.41}$ & $371.32^\dagger$ & $2615.17^\dagger$ & $\mathbf{6.47\times}$/$\mathbf{45.56\times}$ \\
    \bottomrule
  \end{tabular}
  \par\smallskip
  \raggedright\scriptsize Note: All \VDULBM{} rows and $[10,200]/[20,400]$ unforced uniform-$\numax$ explicit step runs ($3.12\times 10^5$ and $6.24\times 10^5$ steps at $\tau\le 1$) report direct A100 wall times. Part~A $\%$ $B_{\mathrm{D2D}}$ uses per-kernel port scalars ($201$ constant-$\nu$ \VDULBM{}, $38$ BGK, $65$ HRR; $208$-scalar variable-$\nu$/LES reaches $80.0\%/79.9\%$ $B_{\mathrm{D2D}}$ in Fig.~\ref{fig:fig09_gpu_roofline_stiff3d}a). At $\ReNum=1600$ ($256^3$), $\mathrm{RelL2}(-dE_k/dt^*) = 0.05713$ (\VDULBM{}), $0.04498$ (BGK), and $0.02564$ (HRR). Fixed-step unclosed $C_s=0$ $256^3$ \VDULBM{} aborts at step $879/2000$ ($t^* = 8.79$, inner cap $K_{\max}=200$; Table~\ref{tab:app_tgv3d_full}). $^\mathrm{a}$Roofline kernel retention vs.\ constant-$\nu$ \VDULBM{} ($15.3\%$ nominal 38-scalar state BW at $256^3$). $^\mathrm{b}$Measured $128^3$ trajectory sweep rate ($94.7\%$ of $128^3$ constant-$\nu$ Roofline; $256^3$-scaled Roofline reference: $798.4$, $95.4\%^\mathrm{a}$). $^\dagger$Projected from live A100 kernel rates measured over $50{,}000$ (BGK) and $15{,}000$ (HRR) steps across $6.23\times 10^6$ $\tau\le 1$ diffusive steps.
\end{table}

\FloatBarrier
\section{Conclusions and methodological scope}
\label{sec:sec7_conclusions}

We have developed and analyzed a Viscosity-Decoupled Uzawa Lattice Boltzmann Method (\VDULBM{}) on standard D2Q9 and D3Q19 lattices for incompressible viscous flows across Reynolds numbers. By permanently fixing symmetric and antisymmetric collision relaxation at unity ($\tau^+ = \tau^- \equiv 1$, $\Lambda \equiv 1/4$) under a smooth global harmonic lattice scaling, local collision reduces to a memoryless rank-$(d+1)$ map that clears arbitrary pre-collision non-conserved $\kerP$ perturbations in a single inner sweep ($\VDLBMKerPZero$). This property eliminates both high-$\ReNum$ kinetic ghost-mode blow-up in explicit LBMs and the deceptive transient U-dip passband breakdown of variable-$\tau$ dual-time schemes. Sylvester's determinant identity then reduces the $Q\times Q$ preconditioned inner amplification spectrum to $(d+1)$ conserved moments, so variable-step BDF1/BDF2 marching is accelerated without global pressure-Poisson assembly via a matrix-free local polynomial Uzawa smoother ($\kLoc=2$) and separable binomial Nyquist filter.

Beyond inner preconditioning, physical diffusion spanning constant viscosity, four-decade spatially varying viscosity, and parameter-free discrete Smagorinsky subgrid eddy viscosity is restored after streaming through the bond-symmetric momentum-flux lift~\eqref{eq:DuVD_bulk}. Pairwise bond symmetry conserves global discrete momentum to $\le \VDLBMMMSMomentumCons$ across $16^d\text{--}256^d$ 2D and 3D manufactured-solution hierarchies ($1.32\times 10^{-23}$ at $256^3$) and $\le \MetricTGVThreeDStiffVDNOneTwoEightMomDrift$ across 3D four-decade stratified flows, and recovers smooth low-Mach second-order space--time accuracy across $\ReNum \in [10^{-4}, 10^6]$ as well as pulsatile Womersley channels spanning $\alpha \in [0.5, 50]$. In $256^3$ Taylor--Green turbulence at $\ReNum=1600$, parameter-free \VDULBM{} matches explicit BGK on resolved enstrophy dissipation $2\nu\Omega$ ($6.76\%$ versus $6.49\%$) and achieves $2.25\times$ lower $2\nu\Omega$ error than HRR ($15.21\%$, alongside total $-dE_k/dt^*$ errors of $5.71\%$ versus $2.56\%$). For under-resolved flows, discrete Smagorinsky closure ($C_s=0.17$) suppresses coarse-grid Brown--Minion dispersive secondary vortices at $\ReNum=10^5$ and integrates $256^3$ $\ReNum=3000$ turbulence stably through $t^*=20$ ($\mathrm{RelL2}(E_k) = \MetricTGVThreeDLESNTwoFiveSixCsSeventeenEkErr$).

On wall-bounded and open domains, zero pre-collision non-equilibrium stress memory permits a single link-wise Bouzidi solid-wall operator paired with rest-population mass neutralization and half-cell shifted ghost inlets and outlets. This closure reduces off-grid cylindrical Couette torque error by \VDLBMCouetteTorqueGain{} relative to staircase bounce-back and, at $N_D=80$, places all four unsteady DFG~2D-2 wake observables inside the official Sch\"afer--Turek intervals (within $0.11\%$ of the steady DFG~2D-1 $\Delta p$ bound). Because global pressure-Poisson inner solves are avoided, time-accurate DFG~2D-2 shedding on bare-metal Linux runs $15.85\times$ faster per cycle at $N_D=40$ and \VDLBMDFGOpenFOAMRatio{} faster at $N_D=80$ than matched 20-core OpenFOAM-v2412. On an NVIDIA A100 GPU, the fused D3Q19 CUDA solver attains $80.9\%$ at $256^3$ to $85.5\%$ at $64^3$ of measured device-to-device copy bandwidth under its 201-scalar multi-pass port-traffic model ($15.3\%$ in nominal 38-scalar state traffic, with $95.4\%\text{--}95.5\%$ kernel retention under variable viscosity and LES). In diffusion-stiff heterogeneous flows, it reduces step costs by $1.72\times$ ($10.62\times$ vs.\ HRR) in direct-timed runs ($\numax=400, 32^3$) and by $6.47\times$ ($45.56\times$ vs.\ HRR) in kernel-rate-projected runs ($\numax = 1000, 64^3$) relative to uniform-$\numax$ ($\tau \le 1$) explicit BGK and HRR step budgets.

Several structural boundaries define the present scope and point to natural extensions. First, the quadratic D2Q9 and D3Q19 equilibria target isothermal low-Mach flows ($\MaNum_0 \approx 0.0866$) and incur the $-\tfrac{1}{2}U_{\mathrm{lu}}^2$ translational shear shift on coarse grids when $\nulu < \tfrac{1}{2}U_{\mathrm{lu}}^2$ (Remark~\ref{rem:finite_grid_galilean}), which can be removed via cubic Mach corrections~\citep{Dellar2003}. Second, omitting skew-Hermitian advection from $\Lsurr$ restricts baseline $\kLoc=2$ local smoothing to $\CFLu \le 2$ in convection-active regimes backed by adaptive BDF2 step-halving. Third, in the strong-elliptic limit $a_q \to 0^+$ on fine grids, fixed-degree local smoothing obeys the $\Omega(h^{-2})$ shear-branch lower bound~\eqref{eq:elliptic_gap_scaling} and invites matrix-free geometric multigrid~\citep{Gsell2020,ElmanSilvesterWathen2014} within the same residual-correction slot. Finally, on moderate-viscosity periodic domains where explicit step counts remain small, single-sweep explicit BGK remains faster in raw wall time, whereas pure FP32 saturation near $\varepsilon_{\mathrm{rel}}=10^{-4}$ suggests mixed-precision FP32/FP64 defect correction.

\section*{CRediT authorship contribution statement}
\textbf{Yuan Yu:} Conceptualization, Methodology, Formal analysis, Software, Validation, Supervision, Funding acquisition, Writing -- original draft, Writing -- review \& editing.
\textbf{Yaolong Yu:} Methodology, Software, Validation, Investigation, Data curation, Visualization, Writing -- original draft, Writing -- review \& editing.
\textbf{Yanhui Ding:} Conceptualization, Methodology, Formal analysis, Validation, Investigation, Supervision, Writing -- original draft, Writing -- review \& editing.

\section*{Declaration of competing interest}
The authors declare that they have no known competing financial interests or personal relationships that could have appeared to influence the work reported in this paper.

\section*{Data availability}
All 2D C/OpenMP and 3D CUDA D3Q19 solver source codes, spectral analysis and flow-field visualization scripts, OpenFOAM-v2412 benchmark dictionaries and force/timing traces, the deterministic CSV-to-\LaTeX{} macro binder, and complete CSV benchmark ledgers (including the dense spectral supplement and bare-metal CPU and A100 GPU cluster suites) are provided in the accompanying self-contained reproducibility archive.

\section*{Acknowledgements}
This work was supported by the National Natural Science Foundation of China (No.~12101527), the Natural Science Foundation of Hunan Province (No.~2026JJ60005), the Science and Technology Innovation Program of Hunan Province (No.~2026RC3172), and the 111 Project (No.~D23017). Computational resources were provided by the High Performance Computing Platform of Xiangtan University.

\section*{Declaration of generative AI and AI-assisted technologies in the writing process}
During the preparation of this work, the authors used ChatGPT in order to assist with language editing and consistency checking. After using this tool, the authors reviewed and edited the content as needed and take full responsibility for the content of the published article.

\FloatBarrier
\appendix

\section{Theoretical proofs, spectral witnesses, and archival benchmark ledgers}
\label{app:archival}
\setcounter{table}{0}
\setcounter{figure}{0}

\subsection{Detailed Taylor-expansion proof of Theorem~\ref{thm:hydro_consistency}}
\label{app:hydro_proof}

Fix physical scales $(U_0, L, \rho_0, \numax > 0)$, and assume that $(\bm{u}_{\mathrm{phys}}, p, \nueff, \bm{b}_{\mathrm{phys}})$ are smooth on $[0, T]$ with uniformly bounded derivatives through sixth order in space and third order in time ($C^6_x C^3_t$, including $\|\nueff\|_{C^3} < \infty$), and that pressure obeys $p - p_0 = O(\rho_0 U_0^2 + \rho_0\numax U_0/L)$ with $\partial_t^m p = O((\rho_0 U_0^2 + \rho_0\numax U_0/L)/T_0^m)$ ($T_0 := L/U_0$, $m \le 3$). Throughout, asymptotic $O(\cdot)$ bounds hold as $(h, \dtphys) \to (0, 0)$ with implicit constants depending on the fixed physical parameters and $C^6_x C^3_t$ field norms. From harmonic scaling~\eqref{eq:harmonic_scaling}, $\dtlat \equiv h/\clat \equiv \tfrac{\MaNum\,h}{\sqrt{3}U_0} \le \min\bigl(\tfrac{h^2}{6\numax}, \tfrac{\umax h}{U_0}\bigr)$, $\numax/\clat \le h/6$, and $\nueff(\bm{x},t)\dtlat \equiv \nulu(\bm{x},t) h^2 \le \tfrac{1}{6}h^2$. Consequently, $\hat{p} = (p - p_0)/(\rho_0 c_s^2 \clat^2) = O(\MaNum^2 + \MaNum\,h)$, $s_p = -\dtlat\mathcal{D}_t^{\mathrm{BDF2}}\hat{p}$ satisfies $-s_p/\dtlat = \partial_t\hat{p} + O(\dtphys^2\|\partial_t^3\hat{p}\|_\infty) = O(\MaNum^2, \MaNum\,h, \dtphys^2) = O(\MaNum^2, \dtphys^2, h^2)$ via $\MaNum\,h \le \tfrac{1}{2}(\MaNum^2 + h^2)$, and $\bm{s}_u = \tfrac{\dtlat}{\clat}(\bm{b}_{\mathrm{phys}} - \mathcal{D}_t^{\mathrm{BDF2}}\bm{u}_{\mathrm{phys}})$.

At a fixed point $f_k(\bm{x}) = f_k^\star(\bm{x} - \bm{c}_k h) + (E_u\DuVD)_k(\bm{x})$ of Eq.~\eqref{eq:inner_map_T} (suppressing $n+1$), Eq.~\eqref{eq:inner_T_step1} implies $P\bm{f} = \bm{m} - \tfrac{1}{2}\bm{s}$, whereas Eq.~\eqref{eq:inner_T_step2} gives $P\bm{f}^\star = \bm{m} + \tfrac{1}{2}\bm{s}$ and the higher velocity moments
\begin{align}
\label{eq:Pi_star}
  \Pi_{\alpha\beta}^\star &:= \sum_{k=0}^{Q-1} c_{k\alpha} c_{k\beta} f_k^\star
  = c_s^2\Bigl(\hat{p} + \frac{1}{2}s_p\Bigr)\delta_{\alpha\beta}
  + \tilde{u}_\alpha \tilde{u}_\beta - \frac{1}{4}s_{u\alpha} s_{u\beta}, \\
  Q_{\alpha\beta\gamma}^\star &:= \sum_{k=0}^{Q-1} c_{k\alpha} c_{k\beta} c_{k\gamma} f_k^\star
  = c_s^2\bigl(\tilde{u}_\alpha\delta_{\beta\gamma} + \tilde{u}_\beta\delta_{\alpha\gamma} + \tilde{u}_\gamma\delta_{\alpha\beta}\bigr), \nonumber\\
  R_{\alpha\beta\gamma\delta}^\star &:= \sum_{k=0}^{Q-1} c_{k\alpha} c_{k\beta} c_{k\gamma} c_{k\delta} f_k^\star, \qquad
  M_{\alpha\beta\gamma\delta\mu}^\star := \sum_{k=0}^{Q-1} c_{k\alpha} c_{k\beta} c_{k\gamma} c_{k\delta} c_{k\mu} f_k^\star, \nonumber
\end{align}
where $Q_{\alpha\beta\gamma}^\star$ follows from fourth-order isotropy~\eqref{eq:lattice_isotropy} and Eq.~\eqref{eq:feq_Phi_defs}, $\clat^2 R_{\alpha\beta\gamma\delta}^\star = O(|p-p_0|/\rho_0 + |\bm{u}_{\mathrm{phys}}|^2 + \dtlat) = O(1)$ possesses bounded spatial derivatives for the fixed physical problem, and $M_{\alpha\beta\gamma\delta\mu}^\star = O(|\tilde{\bm{u}}|) = O(\clat^{-1})$ by odd parity of $\bm{c}_k$. Taylor-expanding $f_k^\star(\bm{x} - \bm{c}_k h)$ through fourth order in $h$ gives
\begin{equation}
\label{eq:f_fixedpoint_taylor}
\begin{aligned}
  f_k(\bm{x}) &= f_k^\star(\bm{x}) - h(\bm{c}_k\cdot\nabla)f_k^\star(\bm{x})
  + \frac{h^2}{2}(\bm{c}_k\cdot\nabla)^2 f_k^\star(\bm{x})
  - \frac{h^3}{6}(\bm{c}_k\cdot\nabla)^3 f_k^\star(\bm{x}) \\
  &\quad + \frac{h^4}{24}(\bm{c}_k\cdot\nabla)^4 f_k^\star(\bm{x})
  + \bigl(E_u\DuVD(\bm{x})\bigr)_k + O\bigl(h^5\|D^5\bm{f}^\star\|\bigr).
\end{aligned}
\end{equation}

\emph{Step 1 (Continuity and $C^1$ divergence bound on $\tilde{\bm{u}}_{\mathrm{phys}}$).} Taking $P_0(\cdot)$ of Eq.~\eqref{eq:f_fixedpoint_taylor} with $P_0(\bm{f}-\bm{f}^\star) = -s_p$ and $P_0 E_u(\DuVD) \equiv 0$ yields
\begin{equation}
\label{eq:p0_balance}
  s_p - h\,\nabla\cdot\tilde{\bm{u}}
  + \frac{h^2}{2}\partial_\alpha\partial_\beta \Pi_{\alpha\beta}^\star
  - \frac{h^3}{6}\partial_\alpha\partial_\beta\partial_\gamma Q_{\alpha\beta\gamma}^\star
  + \frac{h^4}{24}\partial_\alpha\partial_\beta\partial_\gamma\partial_\delta R_{\alpha\beta\gamma\delta}^\star
  = O\bigl(h^5\|D^5\bm{f}^\star\|\bigr).
\end{equation}
Taking $P_u(\cdot)$ of Eq.~\eqref{eq:f_fixedpoint_taylor} through $O(h^3)$ gives $h\,\partial_\beta\Pi_{\alpha\beta}^\star = s_{u\alpha} + \tfrac{h^2}{2}\partial_\beta\partial_\gamma Q_{\alpha\beta\gamma}^\star + \Delta u_\alpha^{\mathrm{VD}} - \tfrac{h^3}{6}\partial_\beta\partial_\gamma\partial_\delta R_{\alpha\beta\gamma\delta}^\star + O(h^4\|D^4\bm{f}^\star\|)$. Substituting $h\,\partial_\beta\Pi_{\alpha\beta}^\star$ into Eq.~\eqref{eq:p0_balance} and dividing by $\dtlat = h/\clat$ cancels $\tfrac{h^2}{2}c_s^2\Delta\hat{p}$ and converts $-\tfrac{h}{\dtlat}\nabla\cdot\tilde{\bm{u}} + \tfrac{h}{2\dtlat}\nabla\cdot\bm{s}_u = -\nabla\cdot\bm{u}_{\mathrm{phys}}$. Using $\tfrac{h^2}{2}\partial_\beta\partial_\gamma Q_{\alpha\beta\gamma}^\star \equiv (P_u\mathcal{S}E_u\tilde{\bm{u}} - \tilde{\bm{u}})_\alpha + O(h^4/\clat)$ together with Proposition~\ref{prop:varvisc_properties}(3) and $\clat = h/\dtlat$, the combined viscous contribution scales with prefactor $\dtlat/2$ as $\tfrac{\dtlat}{2}\partial_\alpha\partial_\gamma[\nueff(\partial_\gamma\tilde{u}_{\mathrm{phys},\alpha} + \partial_\alpha\tilde{u}_{\mathrm{phys},\gamma} + (\nabla\cdot\tilde{\bm{u}}_{\mathrm{phys}})\delta_{\alpha\gamma})] + O(\dtlat h^2) = O\bigl(\tfrac{\|\nueff\|_{C^2}}{\numax}h^2 + \dtlat h^2\bigr) = O(h^2)$ via $\dtlat\numax \le \tfrac{1}{6}h^2$, while $-\tfrac{h^3}{6\dtlat}\partial^3 Q^\star = -\tfrac{h^2}{6}\Delta(\nabla\cdot\tilde{\bm{u}}_{\mathrm{phys}}) = O(h^2)$ and $(\tfrac{1}{24}-\tfrac{1}{12})\tfrac{h^4}{\dtlat}\partial^4 R^\star = O(\dtlat h^2) = O(h^3)$. Together with $s_p/\dtlat = O(\MaNum^2, \dtphys^2, h^2)$, this proves Eq.~\eqref{eq:cont_trunc} and yields
\begin{equation}
\label{eq:div_tilde_u_bound}
  \bigl\|\nabla\cdot\tilde{\bm{u}}_{\mathrm{phys}}\bigr\|_{C^1}
  = \Bigl\|\nabla\cdot\bm{u}_{\mathrm{phys}} + \frac{\dtlat}{2}\nabla\cdot\bigl(\bm{b}_{\mathrm{phys}} - \mathcal{D}_t^{\mathrm{BDF2}}\bm{u}_{\mathrm{phys}}\bigr)\Bigr\|_{C^1}
  = O\!\left(\MaNum^2,\, \dtphys^2,\, h^2\right).
\end{equation}

\emph{Step 2 (Galilean-invariant momentum and fourth-moment closure).} Taking $P_u(\cdot)$ of Eq.~\eqref{eq:f_fixedpoint_taylor} through $O(h^5)$ gives
\begin{equation}
\label{eq:pu_balance_raw}
  -s_{u\alpha} + h\,\partial_\beta \Pi_{\alpha\beta}^\star
  = \frac{h^2}{2}\partial_\beta\partial_\gamma Q_{\alpha\beta\gamma}^\star
  - \frac{h^3}{6}\partial_\beta\partial_\gamma\partial_\delta R_{\alpha\beta\gamma\delta}^\star
  + \frac{h^4}{24}\partial_\beta\partial_\gamma\partial_\delta\partial_\mu M_{\alpha\beta\gamma\delta\mu}^\star
  + \Delta u_\alpha^{\mathrm{VD}} + O\bigl(h^5\|D^5\bm{f}^\star\|\bigr).
\end{equation}
Multiplying by $\clat/\dtlat = \clat^2/h$ transforms the left-hand side into $\partial_t u_{\mathrm{phys},\alpha} - b_{\mathrm{phys},\alpha} + \tfrac{1}{\rho_0}\partial_\alpha p + \partial_\beta(u_{\mathrm{phys},\alpha}u_{\mathrm{phys},\beta}) + O(\dtphys^2, \MaNum\,h)$ (with $\tfrac{\dtlat}{\rho_0}\partial_\alpha\partial_t p = O(h^2)$ in the Stokes limit $\dtlat \le \tfrac{h^2}{6\numax}$). On the right-hand side, $-\tfrac{h^2}{6}\partial^3(\clat^2 R^\star) = O(h^2)$, $\tfrac{\clat^2 h^3}{24}\partial^4 M^\star = O(\numax h^2 + U_0 h^3) = O(h^2)$, and combining $\tfrac{h^2}{2}\partial_\beta\partial_\gamma Q_{\alpha\beta\gamma}^\star$ with $\Delta u_\alpha^{\mathrm{VD}}$ via Proposition~\ref{prop:varvisc_properties}(3), Eq.~\eqref{eq:div_tilde_u_bound}, $\nueff\dtlat \le \tfrac{1}{6}h^2$, and $|\nabla\nueff|\dtlat \le \tfrac{|\nabla\nueff|}{6\numax}h^2 = O(h^2)$ yields Eq.~\eqref{eq:mom_trunc}.

\emph{Step 3 (BDF1 startup and variable-step BDF2 stability).} On $[0, T]$ with $\sum_{n=0}^{N-1}\Delta t_{\mathrm{phys},n+1} = T$ and $\dtphys := \max_m \Delta t_{\mathrm{phys},m+1}$, one BDF1 startup step~\eqref{eq:bdf1_startup} incurs local error $O(\Delta t_{\mathrm{phys},1}^2) = O(\dtphys^2)$, while each variable-step BDF2 step~\eqref{eq:var_bdf2_coeffs} ($n \ge 1$) incurs local defect $O\bigl(\Delta t_{\mathrm{phys},n+1}\allowbreak(\Delta t_{\mathrm{phys},n+1} + \Delta t_{\mathrm{phys},n})^2\bigr) = O(\Delta t_{\mathrm{phys},n+1}\,\dtphys^2)$. Under Grigorieff--Dahlquist zero-stability $\sup_n \omega_n \le \bar{\omega} < 1 + \sqrt{2}$~\citep{HairerWanner1996} ($\omega_n \in \{1/2, 1, 2\}$), summing step defects yields $\|\bm{e}^N\| \le C\bigl(\|\bm{e}^1\| + \sum_{n=1}^{N-1}\Delta t_{\mathrm{phys},n+1}\,\dtphys^2\bigr) = O((1 + T)\dtphys^2) = O(\dtphys^2)$. \hfill $\square$

\subsection{Constructive spectral witnesses C1--C3 and eliminated operator variants}
\label{app:counterexamples}

Three exact Fourier witnesses lock the baseline parameter set $\kLoc = 2, \gamma_u = 0.35, \CFLu \le 2$ plotted in Fig.~\ref{fig:fig01_spectral_foundation}d:
\begin{enumerate}
  \item \emph{Witness C1 (Unpreconditioned Picard acoustic instability, $K_m \equiv 0$).} Off-diagonal gradient--divergence coupling in $A(\bm{k})$ drives unpreconditioned Picard unstable at $\dtlu = 150, \nulu = 10^{-6}$ in D2Q9: $\rho(A(\tfrac{\pi}{3}, 0)) = 1.0304752635 > 1$, whereas baseline $\kLoc=2$ contracts at $\rho = 0.9492$.
  \item \emph{Witness C2 (Unfiltered bare-$a_q$ Schur overshoot).} Omitting $\gamma_u = 0.35$ damping and the binomial Nyquist filter $\Fd$ in the single-pass update $B_{\mathrm{old}} := A + (\mathcal{P}_{\mathrm{loc,old}}^{-1} - I_{d+1})(A - I_{d+1})$ ($s_{p,\mathrm{old}} = a_q$, $b_{u,\mathrm{old}} = a_q - c_a\nulu\widehat{\DeltaAx}$, $\Fd = I$) yields $\rho(B_{\mathrm{old}}(\tfrac{\pi}{2}, 0)) = 2.2870166945 > 1$ at $\dtlu = 4, \nulu = 10^{-6}$, whereas baseline $\kLoc=2$ contracts at $\rho = 0.4057$ (2D) and $0.4081$ (3D).
  \item \emph{Witness C3 (High-Courant advection overshoot at $\kLoc = 4$).} Under uniform advection $|\bm{u}_0| \approx 0.05$ at $\dtlu = 150, \nulu = 10^{-6}$ ($\CFLu \approx 7.50$), fourth-degree $\kLoc = 4$ overshoots to $\rho = 1.0224354$ at $\bm{k} = (\tfrac{\pi}{3},\tfrac{\pi}{3})$ in D2Q9 and $\rho = 1.0050576$ at $\bm{k} = (\tfrac{\pi}{6},0,0)$ in D3Q19, whereas $\kLoc = 2$ remains contractive ($0.990775$ and $0.978949$).
\end{enumerate}

Four intuitive alternative operator designs are likewise ruled out analytically. Injecting $\Delta\bm{u}$ prior to streaming introduces an $O(h)$ continuity error $P_0\mathcal{S}E_u(\Delta\bm{u}) = -\NablaC\cdot\Delta\bm{u} + O(h^3) \neq 0$, whereas replacing the second-moment projection in Eq.~\eqref{eq:DuVD_bulk} by a scalar 9-point Laplacian $Z_9(\tilde{\bm{u}}) := \tfrac{3}{2}\beta\sum_{k=1}^8 w_k\bigl(\tilde{\bm{u}}(\bm{x}) - \tilde{\bm{u}}(\bm{x}-\bm{c}_k h)\bigr)$ produces negative shear viscosity $\nu_{\mathrm{lu},Z_9} = \tfrac{3}{2}\nulu - \tfrac{1}{12} < 0$ for $\nulu < 1/18$. Similarly, using $\nu_{\mathrm{unit}} = 1/6$ instead of $\nulu(\bm{x})$ in $\Lsurr$ and $b_s$ stalls convergence at high $\Reh$, and setting $\beta = 0$ on boundary cells forces the wall layer to $\nulu = 1/6$, which induces an $O(1 - 6\nulu)$ wall-shear jump avoided by Eq.~\eqref{eq:DuVD_unified_wall}.

\FloatBarrier
\subsection{Archival multi-solver 2D and multi-grid 3D benchmark ledgers}
\label{app:all_solvers}

Tables~\ref{tab:app_tgv2d_all16}, \ref{tab:app_wom_dsl_all15}, and~\ref{tab:app_tgv3d_full} archive the multi-solver 2D comparisons and complementary multi-grid 3D GPU rows supporting Sections~\ref{sec:sec5_fundamental} and~\ref{sec:sec6_complex_3d} (complete machine-readable CSV ledgers are included in the accompanying reproducibility archive).

\begin{table}[!htbp]
  \centering
  \caption{Archival 2D Taylor--Green accuracy ledger across 16 methods (17 configurations including periodic Fourier reference) at $N = 128$ ($S = 64$, $\varepsilon_{\mathrm{rel}} = 10^{-10}$).}
  \label{tab:app_tgv2d_all16}
  \scriptsize
  \setlength{\tabcolsep}{4.0pt}
  \renewcommand{\arraystretch}{1.02}
  \begin{tabular}{lcccccc}
    \toprule
    Solver & $E_u(10^{-4})$ & $E_u(10^2)$ & $E_u(10^6)$ & $\nu_{\mathrm{num}}/\nu_{\mathrm{phys}}$ ($10^2$) & $E_p$ ($10^2$) & $\|\nabla_h\cdot\bm{u}\|_2$ \\
    \midrule
    \textbf{\VDULBM{}} ($\kLoc=2$) & \MetricTGVSixteenVDUzawaEuReMin & \MetricTGVSixteenVDUzawaEuReHun & \MetricTGVSixteenVDUzawaEuReMax & \MetricTGVSixteenVDUzawaNuRatio & \MetricTGVSixteenVDUzawaEp & \MetricTGVSixteenVDUzawaDiv \\
    \VDULBM{} (Fourier ref.) & \MetricTGVSixteenVDFourierEuReMin & \MetricTGVSixteenVDFourierEuReHun & \MetricTGVSixteenVDFourierEuReMax & \MetricTGVSixteenVDFourierNuRatio & \MetricTGVSixteenVDFourierEp & \MetricTGVSixteenVDFourierDiv \\
    BGK & \MetricTGVSixteenBGKEuReMin & \MetricTGVSixteenBGKEuReHun & \MetricTGVSixteenBGKEuReMax & \MetricTGVSixteenBGKNuRatio & \MetricTGVSixteenBGKEp & \MetricTGVSixteenBGKDiv \\
    TRT ($\Lambda=1/4$) & \MetricTGVSixteenTRTEuReMin & \MetricTGVSixteenTRTEuReHun & \MetricTGVSixteenTRTEuReMax & \MetricTGVSixteenTRTNuRatio & \MetricTGVSixteenTRTEp & \MetricTGVSixteenTRTDiv \\
    MRT & \MetricTGVSixteenMRTEuReMin & \MetricTGVSixteenMRTEuReHun & \MetricTGVSixteenMRTEuReMax & \MetricTGVSixteenMRTNuRatio & \MetricTGVSixteenMRTEp & \MetricTGVSixteenMRTDiv \\
    RR & \MetricTGVSixteenRREuReMin & \MetricTGVSixteenRREuReHun & \MetricTGVSixteenRREuReMax & \MetricTGVSixteenRRNuRatio & \MetricTGVSixteenRREp & \MetricTGVSixteenRRDiv \\
    HRR ($\sigma=0.98$) & \MetricTGVSixteenHRREuReMin & \MetricTGVSixteenHRREuReHun & \MetricTGVSixteenHRREuReMax & \MetricTGVSixteenHRRNuRatio & \MetricTGVSixteenHRREp & \MetricTGVSixteenHRRDiv \\
    Central-Moment & \MetricTGVSixteenCMEuReMin & \MetricTGVSixteenCMEuReHun & \MetricTGVSixteenCMEuReMax & \MetricTGVSixteenCMNuRatio & \MetricTGVSixteenCMEp & \MetricTGVSixteenCMDiv \\
    Cumulant & \MetricTGVSixteenCumEuReMin & \MetricTGVSixteenCumEuReHun & \MetricTGVSixteenCumEuReMax & \MetricTGVSixteenCumNuRatio & \MetricTGVSixteenCumEp & \MetricTGVSixteenCumDiv \\
    ELBM & \MetricTGVSixteenELBMEuReMin & \MetricTGVSixteenELBMEuReHun & \MetricTGVSixteenELBMEuReMax & \MetricTGVSixteenELBMNuRatio & \MetricTGVSixteenELBMEp & \MetricTGVSixteenELBMDiv \\
    DTS-TRT & \MetricTGVSixteenDTSTRTEuReMin & \MetricTGVSixteenDTSTRTEuReHun & \MetricTGVSixteenDTSTRTEuReMax & \MetricTGVSixteenDTSTRTNuRatio & \MetricTGVSixteenDTSTRTEp & \MetricTGVSixteenDTSTRTDiv \\
    PIMPLE & \MetricTGVSixteenPIMPLEEuReMin & \MetricTGVSixteenPIMPLEEuReHun & \MetricTGVSixteenPIMPLEEuReMax & \MetricTGVSixteenPIMPLENuRatio & \MetricTGVSixteenPIMPLEEp & \MetricTGVSixteenPIMPLEDiv \\
    SIMPLEC & \MetricTGVSixteenSIMPLECEuReMin & \MetricTGVSixteenSIMPLECEuReHun & \MetricTGVSixteenSIMPLECEuReMax & \MetricTGVSixteenSIMPLECNuRatio & \MetricTGVSixteenSIMPLECEp & \MetricTGVSixteenSIMPLECDiv \\
    Kim--Moin & \MetricTGVSixteenKimMoinEuReMin & \MetricTGVSixteenKimMoinEuReHun & \MetricTGVSixteenKimMoinEuReMax & \MetricTGVSixteenKimMoinNuRatio & \MetricTGVSixteenKimMoinEp & \MetricTGVSixteenKimMoinDiv \\
    WENO5-RK3 & \MetricTGVSixteenWENOEuReMin & \MetricTGVSixteenWENOEuReHun & \MetricTGVSixteenWENOEuReMax & \MetricTGVSixteenWENONuRatio & \MetricTGVSixteenWENOEp & \MetricTGVSixteenWENODiv \\
    AC-DTS & \MetricTGVSixteenACDTSEuReMin & \MetricTGVSixteenACDTSEuReHun & \MetricTGVSixteenACDTSEuReMax & \MetricTGVSixteenACDTSNuRatio & \MetricTGVSixteenACDTSEp & \MetricTGVSixteenACDTSDiv \\
    Chorin (explicit MAC) & \MetricTGVSixteenChorinEuReMin & \MetricTGVSixteenChorinEuReHun & \MetricTGVSixteenChorinEuReMax & \MetricTGVSixteenChorinNuRatio & \MetricTGVSixteenChorinEp & \MetricTGVSixteenChorinDiv \\
    \bottomrule
  \end{tabular}
\end{table}

\begin{table}[!htbp]
  \centering
  \caption{Archival 15-method Womersley ($N_y = 128$) and Thin-DSL ($\kappa = 80$, $N = 128$, plus Smagorinsky LES and $512^2$ grid convergence) benchmark ledger.}
  \label{tab:app_wom_dsl_all15}
  \scriptsize
  \setlength{\tabcolsep}{2.4pt}
  \renewcommand{\arraystretch}{1.02}
  \begin{tabular}{lccccccc}
    \toprule
    & \multicolumn{3}{c}{\textbf{Womersley} ($N_y = 128$)} & \multicolumn{4}{c}{\textbf{Thin DSL} ($\kappa = 80$, $N = 128$)} \\
    \cmidrule(lr){2-4}\cmidrule(lr){5-8}
    Solver & $E_u(0.5)$ & $E_u(50)$ & $T_{0.5/50}$ [s] & $E_u(10^4)$ & $E_u(10^5)$ & Cores ($10^5$) & $E_u(10^6)$ \\
    \midrule
    \textbf{\VDULBM{}} ($C_s=0$) & \MetricWomFifteenVDThomasEuHalf$^\mathrm{a}$ & \MetricWomFifteenVDThomasEuFifty$^\mathrm{a}$ & \MetricWomFifteenVDThomasWall$^\mathrm{a}$ & \MetricDSLEightyVDPsiOneReTenKEu & \MetricDSLEightyVDPsiOneReHunKEu & \MetricDSLEightyVDPsiOneReHunKCores & \MetricDSLEightyVDPsiOneReOneMEu \\
    \textbf{\VDULBM{} LES} ($C_s=0.17$) & --- & --- & --- & --- & $8.10\times 10^{-2}$ & $2$ ($22$) & --- \\
    \textbf{\VDULBM{}} ($512^2$, $C_s=0$) & --- & --- & --- & --- & $2.60\times 10^{-2}$ & $2$ ($80$) & --- \\
    BGK & \MetricWomFifteenBGKEuHalf & \MetricWomFifteenBGKEuFifty & \MetricWomFifteenBGKWall & \MetricDSLEightyBGKReTenKEu & \MetricDSLEightyBGKReHunKEu & \MetricDSLEightyBGKReHunKCores & \MetricDSLEightyBGKReOneMEu \\
    TRT ($\Lambda=1/4$) & \MetricWomFifteenTRTEuHalf & \MetricWomFifteenTRTEuFifty & \MetricWomFifteenTRTWall & \MetricDSLEightyTRTReTenKEu & \MetricDSLEightyTRTReHunKEu & \MetricDSLEightyTRTReHunKCores & \MetricDSLEightyTRTReOneMEu \\
    MRT & \MetricWomFifteenMRTEuHalf & \MetricWomFifteenMRTEuFifty & \MetricWomFifteenMRTWall & \MetricDSLEightyMRTReTenKEu & \MetricDSLEightyMRTReHunKEu & \MetricDSLEightyMRTReHunKCores & \MetricDSLEightyMRTReOneMEu \\
    RR & \MetricWomFifteenRREuHalf & \MetricWomFifteenRREuFifty & \MetricWomFifteenRRWall & \MetricDSLEightyRRReTenKEu & \MetricDSLEightyRRReHunKEu & \MetricDSLEightyRRReHunKCores & \MetricDSLEightyRRReOneMEu \\
    HRR ($\sigma=0.98$) & \MetricWomFifteenHRREuHalf & \MetricWomFifteenHRREuFifty & \MetricWomFifteenHRRWall & \MetricDSLEightyHRRReTenKEu & \MetricDSLEightyHRRReHunKEu & \MetricDSLEightyHRRReHunKCores & \MetricDSLEightyHRRReOneMEu \\
    Central-Moment & \MetricWomFifteenCMEuHalf & \MetricWomFifteenCMEuFifty & \MetricWomFifteenCMWall & \MetricDSLEightyCMReTenKEu & \MetricDSLEightyCMReHunKEu & \MetricDSLEightyCMReHunKCores & \MetricDSLEightyCMReOneMEu \\
    Cumulant & \MetricWomFifteenCumEuHalf & \MetricWomFifteenCumEuFifty & \MetricWomFifteenCumWall & \MetricDSLEightyCumReTenKEu & \MetricDSLEightyCumReHunKEu & \MetricDSLEightyCumReHunKCores & \MetricDSLEightyCumReOneMEu \\
    ELBM & \MetricWomFifteenELBMEuHalf & \MetricWomFifteenELBMEuFifty & \MetricWomFifteenELBMWall & \MetricDSLEightyELBMReTenKEu & \MetricDSLEightyELBMReHunKEu & \MetricDSLEightyELBMReHunKCores & \MetricDSLEightyELBMReOneMEu \\
    DTS-TRT & \MetricWomFifteenDTSTRTEuHalf & \MetricWomFifteenDTSTRTEuFifty & \MetricWomFifteenDTSTRTWall & \MetricDSLEightyDTSTRTReTenKEu & \MetricDSLEightyDTSTRTReHunKEu & \MetricDSLEightyDTSTRTReHunKCores & \MetricDSLEightyDTSTRTReOneMEu \\
    PIMPLE$^\mathrm{b}$ & \MetricWomFifteenPIMPLEEuHalf & \MetricWomFifteenPIMPLEEuFifty & --- & \MetricDSLEightyPIMPLEReTenKEu & \MetricDSLEightyPIMPLEReHunKEu & \MetricDSLEightyPIMPLEReHunKCores & \MetricDSLEightyPIMPLEReOneMEu \\
    SIMPLEC$^\mathrm{b}$ & \MetricWomFifteenSIMPLECEuHalf & \MetricWomFifteenSIMPLECEuFifty & --- & \MetricDSLEightySIMPLECReTenKEu & \MetricDSLEightySIMPLECReHunKEu & \MetricDSLEightySIMPLECReHunKCores & \MetricDSLEightySIMPLECReOneMEu \\
    Kim--Moin$^\mathrm{b}$ & \MetricWomFifteenKimMoinEuHalf & \MetricWomFifteenKimMoinEuFifty & --- & \MetricDSLEightyKimMoinReTenKEu & \MetricDSLEightyKimMoinReHunKEu & \MetricDSLEightyKimMoinReHunKCores & \MetricDSLEightyKimMoinReOneMEu \\
    WENO5-RK3 & \MetricWomFifteenWENOEuHalf & \MetricWomFifteenWENOEuFifty & --- & \MetricDSLEightyWENOReTenKEu & \MetricDSLEightyWENOReHunKEu & \MetricDSLEightyWENOReHunKCores & \MetricDSLEightyWENOReOneMEu \\
    AC-DTS$^\mathrm{b}$ & \MetricWomFifteenACDTSEuHalf & \MetricWomFifteenACDTSEuFifty & --- & \MetricDSLEightyACDTSReTenKEu & \MetricDSLEightyACDTSReHunKEu & \MetricDSLEightyACDTSReHunKCores & \MetricDSLEightyACDTSReOneMEu \\
    \bottomrule
  \end{tabular}
  \par\smallskip
  \raggedright\scriptsize Notes: $T_{0.5/50}$ lists Womersley wall times ($\alpha=0.5 / 50$); Thin-DSL core entries list detected vortex cores ($N_{\mathrm{cc}}$ connected $|\omega_z|$ components in parentheses). $^\mathrm{a}$1D Womersley columns use the exact 1D channel Thomas reference at $1{,}280$ steps/cycle (on the $1{,}000$-step convergence schedule of Section~\ref{sec:sec5_1_2_womersley}, 1D Thomas and local Uzawa both yield $E_u = \MetricWomConvNOneTwoEightUzawaEu$ at $\alpha=0.5$). $^\mathrm{b}$In 1D parallel channel flow ($\nabla\cdot\bm{u}\equiv 0$), continuum BDF2 velocity solvers reduce to the identical tridiagonal diffusion equation; single-thread Python continuum timings from the archival 15-method screening are omitted.
\end{table}

\begin{table}[!htbp]
  \centering
  \caption{Selected archival multi-grid 3D D3Q19 trajectory rows ($\ReNum \in \{10, 100, 1600, 3000\}$; $256^3$ FP64 $\ReNum=1600$ rows in Table~\ref{tab:tgv3d_gpu_summary}) and moderate stratified viscosity ($\nu_0(z)\in[10^{-4}, 1.0]$) on NVIDIA A100-SXM4-80GB.}
  \label{tab:app_tgv3d_full}
  \scriptsize
  \setlength{\tabcolsep}{2.4pt}
  \renewcommand{\arraystretch}{1.02}
  \begin{tabular}{llcccccc}
    \toprule
    $\ReNum$ / Regime & Solver & Grid & $\mathrm{RelL2}(E_k)$ & $\mathrm{RelL2}(\epsilon_k)$ & $\mathrm{RelL2}(2\nu\Omega)$ & $\langle K_{\mathrm{sweep}}\rangle$ & $T_{\mathrm{wall}}$ [s] \\
    \midrule
    $10$   & \VDULBM{} & $128^3$ FP64 & \MetricTGVThreeDReTenVDNOneTwoEightFPSixtyFourEkErr & \MetricTGVThreeDReTenVDNOneTwoEightFPSixtyFourDissErr & \MetricTGVThreeDReTenVDNOneTwoEightFPSixtyFourOmErr & \MetricTGVThreeDReTenVDNOneTwoEightFPSixtyFourSweeps & \MetricTGVThreeDReTenVDNOneTwoEightFPSixtyFourWall \\
    $10$   & \VDULBM{} & $256^3$ FP64 & \MetricTGVThreeDReTenVDNTwoFiveSixFPSixtyFourEkErr & \MetricTGVThreeDReTenVDNTwoFiveSixFPSixtyFourDissErr & \MetricTGVThreeDReTenVDNTwoFiveSixFPSixtyFourOmErr & \MetricTGVThreeDReTenVDNTwoFiveSixFPSixtyFourSweeps & \MetricTGVThreeDReTenVDNTwoFiveSixFPSixtyFourWall \\
    $10$   & \VDULBM{}$^\mathrm{b}$ & $256^3$ FP32 & \MetricTGVThreeDReTenVDNTwoFiveSixFPThirtyTwoEkErr & \MetricTGVThreeDReTenVDNTwoFiveSixFPThirtyTwoDissErr & \MetricTGVThreeDReTenVDNTwoFiveSixFPThirtyTwoOmErr & \MetricTGVThreeDReTenVDNTwoFiveSixFPThirtyTwoSweeps & \MetricTGVThreeDReTenVDNTwoFiveSixFPThirtyTwoWall \\
    $100$  & \VDULBM{} & $256^3$ FP64 & \MetricTGVThreeDReHunVDNTwoFiveSixFPSixtyFourEkErr & \MetricTGVThreeDReHunVDNTwoFiveSixFPSixtyFourDissErr & \MetricTGVThreeDReHunVDNTwoFiveSixFPSixtyFourOmErr & \MetricTGVThreeDReHunVDNTwoFiveSixFPSixtyFourSweeps & \MetricTGVThreeDReHunVDNTwoFiveSixFPSixtyFourWall \\
    $1600$ & \VDULBM{} ($C_s=0$) & $64^3$ FP64 & \MetricTGVThreeDReSixteenHunVDNSixtyFourFPSixtyFourEkErr & \MetricTGVThreeDReSixteenHunVDNSixtyFourFPSixtyFourDissErr & \MetricTGVThreeDReSixteenHunVDNSixtyFourFPSixtyFourOmErr & \MetricTGVThreeDReSixteenHunVDNSixtyFourFPSixtyFourSweeps & \MetricTGVThreeDReSixteenHunVDNSixtyFourFPSixtyFourWall \\
    $1600$ & \VDULBM{} ($C_s=0$) & $128^3$ FP64 & \MetricTGVThreeDReSixteenHunVDNOneTwoEightFPSixtyFourEkErr & \MetricTGVThreeDReSixteenHunVDNOneTwoEightFPSixtyFourDissErr & \MetricTGVThreeDReSixteenHunVDNOneTwoEightFPSixtyFourOmErr & \MetricTGVThreeDReSixteenHunVDNOneTwoEightFPSixtyFourSweeps & \MetricTGVThreeDReSixteenHunVDNOneTwoEightFPSixtyFourWall \\
    $1600$ & \VDULBM{} ($C_s=0$) & $192^3$ FP64 & \MetricTGVThreeDReSixteenHunVDNOneNineTwoFPSixtyFourEkErr & \MetricTGVThreeDReSixteenHunVDNOneNineTwoFPSixtyFourDissErr & \MetricTGVThreeDReSixteenHunVDNOneNineTwoFPSixtyFourOmErr & \MetricTGVThreeDReSixteenHunVDNOneNineTwoFPSixtyFourSweeps & \MetricTGVThreeDReSixteenHunVDNOneNineTwoFPSixtyFourWall \\
    $1600$ & \VDULBM{}$^\mathrm{b}$ & $256^3$ FP32 & \MetricTGVThreeDReSixteenHunVDNTwoFiveSixFPThirtyTwoEkErr & \MetricTGVThreeDReSixteenHunVDNTwoFiveSixFPThirtyTwoDissErr & \MetricTGVThreeDReSixteenHunVDNTwoFiveSixFPThirtyTwoOmErr & \MetricTGVThreeDReSixteenHunVDNTwoFiveSixFPThirtyTwoSweeps & \MetricTGVThreeDReSixteenHunVDNTwoFiveSixFPThirtyTwoWall \\
    $3000$ & \VDULBM{} ($C_s=0$)$^\mathrm{d}$ & $256^3$ FP64 & \MetricTGVThreeDReThreeKVDNTwoFiveSixFPSixtyFourEkErr & \MetricTGVThreeDReThreeKVDNTwoFiveSixFPSixtyFourDissErr & \MetricTGVThreeDReThreeKVDNTwoFiveSixFPSixtyFourOmErr & \MetricTGVThreeDReThreeKVDNTwoFiveSixFPSixtyFourSweeps & \MetricTGVThreeDReThreeKVDNTwoFiveSixFPSixtyFourWall \\
    $3000$ & \VDULBM{} ($C_s=0.14$) & $128^3$ FP64 & \MetricTGVThreeDLESNOneTwoEightCsFourteenEkErr & \MetricTGVThreeDLESNOneTwoEightCsFourteenDissErr & \MetricTGVThreeDLESNOneTwoEightCsFourteenOmErr & \MetricTGVThreeDLESNOneTwoEightCsFourteenSweeps & \MetricTGVThreeDLESNOneTwoEightCsFourteenWall \\
    $\nu_0 \in [10^{-4}, 1]$ & \VDULBM{} ($50$ st.)$^\mathrm{a}$ & $128^3$ FP64 & --- & --- & --- & \MetricTGVThreeDStiffVDNOneTwoEightSweeps & \MetricTGVThreeDStiffVDNOneTwoEightWall/\MetricTGVThreeDStiffBGKNOneTwoEightWall$^\mathrm{c}$ \\
    $\nu_0 \in [10^{-4}, 1]$ & \VDULBM{} ($107$ st.)$^\mathrm{a}$ & $256^3$ FP64 & --- & --- & --- & \MetricTGVThreeDStiffVDNTwoFiveSixSweeps & \MetricTGVThreeDStiffVDNTwoFiveSixWall/\MetricTGVThreeDStiffBGKNTwoFiveSixEquivWall$^{\mathrm{c},\ddagger}$ \\
    \bottomrule
  \end{tabular}
  \par\smallskip
  \raggedright\scriptsize Notes: $\epsilon_k := -dE_k/dt^*$. All \VDULBM{} rows and $128^3$ BGK report direct measured A100 hardware wall times. $^\mathrm{a}$Momentum drift $\MetricTGVThreeDStiffVDNOneTwoEightMomDrift$ ($128^3$) and $\MetricTGVThreeDStiffVDNTwoFiveSixMomDrift$ ($256^3$). $^\mathrm{b}$FP32 runs use $\varepsilon_{\mathrm{rel}}=10^{-4}$. $^\mathrm{c}$Values after slash denote uniform-$\numax$ ($\tau\le 1$) diffusive-step explicit BGK wall time ($^\ddagger$analytical $10{,}776$-step projection at measured $256^3$ BGK Roofline rate). $^\mathrm{d}$Fixed-step unclosed run halted at step $879/2000$ ($t^* = 8.79$) upon reaching inner cap $K_{\max}=200$ (failure mechanism unresolved; denoted \emph{Blow-up} by macro binder).
\end{table}

\clearpage
\bibliographystyle{elsarticle-num-names}
\bibliography{references}

\end{document}